\documentclass[aps,prb,twocolumn,noshowpacs,superscriptaddress, floatfix, nofootinbib, letterpaper]{revtex4} 

\usepackage{amsfonts}
\usepackage{subfigure}
\usepackage{latexsym}
\usepackage{datetime}

\usepackage{mathrsfs}

\usepackage{graphicx}
\usepackage{amsmath}
\usepackage{amsthm}
\usepackage{amstext}
\usepackage{amssymb}
\usepackage{mathrsfs}
\usepackage{amsfonts}
\usepackage{amsbsy} 
\usepackage{dcolumn}
\usepackage{bm}
\usepackage{multirow}

\usepackage[usenames, dvipsnames]{color}
\usepackage[svgnames]{xcolor}
\usepackage[colorlinks,citecolor=RoyalBlue, urlcolor=RoyalBlue, linkcolor=RoyalBlue ]{hyperref} 

\usepackage{verbatim} 

\newcommand{\be}{\begin{equation}}
\newcommand{\ee}{\end{equation}}
\newcommand{\bea}{\begin{eqnarray}}
\newcommand{\eea}{\end{eqnarray}}
\newcommand{\barr}{\begin{array}}
\newcommand{\earr}{\end{array}}

\long\def\begincomment#1\endcomment{}

\newcommand{\diag}{\mathop{\mathrm{diag}}}

\newcommand{\GSD}{\mathop{\mathrm{GSD}}}

\newcommand{\E}{\mathop{\mathrm{E}}}
\newcommand{\U}{\text{U}}
\newcommand{\A}{\mathop{\mathrm{A}}}
\newcommand{\B}{\mathop{\mathrm{B}}}
\newcommand{\tr}{\mathop{\mathrm{tr}}}
\newcommand{\Hom}{\mathop{\mathrm{Hom}}}

\newcommand{\Z}{\mathbb{Z}}
\newcommand{\C}{\mathbb{C}}
\newcommand{\bfW}{\mathbf{W}}
\newcommand{\bfV}{\mathbf{V}}
\newcommand{\cM}{\mathcal{M}}

\newcommand{\ii}{\mathrm{i}}
\newcommand{\dd}{\mathrm{d}}

\newcommand{\frm}[1]{\begin{center}\fbox{\parbox{3.3in}{\parindent=0pt #1}}\end{center}}

\newcommand{\ti}{\text{i}}
\newcommand{\tF}{\text{F}}
\newcommand{\tG}{\text{G}}

\usepackage{soul}

\usepackage{tikz}

\usepackage{enumitem,xcolor,cleveref}

\usetikzlibrary{matrix,shapes,arrows,positioning,chains, calc, decorations.pathmorphing,decorations.markings}

\usepackage{pifont}
\newcommand{\cmark}{\ding{51}}%

\newcommand{\Refe}[1]{Ref.~\onlinecite{#1}}
\newcommand{\Fig}[1]{Fig.~\ref{#1}}

\begin{document}

\title{
{\fontsize{11}{12}\selectfont{Non-Perturbative Regularization of 
1+1D Anomaly-Free Chiral Fermions and Bosons:}}\\[2mm]
{\fontsize{11}{11}\selectfont{On the equivalence of anomaly matching conditions and boundary 
gapping rules}} 
}


\author{Juven Wang}  \email{juven@ias.edu}  
\affiliation{Department of Physics, Massachusetts Institute of Technology, Cambridge, MA 02139, USA}
\affiliation{Perimeter Institute for Theoretical Physics, Waterloo, ON, N2L 2Y5, Canada}
\affiliation{School of Natural Sciences, Institute for Advanced Study, Einstein Drive, Princeton, NJ 08540, USA}
\affiliation{Center of Mathematical Sciences and Applications, Harvard University, MA 02138, USA}
\author{Xiao-Gang Wen}  \email{xgwen@mit.edu} 
\affiliation{Perimeter Institute for Theoretical Physics, Waterloo, ON, N2L 2Y5, Canada}
\affiliation{Department of Physics, Massachusetts Institute of Technology, Cambridge, MA 02139, USA}
\affiliation{Institute for Advanced Study, Tsinghua University, Beijing,
100084, P. R. China}




\begin{abstract} A non-perturbative lattice regularization of chiral fermions and bosons 
with anomaly-free chiral symmetry $G$ in 1+1D spacetime is proposed.  
More precisely,
we ask ``whether there is a \emph{local} \emph{short-range} quantum Hamiltonian with a \emph{finite} Hilbert space for a \emph{finite} system 
 realizing \emph{onsite symmetry} $G$ defined on a 1D spatial lattice with continuous time, such that its low energy physics produces a 1+1D anomaly-free chiral matter theory of symmetry $G$?''
In particular, 
we propose that the
chiral fermion theory with chiral U(1) 
3$_L$-5$_R$-4$_L$-0$_R$ symmetry, with two left-moving
fermions of charge-3 and 4, and two right-moving fermions of charge-5
and 0 at IR low energy, can emerge from a 1D UV spatial lattice with a chiral U(1) symmetry, if we include \emph{properly designed}
multi-fermion interactions with intermediate strength (i.e., the dimensionless coupling constant is naturally order 1). 
In general, we propose that any 1+1D U(1)-anomaly-free chiral matter theory 
can be defined as a finite
system on a 1D lattice with onsite symmetry 
by using a quantum Hamiltonian with continuous time,
but without suffering from Nielsen-Ninomiya theorem's fermion-doubling,
if we include properly designed 
interactions between matter fields. 
We propose how to design such interactions by looking for extra symmetries via bosonization/fermionization. 
We comment on the new ingredients and the differences of ours  
compared to Ginsparg-Wilson fermion, Eichten-Preskill and Chen-Giedt-Poppitz (CGP) models, and suggest modifying CGP model to have successful mirror-decoupling.  
{Since a lattice onsite internal symmetry can be 
gauged, we thus can further define a non-perturbative regularization of any anomaly-free U(1) chiral gauge theory in 1+1D.}
As an additional remark, we show 
a topological non-perturbative proof of 
the equivalence relation between the $G$-symmetric
't Hooft anomaly cancellation conditions and the $G$-symmetry-preserving 
gapping rules (e.g. Haldane's stability conditions for Luttinger liquid) for multiple U(1) symmetries.
We 
expect
that our result holds universally regardless of spatial Hamiltonian or spacetime Lagrangian/path integral formulation of quantum theory. 
Numerical tests on our proposal are demanding tasks but highly desirable for future work.
%

\end{abstract}
\maketitle

\tableofcontents

\section{Introduction and Summary}

Regulating and defining a chiral fermion field theory is a very important
problem, since the Standard Model is one such
theory with the parity symmetry maximally violated in the weak force.\cite{Lee:1956qn,
Glashow:1961tr, Salam:1964ry, Weinberg:1967tq,
Donoghue:1992dd} However, the Nielsen-Ninomiya's fermion-doubling
problem\cite{Nielsen:1980rz,Nielsen:1981xu,Nielsen:1981hk,Luscher:2000hn,Kaplan:2009yg}
makes it very difficult to define chiral fermions (in an even-dimensional
spacetime) on the lattice.  
Many previous research works attempt to solve this important problem:
These include the standard lattice gauge theory method,\cite{K7959} 
the domain-wall fermion,\cite{K9242,S9390} 
and the overlap fermion.\cite{L9995,N0103,S9947,Luscher:2000hn}


There is also the mirror fermion approach\cite{EP8679,M9259,BMP0628,GP0776} which starts with a lattice
model containing chiral fermions in one original \emph{light sector} coupled to gauge theory, \emph{and} its chiral
conjugated as the \emph{mirror sector}.  Then, one tries to include direct interactions or
boson mediated interactions\cite{S8631,S9252} between fermions to gap out the
mirror sector only.  
The later works either end up with unsuccessful attempts 
\cite{GPR9396,L9418,CGP1247} or argue that it is almost impossible to gap out
(i.e. fully open the energy gaps of) the mirror sector without  breaking the
gauge symmetry in some mirror fermion models.\cite{BD9216}

The previous 
unsuccessful lattice-gauge approaches may  
either (i) assume
non-interacting lattice fermions, apart from the interaction to the lattice
gauge field, in the conventional standard lattice gauge theory, 
or (ii) introduce certain inappropriate 
interactions between fermions that cause their attempt to gapping the mirror gapless modes failed,
such as the Wilson-Yukawa approach from 
Ref.~\onlinecite{S9252, S8631, Smit:1989tz} and later works.\cite{Bock:1992gp}

In this work, we \emph{propose} that the lattice 
mirror fermion approach actually works if we include properly-designed direct fermion-fermion interaction with appropriate intermediate strength 
({\it i.e.} the dimensionless coupling constants are of an order
1\cite{Wen:2013oza,Wen:2013ppa}).\footnote{In the present work,
we do not yet have explicit numerical results to show the gapping of mirror fermion sectors for the interacting system.
We warn the readers that a real numerical test for our interacting system still is desirable for future works.
The numerical simulations for the interacting system that we proposed is a highly non-trivial and very
demanding task. 
We provide evidence but 
do not have a rigorous proof that our proposed lattice model will lead to the proposed continuum field theory.
We also do not yet know the phase diagram structure by tuning the interaction strength, it is a much challenging question for the future.\\[2mm]
After the completion of this work, Ref.~\onlinecite{PomataWangWei2018, ZengZhuWangYou2202.12355} attempt to numerically simulate the 
1+1D U(1) 3$_L$-5$_R$-4$_L$-0$_R$ chiral fermion lattice model proposed in this article.
Ref.~\onlinecite{PomataWangWei2018} finds some evidence of numerical simulations supporting our model.
Ref.~\onlinecite{ZengZhuWangYou2202.12355} 
is able to show the successful gapping of the mirror fermion and their decoupling from the gapless chiral fermion,
thus this numerical evidence supports our lattice construction of the 1+1D chiral fermion model.
} 

In other words, a general framework of the
mirror fermion approach actually works for constructing a lattice chiral
fermion theory, at least in 1+1D, as long as we design proper interactions between gapless modes.
Specifically, any anomaly-free chiral
fermion/boson field theory can be defined as a finite quantum system on a 1D
lattice where the internal chiral symmetry is realized as an onsite global symmetry,
provided that we allow lattice fermion/boson to have {\it interactions}, {\it
instead of being free}.  (Here, the ``chiral'' theory here means that it
``breaks parity $P$ symmetry.'' Our 1+1D chiral fermion theory breaks parity
$P$ and time reversal $T$ symmetry. See Appendix \ref{appendixA} for $C,P,T$
symmetry in 1+1D.) Our insight comes from
Ref.\,\onlinecite{Wen:2013oza,Wen:2013ppa}, where the connection between 
(1) 't Hooft anomalies of global symmetry $G$,\cite{'tHooft:1979bh}
(2) dynamical gauge anomalies by gauging $G$, 
and (3) symmetry-protected topological states (SPTs or SPT order) protected by $G$ \cite{Chen:2011pg} (in
one-higher dimension) is found.



To welcome our readers fully appreciate our logic, we shall first define our important basic concepts clearly:\\
%

\noindent
$(\diamond 1)$ \emph{Onsite symmetry}\cite{Chen:2011pg,2011PhRvB..84w5141C}
means that the overall symmetry transformation operator $U(g)$ of symmetry group $G$ can be defined as the tensor
product of each single site's symmetry transformation $U_i(g)$, via $U(g)= \otimes _i U_i(g)$ with $g \in G$.
\emph{Nonsite symmetry}: means $U(g)_{\text{non-onsite}} \neq \otimes _i U_i(g)$.\\ 

\noindent
$(\diamond 2)$ \emph{Local Hamiltonian with short-range interactions} 
means that the non-zero amplitude of matter (fermion/boson) hopping/interactions in finite time 
has a \emph{finite} range propagation, and cannot be an \emph{infinite} range. 
Strictly speaking, the quasi-local \emph{exponential decay} (of kinetic hopping/interactions) is \emph{non-local} and \emph{not short-range}.\\

\noindent
$(\diamond 3)$ \emph{finite(-Hilbert-space) system} means that the dimension of Hilbert space is finite 
if the system has finite lattice sites (e.g. on a finite-size cylinder).\\

Nielsen-Ninomiya theorem\cite{Nielsen:1980rz,Nielsen:1981xu,Nielsen:1981hk}
states that the attempt to regularize chiral fermion on a lattice
as a local \emph{free non-interacting} fermion model with fermion number conservation
({\it i.e.} with U(1) symmetry\cite{U(1)sym}) 
has fermion-doubling problem\cite{Nielsen:1980rz,Nielsen:1981xu,Nielsen:1981hk,Luscher:2000hn,Kaplan:2009yg} in an even-dimensional spacetime.
To apply this no-go theorem, however, the symmetry 
is assumed to be an onsite symmetry. 

Ginsparg-Wilson
fermion approach 
copes with this
no-go theorem by solving
Ginsparg-Wilson (GW) relation\cite{Ginsparg:1981bj, {Wilson:1974sk}} based on the quasi-local Neuberger-Dirac 
operator,\cite{Neuberger:1997fp,Neuberger:1998wv,Hernandez:1998et} 
where \emph{quasi-local is strictly non-local}.
In this work, we show 
that the quasi-localness of Neuberger-Dirac 
operator in the GW fermion approach imposes a \emph{non-onsite}\cite{Chen:2011pg,Chen:2012hc,Santos:2013uda} U(1) symmetry, instead of an onsite 
symmetry. 
(While here we simply summarize the result, one can read the details of onsite and non-onsite symmetry, and its relation to GW fermion in Appendix \ref{appendixB}.)
For our specific approach for the mirror-fermion decoupling, we
\emph{will not} implement the GW fermions (of non-onsite symmetry) construction,
instead, we will use lattice fermions with onsite symmetry but with particular properly-designed interactions. 
Comparing GW fermion to our approach, we see that
\begin{itemize}
\item {\bf Ginsparg-Wilson (GW) fermion approach} obtains ``{\it chiral
fermions from a local free fermion lattice model with non-onsite $\U(1)$ symmetry} ({\it without fermion doublers}).''
(Here one regards Ginsparg-Wilson fermion applying the Neuberger-Dirac operator, which is strictly non-onsite and non-local.)

\item {\bf Our approach} obtains ``{\it chiral fermions from  local interacting
fermion lattice model with onsite $U(1)$ symmetry} ({\it without fermion doublers}, {\it or gapping fermion doublers}),
{\it if and only if all $\U(1)$ 
anomalies are cancelled}.''
\end{itemize}

Also, the conventional GW fermion approach discretizes the
Lagrangian/the action on the spacetime lattice, while  we use a local short-range quantum
Hamiltonian on 1D spatial lattice with a continuous time. Such a
distinction causes some difference.  For example, it is known that
Ginsparg-Wilson fermion \emph{can} implement a single Weyl fermion for the free case
without gauge field on a 1+1D space-time-lattice due to the works of Neuberger,
L\"uscher, etc.  
Our approach \emph{cannot} implement a single Weyl fermion on a 1D space-lattice within local short-range Hamiltonian. 
(However, only if we are allowed to introduce a non-local infinite-range hopping Hamiltonian term,
our approach can implement a single Weyl fermion.)\\ 

\color{black}

\noindent
{{\bf Comparison to Eichten-Preskill and Chen-Giedt-Poppitz models}}:
Due to the past 
investigations, a majority of the high-energy lattice
community believes that the mirror-fermion decoupling (or lattice gauge
approach) fails to realize chiral fermion or chiral gauge theory.  
Thus one may challenge our approach by asking ``how is our mirror-fermion
decoupling model 
different from Eichten-Preskill
and Chen-Giedt-Poppitz models?''  And ``why does the recent numerical attempt of
Chen-Giedt-Poppitz fail?\cite{CGP1247}''
We stress again that, 
our approach provides
\emph{properly designed fermion interaction terms} to make things work: gapping the mirror-world chiral fermions, 
due to the recent understanding to {\bf topological gapped boundary conditions}\cite{{h95},{Kapustin:2010hk},{Wang:2012am},{Levin:2013gaa}}:
\begin{itemize}
\item {\bf Eichten-Preskill (EP)}\cite{EP8679} proposes a generic idea of the
mirror-fermion approach for the chiral gauge theory.  There the
\emph{perturbative} analysis on the \emph{weak-coupling and strong-coupling}
expansions are used to demonstrate whether mirror-fermion decoupling phases can
exist in the phase diagram.  The action is discretized on the spacetime
lattice.  In EP 
approach, one tries to gap out the
mirror-fermions via the mass term of composite fermions that do not break the
(gauge) symmetry on lattice.  The  mass term of composite fermions are actually
fermion interacting terms.  So in EP 
approach, one tries to gap
out the mirror-fermions via the direct fermion interaction that do not  break
the (gauge) symmetry on lattice.  
However, \emph{including all the symmetry-preserving interactions} (\emph{or symmetric interactions}) 
\emph{may not be compatible with gapping mirror sectors}. 
Even when the mirror sector is anomalous, one can
still add the direct fermion interaction that do not  break the (gauge)
symmetry.  So the presence of symmetric direct fermion interaction may or may
not be able to gap out the  mirror sector.  When the mirror sector is
anomaly-free, we will show in our work, some symmetric interactions are
\emph{helpful} for gapping out the mirror sectors, while other symmetric interactions
are \emph{harmful}. The key issue for us is to design the proper interaction to gap out
the mirror sector.

\item {\bf Chen-Giedt-Poppitz (CGP)}\cite{CGP1247} follows the EP
general framework to deal with a 3-4-5 anomaly-free model with a single U(1)
symmetry.  All the U(1) symmetry-allowed Yukawa-Higgs terms are introduced to
mediate multi-fermion interactions.  The Ginsparg-Wilson fermion and the
Neuberger's overlap Dirac operator are implemented, the fermion actions are
discretized on the spacetime lattice.  Again, the interaction terms are
designed only based on symmetry, which contain both helpful and harmful
terms toward gapping mirror fermions, as we will show.

\item {\bf Our model} in general 
belongs to the
mirror-fermion-decoupling idea.  The anomaly-free model we proposed is named as
the 3$_L$-5$_R$-4$_L$-0$_R$ model.  Our 3$_L$-5$_R$-4$_L$-0$_R$ is in reality
different from Chen-Giedt-Poppitz's 3-4-5 model, since we implement:\\

\noindent
\underline{(i) {\bf an onsite-symmetry local lattice model}}: Our lattice Hamiltonian
is built on 1D spatial lattice with \emph{on-site} U(1) symmetry. 
We \emph{neither} implement the GW 
fermion \emph{nor} the
Neuberger-Dirac operator (both have strictly non-onsite and non-local symmetries).
\\

\noindent
\underline{(ii) {\bf a particular set of interaction terms}}  \underline{ \bf{with proper strength}}:
Our multi-fermion interaction terms are properly-designed gapping terms
that obey not only the symmetry but also certain Lagrangian subgroup algebra. 
Those interaction terms are called \emph{helpful} gapping terms, satisfying {\bf Boundary Fully Gapping Rules}.
We will show that the Chen-Giedt-Poppitz's Yukawa-Higgs terms induce extra 
multi-fermion interaction terms which \emph{do not} satisfy
{\bf Boundary Fully Gapping Rules}. 
Those extra terms are
incompatible \emph{harmful} terms,
competing with the \emph{helpful} gapping terms and 
causing the preformed energy gap (which is not a usual quadratic mass gap)
unstable so preventing the mirror sector from being gapped out.
(This can be one of the reasons for the failure of mirror-decoupling in Ref.\onlinecite{CGP1247}.)
We stress that, due to a \emph{topological non-perturbative} reason, 
only a particular set of ideal interaction terms are helpful
to fully gap the mirror sector. Adding more or removing interactions can cause
the energy gap unstable thus the phase flowing to gapless states.
In addition, we stress that only when the helpful interaction terms are in a proper range, \emph{intermediate strength} for dimensionless coupling of order 1,
can they fully gap the mirror sector, and yet not gap the original sector (details in Sec.\ref{sec:mapping-strong-gapping}).
Throughout our work, when we say strong coupling for our model, we really mean intermediate(-strong) coupling in an appropriate range.
In CGP model, however, their strong coupling may be \emph{too strong} (with their kinetic term neglected);
which can be another reason for the failure of mirror-decoupling.\cite{CGP1247}

\underline{(iii) {\bf extra symmetries}}: For our model, a total even number $N$
of left/right moving Weyl fermions ($N_L=N_R=N/2$), we will add only $N/2$
linear-independent helpful gapping terms under the constraint of the Lagrangian subgroup algebra and {\bf
Boundary Fully Gapping Rules}.  As a result, the full symmetry of our lattice model is
U(1)$^{N/2}$ (where the gapping terms break U(1)$^{N}$ down to U(1)$^{N/2}$).
For the case of our 3$_L$-5$_R$-4$_L$-0$_R$ model, the full U(1)$^2$ symmetry
has two sets of U(1) charges, $\U(1)_{\text{1st}}$ 3-5-4-0 and $\U(1)_{\text{2nd}}$ 0-4-5-3, both are anomaly-free and 
mixed-anomaly-free.  Although the physical consideration
only requires the interaction terms to have
on-site $\U(1)_{\text{1st}}$ symmetry, 
looking for interaction terms with extra U(1) symmetry can
help us to identify the helpful gapping terms and design the proper lattice interactions.
CGP model has only
a single $\U(1)_{\text{1st}}$ symmetry. Here we suggest improving that model by removing all
the interaction terms that break the $\U(1)_{\text{2nd}}$ symmetry (thus adding all
possible terms that preserve the two U(1) symmetries) with an
intermediate strength.  
\end{itemize}

The plan and a short summary (see Fig.~\ref{fig:flow}) of our paper are the following. 
In Sec.\ref{sec3-5-4-0} we first consider a 3$_L$-5$_R$-4$_L$-0$_R$
anomaly-free chiral fermion field theory model, with a full $\U(1)^2$ symmetry:
A first 3-5-4-0 $\U(1)_{\text{1st}}$ symmetry
for two left-moving fermions of charge-3 and charge-4, and for
two right-moving fermions of charge-5 and charge-0.
And a second  0-4-5-3 $\U(1)_{\text{2nd}}$ symmetry
for two left-moving fermions of charge-0 and charge-5, and for
two right-moving fermions of charge-4 and charge-3.
If we wish to have a
\emph{single} $\U(1)_{\text{1st}}$ symmetry, we can weakly break the $\U(1)_{\text{2nd}}$ symmetry by adding tiny local $\U(1)_{\text{2nd}}$-symmetry breaking term.

We claim that this model can be put on the lattice with an
onsite $\U(1)$ symmetry, but without fermion-doubling problem.  
We construct a 2+1D lattice model by simply using four layers of the zeroth Landau
levels(or more precisely, four filled bands with Chern numbers\cite{Thouless:1982zz} $-1,+1,-1,+1$ on
a lattice\cite{Haldane:1988zza,{Wen:1990fv}}) which produces charge-3 left-moving, charge-5 
right-moving, charge-4 left-moving, charge-0 right-moving, totally four
fermionic modes at low energy on one edge.  Therefore, by putting the 2D bulk
spatial lattice on a cylinder with two edges, one can leave edge states on
one edge untouched so they remain chiral and gapless, while turning on
 interactions to gap out the mirror edge states on the other edge
with a large energy gap (which is not a usual quadratic mass gap).  

In Sec.\ref{sec:mapping-field-lattice}, 
we provide a correspondence from the continuum field theory to a discrete lattice model.
The numerical result of 
the chiral-$\pi$ flux square lattice with nonzero Chern numbers, supports the free fermion part of our model.
We study the kinetic and interacting part of Hamiltonian with dimensional scaling, energy scale and interaction strength analysis.
In Sec.\ref{anomaly-gap-proof}, 
we justify the mirror edge can be gapped
by analytically bosonizing the fermion theory and confirm the interaction terms obeys ``the boundary fully gapping rules.\cite{h95,Wang:2012am,Levin:2013gaa,{Kapustin:2013nva},{Wang:2013vna},Barkeshli:2013jaa,Barkeshli:2013yta,
Plamadeala:2013zva,Lu:2012dt,Kapustin:2010hk,Hung:2013nla,Lan2014uaaLWW1408.6514}''

To consider a more general model construction, inspired by the insight of SPTs,\cite{Wen:2013oza,Wen:2013ppa,Chen:2011pg} in Sec.\ref{SPT-CS}, we apply the bulk-edge correspondence between Chern-Simons theory and the chiral boson theory.\cite{Elitzur:1989nr,WZW,W,Wen:1995qn,h95,Wang:2012am,Levin:2013gaa,Barkeshli:2013jaa,Lu:2012dt} 
We refine and make connections between the key
concepts in our paper in Sec.\ref{anomaly-hall} and \ref{anomaly-gap}. These are ``the anomaly
factor\cite{Donoghue:1992dd,Fujikawa:2004cx,'tHooft:1979bh,Harvey:2005it}''
and ``effective Hall conductance''
`` 't Hooft anomaly matching condition\cite{'tHooft:1979bh,Harvey:2005it}''  and ``the boundary
fully gapping rules.\cite{h95,Wang:2012am,Levin:2013gaa,Barkeshli:2013jaa,Lu:2012dt,Kapustin:2010hk,Hung:2013nla}''
In Sec.\ref{model},
a non-perturbative lattice definition of 1+1D anomaly-free chiral matter model is given, and many examples of fermion/boson models are provided.
These model constructions are supported by our proof of the equivalence relations between ``the anomaly matching
condition'' and ``the boundary fully gapping rules'' in Appendix \ref{appendixC} and \ref{appendixD}.

In  Fig.~\ref{fig:flow}, we put these various models with various effective energy scales
into a renormalization group (RG) perspective from the UV (ultraviolet: high energy and short distance) to IR (infrared: low energy and long distance):\\ 
$\bullet$ UV lattice Hamiltonian fermion model, \\
$\bullet$ UV continuum (fermion/boson) field theory, and\\ 
$\bullet$ IR fixed-point chiral fermion theory.\\


\begin{widetext}
\onecolumngrid
\begin{figure}[!h]
{\includegraphics[width=.92\textwidth]{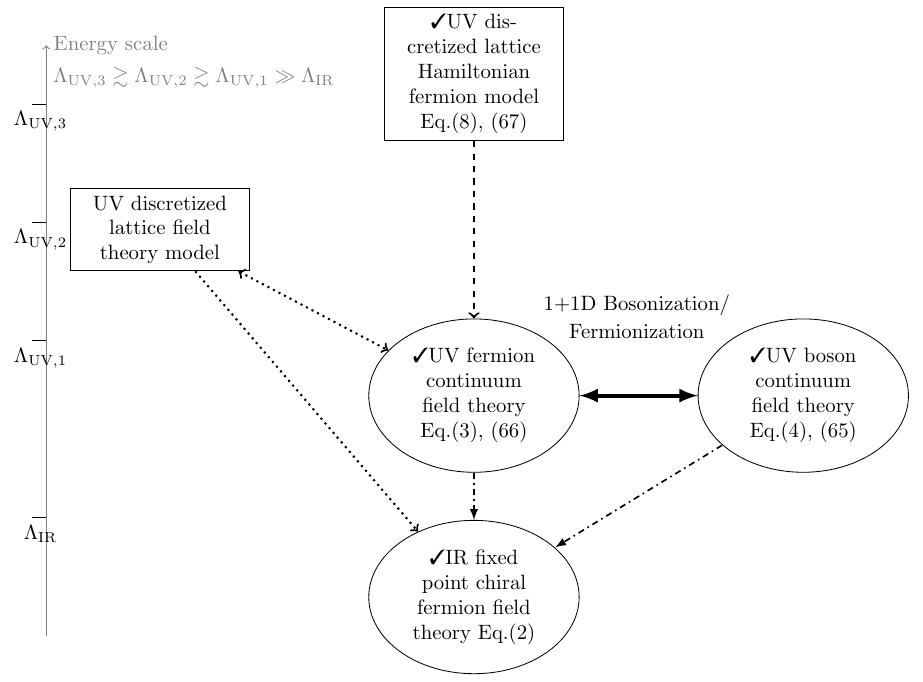}}
\caption{We construct a UV (ultraviolet high-energy) lattice model in Eq.~(\ref{H3540})
and (\ref{eq:H-int-all}), whose energy scale $\Lambda_{3,\text{UV}} \simeq 1/a$.
In contrast, the lattice QCD community usually employs a direct lattice regularization of a continuum field theory at another energy scale $\Lambda_{2,\text{UV}}$.
In this work, we do \emph{not} explore the UV lattice regularization of a field theory model.
However, we consider the models marked with the (\cmark) mark:
The UV continuum field theory, including both the fermionic model (Eq.~(\ref{Lf3-5-4-0}) and (\ref{eq:fermionize-all})) and the bosonic model 
(Eq.~(\ref{b3540}) and (\ref{eq:bosonize-all})) at another energy scale $\Lambda_{1,\text{UV}}$.
The UV continuum field theory does \emph{not} have to be renormalizable in the renormalization group (RG) sense; however, 
we provide a deeper UV completion of this UV continuum field theory by the UV Hamiltonian model at  $\Lambda_{3,\text{UV}}$.
In this work, we set the $\Lambda_{3,\text{UV}} \gtrsim \Lambda_{2,\text{UV}} \gtrsim\Lambda_{1,\text{UV}}$.
Since the energy scale $\Lambda_{3,\text{UV}} \gtrsim \Lambda_{2,\text{UV}} \gtrsim\Lambda_{1,\text{UV}}$ is set about the same, 
the RG flow analysis can be controlled along the way. This includes the controlled RG flow $\dasharrow$. 
Here we do \emph{not} study anything along with the flows of two dotted arrows ($\cdots$), since we do \emph{not} attempt from the UV lattice field theory model 
(which is a conventional model of the lattice QCD community).
We can also analyze along with the two dashed-dotted arrows (-.-.-.): We find that the RG flows to a completely gapped phase for the mirror sector, 
which is known in QFT and  condensed matter literature.\cite{h95,Kapustin:2010hk,Wang:2012am,Levin:2013gaa}
The boldface $\longleftrightarrow$ arrow is based on the standard bosonization/fermionization method in 1+1D.
}
\label{fig:flow}
\end{figure}
\end{widetext}

In contrast, we do not directly implement:\\ 
$\bullet$  UV lattice field theory regularization,\\
which is the conventional method for the lattice QCD community.
In other words,
we do not attempt to \emph{directly discretize} 
``the UV fermion field theory'' (to be shown in Eq.~(\ref{Lf3-5-4-0}) and (\ref{eq:fermionize-all}))
on a lattice in order to obtain
the ``UV lattice field theory model.'' 
Namely, we do not attempt to perform the analysis shown along the dotted arrows ($\cdots$) in Fig.~\ref{fig:flow}.

However, we comment on all the other RG flows and the other correspondences (bosonization/fermionization in 1+1D) shown in
Fig.~\ref{fig:flow}.
 We formulate a UV lattice Hamiltonian model instead (to be shown in Eq.~(\ref{H3540}) and (\ref{eq:H-int-all})) at a higher energy scale
 $\Lambda_{3,\text{UV}} \; (\simeq 1/a)$,
whose emergent effective UV field theory at a lower energy scale  $\Lambda_{1,\text{UV}}$
becomes the 
UV continuum fermionic field theory
(to be shown in Eq.~(\ref{Lf3-5-4-0}) and (\ref{eq:fermionize-all}))
or the UV continuum bosonic field theory
(to be shown in Eq.~(\ref{b3540}) and (\ref{eq:bosonize-all})).
In addition, the emergent IR
 field theory at the deep IR becomes the desired
 ``IR fixed point chiral fermion field theory'' (to be shown in Eq.~(\ref{cf})).
 
%
\noindent
\noindent
{By providing a UV lattice Hamiltonian model (shown in Eq.~(\ref{H3540}) and (\ref{eq:H-int-all}))
whose emergent IR
 field theory at the deep IR becomes the desired
 ``IR fixed point chiral fermion field theory'' (shown in Eq.~(\ref{cf}), we propose that this model would achieve
 our goal: a  non-perturbative regularization of 1+1D anomaly-free chiral fermions and bosons on a lattice.}

In  Appendix \ref{appendixA}, we discuss the $C,P,T$ symmetry in a 1+1 D fermion theory.
In  Appendix \ref{appendixB}, we show that GW 
fermions realize its axial U(1) symmetry by a non-onsite symmetry transformation.
As 
the non-onsite symmetry signals the nontrivial edge states of bulk SPTs,\cite{Chen:2011pg,Chen:2012hc,Santos:2013uda}
thus GW 
fermions can be regarded as gapless edge states of some bulk fermionic SPT states, such as certain 
topological insulators.
We also explain why it is easy to gauge an onsite symmetry (such as our chiral fermion model), and why it is difficult to gauge a non-onsite symmetry (such as GW 
fermions).
Since the lattice on-site symmetry can always be gauged, our result suggests a non-perturbative definition of any anomaly-free chiral gauge theory in 1+1D.
In Appendix \ref{AppendixE}, we provide physical, perturbative and non-perturbative understandings on 
``boundary fully gapping rules.''
In Appendix \ref{AppendixF}, we provide more details and examples about our lattice models. 
With this overall
understanding, in Sec.\ref{summary} we summarize with deeper implications and future directions.\\

\noindent
[Note on the terminology: 
Here in our work, U(1) symmetry may generically imply 
copies of U(1) symmetry such as U(1)$^{\rm n}$, with positive integer ${\rm n}$.  
 Topological {\bf Boundary Fully Gapping Rules} are defined as the rules to open the 
energy gap (which is not necessarily a usual quadratic mass gap)
 of the boundary states.
Topological {\bf Gapped Boundary Conditions} are defined to specify 
certain boundary types 
which are gapped (thus topological).
There are two kinds of usages of \emph{lattices} here discussed in our work:
one is the {{\bf Hamiltonian lattice}} model to simulate the chiral fermions/bosons.
The other \emph{lattice} is the 
{\bf Chern-Simons (representation) lattice} structure of 
Hilbert space, which is a quantized lattice due 
to the level/charge quantization of Chern-Simons theory.]\\

\noindent
{\bf Note added}: After the completion of this present work in 2013, the authors have,
later in 2018,  reconstructed a variant version of the 1+1D lattice model in Ref.~\onlinecite{Wang:2018ugf} 
which low energy realizes a 3$_L$-5$_R$-4$_L$-0$_R$ chiral fermion field theory. 
Ref.~\onlinecite{Wang:2018ugf} lattice model is different 
but based on the same topological non-perturbative proof given in our Appendix \ref{appendixA} and \ref{appendixB}. 
Since our proof holds universally independent from Hamiltonian or Lagrangian/path integral formulation of quantum theory,
the proof implies that the lattice regularization of 1+1D U(1) chiral fermion theory based on proper non-perturbative interactions shall work successfully, regardless
lattice Hamiltonian or lattice Lagrangian/path integral formulations.
Ref.~\onlinecite{ZengZhuWangYou2202.12355} successfully simulates the 
1+1D U(1) 3$_L$-5$_R$-4$_L$-0$_R$ chiral fermion lattice model proposed in this article.

\noindent
\section{3$_L$-5$_R$-4$_L$-0$_R$ Chiral Fermion model:
A New Lattice Proposal  \label{sec3-5-4-0}}

The simplest chiral (Weyl) fermion field theory with U(1) symmetry in $1+1$D is given by the action
\be
\label{WeylS}
S_{\Psi,free}=\int  \dd t \dd x \; \ti \psi^\dagger_{L} (\partial_t-\partial_x) \psi_{L}.
\ee
However, Nielsen-Ninomiya theorem claims that
such a theory cannot be put on a lattice with unbroken onsite U(1) symmetry,
due to 
the fermion-doubling problem.\cite{Nielsen:1980rz,Nielsen:1981xu,Nielsen:1981hk} 
While the Ginsparg-Wilson fermion approach can still implement an anomalous single Weyl fermion 
on the lattice,
our approach cannot (unless we modify local Hamiltonian to
infinite-range hopping non-local Hamiltonian). As we will show, our approach is more restricted, only limited to the anomaly-free theory. 
Let us instead consider an anomaly-free
3$_L$-5$_R$-4$_L$-0$_R$ chiral fermion field theory with an action,
\begin{widetext}
\be
\label{cf}
S_{\Psi_{\A},free}=\int  \dd t \dd x \; \Big(
 \ti \psi^\dagger_{L,3} (\partial_t-\partial_x) \psi_{L,3}
+ \ti\psi^\dagger_{R,5} (\partial_t+\partial_x) \psi_{R,5}
+ \ti\psi^\dagger_{L,4} (\partial_t-\partial_x) \psi_{L,4}
+\ti\psi^\dagger_{R,0} (\partial_t+\partial_x) \psi_{R,0}\Big),
\ee
\end{widetext}
where
$\psi_{L,3}$,
$\psi_{R,5}$,
$\psi_{L,4}$, and
$\psi_{R,0}$ are 1-component Weyl spinor, carrying U(1) charges 3,5,4,0 respectively. The subscript $L$ (or $R$)
indicates left (or right) moving along $-\hat{x}$ (or $+\hat{x}$).  Although this
theory has equal numbers of left and right moving modes, 
it violates the parity 
and time reversal symmetry, 
so it is a chiral theory (details about $C,P,T$ symmetry in Appendix \ref{appendixA}). 
Such a  chiral fermion field theory
is very special because it is free from U(1) anomaly - it satisfies the anomaly matching
condition\cite{Donoghue:1992dd,Fujikawa:2004cx,'tHooft:1979bh,Harvey:2005it} 
in $1+1$D, 
which means $\sum_j q_{L,j}^2-q_{R,j}^2=3^2-5^2+4^2-0^2=0$.
It is crucial to include the ``neutrino'' $\psi_{R,0}$
so the theory is gravitational-anomaly-free, 
because two left-moving and two right-moving Weyl fermions give the zero chiral central charge $c_L-c_R=2-2=0$.
We ask:\\

\noindent
{
{\bf Question\;1 }:
``Whether there is a \emph{local} \emph{finite} 
Hamiltonian realizing the above U(1) 3-5-4-0 symmetry as an onsite symmetry with \emph{short-range interactions} 
defined on a 1D spatial lattice with a continuous time, such that its low energy physics produces the anomaly-free chiral fermion theory Eq.(\ref{cf})?''}



\begin{figure}[!h]
{\includegraphics[width=.48\textwidth]{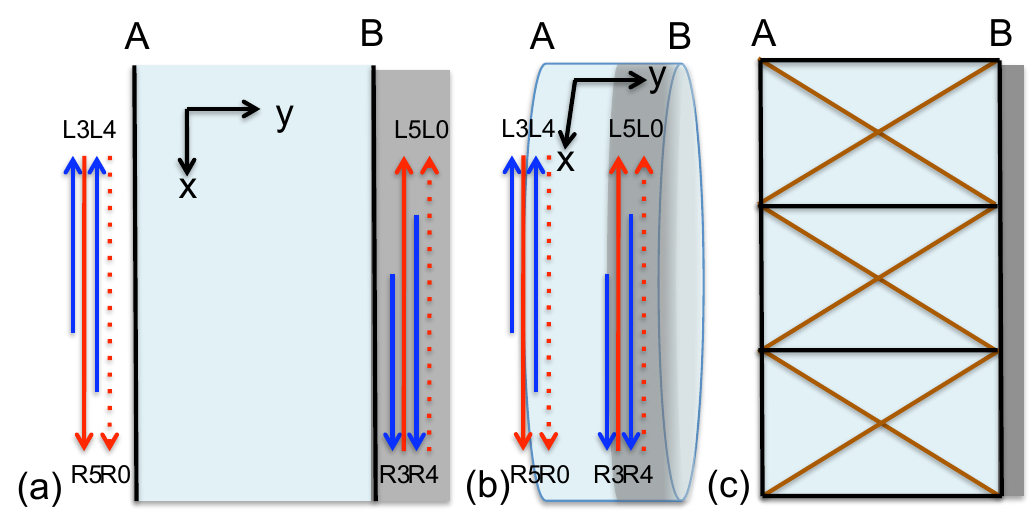}}
\caption{3$_L$-5$_R$-4$_L$-0$_R$ chiral fermion model: (a) The fermions carry U(1) charge $3$,$5$,$4$,$0$ for $\psi_{L,3},$$\psi_{R,5},$$\psi_{L,4},$$\psi_{R,0}$ on the edge A, and also for its mirror partners $\tilde{\psi}_{R,3},$$\tilde{\psi}_{L,5},$$\tilde{\psi}_{R,4},$$\tilde{\psi}_{L,0}$
on the edge B. We focus on the model with a periodic boundary condition along $x$, and a finite-size length along $y$,
effectively as, (b) on a spatial cylinder.
(c) The ladder model on a cylinder 
with the $t$ hopping term along black links, the $t'$ hopping term along brown links. The shadow on the edge B indicates the gapping terms 
with $G_1,G_2$ couplings in Eq.(\ref{H3540}) are imposed.}
\label{3540}
\end{figure}

\begin{widetext}
\onecolumngrid
Yes. We show that the above chiral fermion field theory
can be put
 on a lattice with unbroken onsite U(1) symmetry, 
if we include properly-designed interactions between fermions.
In fact, we propose that the chiral fermion field theory in Eq.(\ref{cf})
appears as the low energy effective theory of the following 2+1D lattice model
on a cylinder (see Fig.\ref{3540}) with a properly designed Hamiltonian.
To derive such a Hamiltonian, we start from thinking the full two-edges fermion theory with the action $S_\Psi$,
where the particularly chosen multi-fermion interactions $S_{\Psi_{\B},interact}$ will be explained:
%
%
\bea 
&S_\Psi&=S_{\Psi_{\A},free}+S_{\Psi_{\B},free}+S_{\Psi_{\B},interact}=\int  \dd t \dd x \; \bigg( \ti\bar{\Psi}_{\A} \Gamma^\mu  \partial_\mu \Psi_{\A}+ \ti\bar{\Psi}_{\B} \Gamma^\mu  \partial_\mu \Psi_{\B}    \label{Lf3-5-4-0} \\
 &\;&+\tilde{g}_{1} \big( (\tilde{\psi}_{R,3} )  (\tilde{\psi}_{L,5} )
 ( \tilde{\psi}^\dagger_{R,4} \nabla_x \tilde{\psi}^\dagger_{R,4}  ) ( \tilde{\psi}_{L,0} \nabla_x \tilde{\psi}_{L,0}  ) +\text{h.c.} \big)  
 + \tilde{g}_{2}   \big( (\tilde{\psi}_{R,3} \nabla_x \tilde{\psi}_{R,3} )  (\tilde{\psi}_{L,5}^\dagger  \nabla_x \tilde{\psi}_{L,5}^\dagger  )( \tilde{\psi}_{R,4}   )( \tilde{\psi}_{L,0}   )+\text{h.c.} \big)\bigg), \nonumber
\eea
\end{widetext}

The notation for fermion fields on the edge A are $\Psi_{\A}=(\psi_{L,3},\psi_{R,5},\psi_{L,4},\psi_{R,0})$,
and fermion fields on the edge B are $\Psi_{\B}=(\tilde{\psi}_{L,5},\tilde{\psi}_{R,3},\tilde{\psi}_{L,0},\tilde{\psi}_{R,4})$.
(Here a left moving mode in $\Psi_{\A}$ corresponds to a right moving mode in $\Psi_{\B}$ because of Landau level/Chern band chirality, the details of lattice model will be explained.)
The gamma matrices in 1+1D are presented in terms of Pauli matrices, with
$\gamma^0=\sigma_x$, $\gamma^1=\ti\sigma_y$, $\gamma^5\equiv\gamma^0\gamma^1=-\sigma_z $,
and $\Gamma^0=\gamma^0\oplus\gamma^0$, $\Gamma^1=\gamma^1\oplus \gamma^1$, $\Gamma^5\equiv\Gamma^0\Gamma^1$ and $\bar{\Psi}_i \equiv \Psi_i^\dagger \Gamma^0$.
The \text{h.c.} is a hermitian conjugation of the aforementioned term.

We emphasize that although the interaction terms in $S_{\Psi_{\B},interact}$ in Eq.(\ref{Lf3-5-4-0}) look to be \emph{irrelevant operators} in a perturbative RG sense 
(so naively people may mistakenly argue that the irrelevant operators cannot drive the gapless phase to gapped phase on the B edge),
we later prove that, based on a non-perturbative and topological analysis in Sec.~\ref{sec:topo-nonp-proof}
and Appendix \ref{AppendixE}, 
at strong coupling (or the intermediate coupling at the lattice scale in Eq.(\ref{H3540}))
the interaction terms can drive the mirror B edge to a gapped phase. 
See also a shorten version of the proof on gapping mirror chiral fermions in Ref.~\onlinecite{Wang:2018ugf}.

In 1+1D, we can do bosonization,\cite{fermionization1} 
where the fermion matter field $\Psi$ turns into bosonic phase field $\Phi$,
more explicitly $\psi_{L,3}\sim e^{\ti \Phi^{\A}_3} $, $\psi_{R,5}\sim e^{\ti \Phi^{\A}_5} $, $\psi_{L,4}\sim e^{\ti \Phi^{\A}_4} $, $\psi_{R,0}\sim e^{\ti \Phi^{\A}_0} $ on A edge,
$\tilde{\psi}_{R,3}\sim  e^{\ti \Phi^{\B}_3} $, $\tilde{\psi}_{L,5} \sim e^{\ti \Phi^{\B}_5} $, $\tilde{\psi}_{R,4} \sim  e^{\ti \Phi^{\B}_4} $, $\tilde{\psi}_{L,0}  \sim  e^{\ti \Phi^{\B}_0}$ on B edge,
up to normal orderings $ :e^{\ti \Phi}:$ and pre-factors,\cite{fermionization2} 
but the 
precise factor is not of our interest since our goal is to obtain its non-perturbative lattice realization.
So Eq.(\ref{Lf3-5-4-0}) becomes

\begin{widetext}
\bea 
S_{\Phi}=S_{\Phi^{\A}_{free}}+S_{\Phi^{\B}_{free}}+S_{\Phi^{\B}_{interact}}=&&
\frac{1}{4\pi}  \int \dd t \dd x  \big(K^{\A}_{IJ}  \partial_t \Phi^{\A}_I   \partial_x \Phi^{\A}_{J} -V_{IJ}  \partial_x \Phi^{\A}_I   \partial_x \Phi^{\A}_{J}\big)+\big(K^{\B}_{IJ}  \partial_t \Phi^{\B}_I   \partial_x \Phi^{\B}_{J} -V_{IJ}  \partial_x \Phi^{\B}_I   \partial_x \Phi^{\B}_{J} \big) \label{b3540}        \;\;\;\nonumber\\
&& +\int \dd t \dd x  \bigg( g_{1}  \cos( \Phi^{\B}_{3}+\Phi^{\B}_{5}-2 \Phi^{\B}_{4}+2\Phi^{\B}_{0})+  g_{2}  \cos( 2\Phi^{\B}_{3}-2\Phi^{\B}_{5}+\Phi^{\B}_{4}+\Phi^{\B}_{0}) \bigg). \;\;\;\;\;\;\;
 \eea
 \end{widetext}
 
Here $I,J$ runs over $3,5,4,0$ while $K^{\A}_{IJ}=-K^{\B}_{IJ}=\diag(1,-1,1,-1)$ and 
$V_{IJ}=\diag(1,1,1,1)$ are diagonal matrices.

{\bf All we have to show is that gapping terms, the cosine terms with ${g}_1,{g}_2$ coupling can gap out all states on the edge B.}

First, let us understand more about the full U(1) symmetry.  What are the U(1) symmetries?
They are transformations of 
$$ \text{fermions } \psi \to \psi \cdot e^{\ti q \theta},\;\;\; \text{bosons }\;\;\; \Phi \to \Phi + q \;\theta$$ 
making the full action invariant.
The original \emph{four} Weyl fermions have an abelian U(1)$^4$ symmetry. 
Under \emph{two} linear-independent interaction terms in $S_{\Psi_{\B},interact}$ (or $S_{\Phi^{\B}_{interact}}$),
the U(1)$^4$ is broken down to a U(1)$^2$ symmetry.
If we denote these $q$ as a charge vector $\mathbf{t}=(q_3,q_5,q_4,q_0)$,
we find there are such two charge vectors 
$${\text{
$\mathbf{t}_1=(3,5,4,0)$  and $\mathbf{t}_2=(0,4,5,3)$}}
$$
for U(1)$_{\text{1st}}$ and U(1)$_{\text{2nd}}$ symmetries respectively.

We emphasize 
that finding those gapping terms in this  U(1)$^2$ anomaly-free theory is not accidental. The {\bf anomaly matching condition}\cite{Donoghue:1992dd,Fujikawa:2004cx,'tHooft:1979bh,Harvey:2005it} here is 
satisfied, for the coefficient of  anomalies $\sum_j q_{L,j}^2-q_{R,j}^2=3^2-5^2+4^2-0^2=0^2-4^2+5^2-3^2=0$, 
and the mixed anomaly: $3 \cdot 0 -5 \cdot 4 + 4 \cdot 5 - 0\cdot 3=0$ which can be formulated as 
\be \label{tKt}
{\boxed{ \mathbf{t}^T_i \cdot (K^{\A}) \cdot \mathbf{t}_j=0}}\;, \;\;\; i,j \in \{1,2\} 
\ee 
with the U(1) charge vector $\mathbf{t}=(3,5,4,0)$, with its transpose $\mathbf{t}^T$.

\;\;\; On the other hand, the {\bf boundary fully gapping rules} (as we will explain, and the full details in Appendix \ref{AppendixE}),\cite{h95,Wang:2012am,Levin:2013gaa,Lu:2012dt} for a theory of Eq.(\ref{b3540}), require two gapping terms, here $g_{1} \cos( \ell_1 \cdot \Phi)+ g_{2}\cos( \ell_2 \cdot \Phi)$,
such that self and mutual statistical angles $\theta_{ij}$\cite{W,Wen:1995qn} defined below among the Wilson-line operators $\ell_i,\ell_j$ are zeros,
\be \label{LKL} 
{\boxed{ \theta_{ij}/(2\pi) \equiv \ell_i^T \cdot (K^{\B})^{-1}  \cdot \ell_j = 0} }\;, \;\;\; i,j \in \{1,2\}.    
\ee
Indeed, here we have 
$$
{\text{
{ $\ell_1=(1,1,-2,2),\quad \ell_2=(2,-2,1,1)$}
}}
$$
satisfying the rules. 
We can alternatively choose:
$$
{\text{
$\ell_1=(3,-5,4,0), \quad \ell_2=(0,4,-5,3)$. 
}}
$$
The $g_{1} \cos( \ell_1 \cdot \Phi)+ g_{2}\cos( \ell_2 \cdot \Phi)$ is symmetric interaction, invariant respect the symmetry
$\Phi \to \Phi + q \;\theta$ for $q$ charges labeled by $\mathbf{t}_1$ and $\mathbf{t}_2$, 
because
\be \label{tKt}
{\boxed{ \mathbf{t}^T_i \cdot \ell_j=0}}\;, \;\;\; i,j \in \{1,2\}. 
\ee
So both U(1)$_{\text{1st}}$ and U(1)$_{\text{2nd}}$
for any $\theta \in [0, 2 \pi)$ are preserved.
We propose that the mirror edge states on the edge B can be fully gapped out by the symmetric interaction.

We will prove the {\bf anomaly matching condition} is equivalent to find a set of gapping terms $g_{a}  \cos(\ell_{a}^{} \cdot\Phi_{}) $, obeying the {\bf boundary fully gapping rules},
detailed in Sec.\ref{anomaly-hall}, \ref{anomaly-gap}, Appendix \ref{appendixC} and \ref{appendixD}. 
Simply speaking,

``{The {\bf anomaly matching condition} (Eq.(\ref{tKt})) in 1+1D 
is {equivalent} to (an if and only if relation)
\\ 
the {\bf boundary fully gapping rules} (Eq.(\ref{LKL})) in 1+1D boundary/2+1D bulk for an equal number of left-right moving modes ($N_L=N_R$, with central charge $c_L=c_R$).}''

We prove this is true at least for U(1) symmetry case, with the bulk theory is a 2+1D SPT state and the boundary theory is in 1+1D.

{\bf We now propose a lattice Hamiltonian model for this 3$_L$-5$_R$-4$_L$-0$_R$ chiral fermion realizing Eq.(\ref{Lf3-5-4-0})} (thus Eq.(\ref{cf}) at the low energy once the Edge B is gapped out).
Importantly, we \emph{do not} discretize the action Eq.(\ref{Lf3-5-4-0}) on the spacetime lattice.
We \emph{do not} use Ginsparg-Wilson (GW) fermion \emph{nor} the Neuberger-Dirac operator.
GW and Neuberger-Dirac scheme contains \emph{non-onsite symmetry} (details in Appendix \ref{appendixB}) which cause the lattice \emph{difficult to be gauged} to 
a chiral gauge theory.
Instead, the key step is that we 
implement the \emph{on-site symmetry} lattice fermion model. 
The \emph{free kinetic part} is a fermion-hopping model which has a \emph{finite 2D bulk energy gap} 
but with \emph{gapless 1D edge states}. This can be done by using any {\bf lattice Chern insulator}.

We stress that  {\bf any} lattice 
with onsite-symmetry shall work, and  
we design one as in Fig.\ref{3540}. 
{(In fact, we later design another variant version of 1D lattice model in Ref.~\onlinecite{Wang:2018ugf}.)}
Our full Hamiltonian with two interacting $G_1$ and $G_2$ gapping terms is
\begin{widetext}
\bea 
\label{H3540}
H&=&\sum_{q=3,5,4,0} 
\bigg(  \sum_{\langle i, j \rangle}
\big(t_{ij,q}\; \hat{f}^\dagger_{q}(i)
\hat{f}_{q}(j)+\text{h.c.}\big) + \sum_{\langle\langle i, j
\rangle\rangle} \big(  t'_{ij,q}
\;\hat{f}^\dagger_{q}(i) \hat{f}_{q}(j)+\text{h.c.}\big) \bigg)
\\ 
&+& G_{1} \sum_{j \in \B} \bigg(
\big(\hat{f}_3(j)\big)^1
\big(\hat{f}_5(j)\big)^1
\big(\hat{f}^\dagger_4(j)_{pt.s.}\big)^{2}
\big(\hat{f}_0(j)_{pt.s.}\big)^2
+\text{h.c.} \bigg) 
+ G_{2} \sum_{j \in \B} \bigg( 
\big(\hat{f}_3(j)_{pt.s.}\big)^2
\big(\hat{f}^\dagger_5(j)_{pt.s.}\big)^2
\big(\hat{f}_4(j)\big)^1
\big(\hat{f}_0(j)\big)^1 
+\text{h.c.}
\bigg)
\nonumber 
\eea
\end{widetext}
\noindent
where $\sum_{j \in \B}$ sums over the lattice points on the right boundary (the edge B in Fig.\ref{3540}),
and the fermion operators $\hat f_{3}$, $\hat f_{5}$, $\hat f_{4}$, $\hat
f_{0}$ carry a U(1)$_{\text{1st}}$ charge 3,5,4,0 
and another U(1)$_{\text{2nd}}$ charge 0,4,5,3 respectively.   
The lattice fermion operators are \emph{non-relativistic}, satisfying canonical anticommutation relations:
$$\{ \hat{f}_{q}(i),
\hat{f}^\dagger_{q'}(j)  \}=\delta_{(i,j)}\delta_{(q,q')}.$$
We emphasize that this lattice model has \emph{onsite} U(1)$^2$ symmetry, since this Hamiltonian, including interaction terms, is invariant 
under a global U(1)$_{\text{1st}}$ transformation \emph{on each site} for any $\theta$ angle:
$\hat f_{3} \to \hat f_{3} e^{\ti 3\theta}$, $\hat f_{5} \to \hat f_{5} e^{\ti 5\theta}$, $\hat f_{4} \to \hat f_{4} e^{\ti 4\theta}$, $\hat f_{0} \to \hat f_{0}$,
and invariant under another global U(1)$_{\text{2nd}}$ transformation for any $\theta$ angle:
$\hat f_{3} \to \hat f_{3} $, $\hat f_{5} \to \hat f_{5} e^{\ti 4\theta}$, $\hat f_{4} \to \hat f_{4} e^{\ti 5\theta}$, $\hat f_{0} \to \hat f_{0} e^{\ti 3\theta}$.
The U(1)$_{\text{1st}}$ charge is the reason why it is named as 3$_L$-5$_R$-4$_L$-0$_R$ 
model.

As for notations, 
$\langle i, j \rangle$ stands for nearest-neighbor hopping along black links
and $\langle\langle i, j \rangle\rangle$ stands for next-nearest-neighbor hopping along brown links in Fig.\ref{3540}.
Here $pt.s.$  
stands for point-splitting. 
For example, $(\hat
f_3(j)_{pt.s.})^2\equiv   \hat f_{3}(j) \hat f_{3}(j+\hat
x)$.
We stress that the full Hamiltonian (including interactions) Eq.(\ref{H3540}) is \emph{short-range and local}, 
because each term only involves coupling within finite number of neighbor sites.
The hopping amplitudes $t_{ij,3}=t_{ij,4}$ and $t'_{ij,3}=t'_{ij,4}$
produce the energy bands with a Chern number $-1$, while the hopping amplitudes
$t_{ij,5}=t_{ij,0}$ and $t'_{ij,5}=t'_{ij,0}$ produce the energy bands with a Chern number
$+1$ (see {{Sec.\ref{numeric}}}).\cite{Thouless:1982zz,Haldane:1988zza,Parameswaran:2013pca,Tang et al.(2011),Sun et
al.(2011),Neupert et al.(2011)} The ground state is obtained by filling the
above four bands.
The fermionic particle or hole excitations near the filled energy bands become the \emph{relativistic} fermions.

As Eq.(\ref{H3540}) contains U(1)$_{\text{1st}}$ and an accidental extra U(1)$_{\text{2nd}}$ symmetry, we shall ask:\\

{ {\bf Question 2}:
``Whether there is a \emph{local}  \emph{finite}  Hamiltonian realizing \emph{only} a chiral U(1) 3-5-4-0
symmetry as an onsite symmetry with \emph{short-range interactions}
defined on a 1D spatial lattice with a continuous time, such that its low energy physics produces the anomaly-free chiral fermion theory Eq.(\ref{cf})?''}\\

Yes, by adding a small local perturbation to break U(1)$_{\text{2nd}}$ 0-4-5-3 symmetry, we can achieve a faithful
U(1)$_{\text{1st}}$ 3-5-4-0 symmetry chiral fermion theory of Eq.(\ref{cf}).
For example, we can adjust  Eq.(\ref{H3540})'s $H \to H +\delta H $ by adding: 
\bea
&&\delta H = G_{\rm tiny}' \sum_{j \in \B}  
\Big( \big(\hat{f}_3(j)_{pt.s.}\big)^3
\big(\hat{f}^\dagger_5(j)_{pt.s.}\big)^1
\big(\hat{f}^\dagger_4(j)\big)^1 + \text{h.c.} \Big) \nonumber\\
&& \Leftrightarrow 
\tilde{g}_{\rm tiny}'   \big( (\tilde{\psi}_{L,3} \nabla_x \tilde{\psi}_{L,3} \nabla_x^2 \tilde{\psi}_{L,3} )  (\tilde{\psi}_{R,5}^\dagger )( \tilde{\psi}_{L,4}^\dagger )+\text{h.c.} \big) \nonumber \\
&& \Leftrightarrow g_{\rm tiny}'  \cos( 3\Phi^{\B}_{3}-\Phi^{\B}_{5}- \Phi^{\B}_{4}) \equiv 
g_{\rm tiny}'  \cos( \ell_{}' \cdot \Phi^{\B}).
\eea
Here we have $\ell_{}'=(3,-1,-1,0)$. The $g_{\rm tiny}'  \cos( \ell_{}' \cdot \Phi^{\B})$ is not designed to
be a gapping term (its self and mutual statistics 
 happen to be nontrivial:
  ${\ell_{}'^T\cdot (K^{\B})^{-1}  \cdot \ell_{}' \neq 0}$, 
 ${\ell_{}'^T \cdot (K^{\B})^{-1}  \cdot \ell_2 \neq  0}$), but this tiny perturbation term  
 is meant to preserve U(1)$_{\text{1st}}$ 3-5-4-0 symmetry only,
 thus 
 \bea
 {
 {\ell_{}'^T \cdot \mathbf{t}_1  ={\ell_{}'^T \cdot (K^{\B})^{-1}  \cdot \ell_1 =  0}}}.
\eea
We must set
$(|G'_{{\rm tiny}} |/|G|) \ll 1$ with $|G_1| \sim |G_2| \sim |G|$ about the same magnitude, 
so that
 the tiny local perturbation will not destroy the 
energy gap (not a usual quadratic mass gap for fermions).

Without the interaction, i.e. $G_1=G_2=0$, the edge excitations of the above four
bands produce the chiral fermion theory Eq.(\ref{cf}) on the left edge A
and the mirror partners 
on the right edge B. 
So the total low energy effective theory is non-chiral.
In Sec.\ref{numeric}, we will provide an explicit lattice model for this free fermion theory.

However, by turning on the intermediate-strength 
interaction $G_1,G_2\neq 0$, we claim the interaction terms can fully gap out the edge excitations on the right mirror edge B as in
Fig.\ref{3540}. To find those gapping terms is not accidental - it is guaranteed by our proof (see Sec.\ref{anomaly-hall}, \ref{anomaly-gap}, Appendix \ref{appendixC} and \ref{appendixD}) of the equivalence between
{\bf the anomaly matching condition}\cite{Donoghue:1992dd,Fujikawa:2004cx,'tHooft:1979bh,Harvey:2005it} 
(as ${ \mathbf{t}^T_i \cdot (K)^{-1} \cdot \mathbf{t}_j=0}$ of Eq.(\ref{tKt}) 
) and {\bf the boundary fully gapping rules}\cite{h95,Wang:2012am,Levin:2013gaa,Barkeshli:2013jaa,Lu:2012dt,Kapustin:2010hk,Hung:2013nla} (here $G_1,G_2$ terms can gap out the edge) in $1+1$ D.
The low energy effective theory of the interacting lattice
model with only gapless states on the edge A 
is the chiral fermion theory in Eq.(\ref{cf}).  Since the width of the
cylinder is finite, the lattice model Eq.(\ref{H3540}) is actually a 1+1D
lattice model, which gives a non-perturbative lattice definition of the chiral fermion
theory Eq.(\ref{cf}).  Indeed, the Hamiltonian and the lattice need not to be restricted merely to Eq.(\ref{H3540}) and Fig.\ref{3540},
we stress that any on-site symmetry lattice model produces four bands with the desired Chern numbers would work.
We emphasize again that the U(1) symmetry is realized as
an onsite symmetry\cite{Chen:2011pg,2011PhRvB..84w5141C} in our lattice model. It is easy to gauge such an onsite
U(1) symmetry (explained in Appendix \ref{appendixB}) to obtain a chiral fermion theory coupled to a U(1) gauge field.

\section{From a continuum field theory to a discrete lattice model} 
\label{sec:mapping-field-lattice}

We now comment about the mapping from a continuum field theory of the action Eq.(\ref{cf}) 
to a discretized space Hamiltonian Eq.(\ref{H3540}) with a continuous time.
We \emph{do not} use \emph{Ginsparg-Wilson scheme}, and our gapless edge states are \emph{not} derived from the discretization of spacetime action. 
Instead, we will show that the Chern insulator Hamiltonian in Eq.(\ref{H3540}) as we described can provide essential gapless edge states for a free theory (without interactions $G_1$ and $G_2$).  We may view that the continuum field theory is an IR fixed point of the UV lattice Hamiltonian under RG flow, see \Fig{kx_chiral}.
But the continuum field theory is \emph{not} directly derived from taking the lattice constant $a \to 0$:
we see that the lattice fermion $\hat{f}_{q}$ is a \emph{non-relativistic} operator satisfying the canonical anticommutation relations,
while the fermion field $\psi$ is a \emph{relativistic} spacetime Weyl spinor.


{\bf Energy and Length Scales}:
We consider a finite 1+1D quantum system with a periodic length scale $L$ for the compact circle of the cylinder in Fig.\ref{3540}.
The finite size width of the cylinder is $w$. The lattice constant is $a$. The energy gap (not a usual quadratic mass gap) 
we wish to generate on the mirror edge is $\Delta_m$, 
which causes a two-point correlator has an exponential decay: 
\be
\langle \psi^\dagger(r) \psi(0) \rangle \sim \langle e^{-\ti \Phi(r)} e^{\ti \Phi(0)}\rangle  \sim \exp(-|r|/\xi)
\ee
with a correlation length scale $\xi$.
The expected length scales follow that
\be \label{eq:lengthscale}
a < \xi \ll  w \ll  L.
\ee
The 1D system size $L$ is larger than the width $w$, the width $w$  is larger than the correlation length $\xi$,
the correlation length $\xi$ is larger than the lattice constant $a$.

\subsection{Free kinetic part and the edge states of a Chern insulator}

\subsubsection{Kinetic part mapping and RG analysis} \label{subsubsec:kinetic}

The {\bf kinetic part} of the lattice Hamiltonian
contains the nearest neighbor hopping term  
$\sum_{\langle i, j \rangle}$ 
$\big(t_{ij,q}$  $\hat{f}^\dagger_{q}(i)
\hat{f}_{q}(j)+\text{h.c.}\big)
$ together with the
next-nearest neighbor hopping term
$\sum_{\langle\langle i, j
\rangle\rangle} \big(  t'_{ij,q}
\;\hat{f}^\dagger_{q}(i) \hat{f}_{q}(j)+\text{h.c.}\big)$, 
which generate the leading order field theory kinetic term 
via
\be \label{subsec:free_map}
t_{ij} \hat{f}^\dagger_{q}(i) \hat{f}_{q}(j) \sim a\; \ti \psi_q^\dagger \partial_x \psi_q + \dots ,
\ee
here hopping constants $t_{ij},t_{ij}'$ with a dimension of energy  $[t_{ij}] = [t_{ij}'] = 1$, 
and $a$ is the lattice spacing with a value $[a]=-1$. Thus,
$[\hat{f}_{q}(j)]=0$ and $[\psi_q] =\frac{1}{2}$.
The map from 
\be \label{eq:subf}
f_q  \to \sqrt{a}\, \psi_q + \dots
\ee
contains subleading terms. 
Subleading terms $\dots$ 
potentially contain higher derivative $\nabla^n_x \psi_q$ 
are only {subleading perturbative effects}
\be 
f_q \to \sqrt{a}\,( \psi_q + \dots + a^n \, \alpha_{\text{small}} \nabla^n_x \psi_q 
+ \dots )\nonumber
\ee
with small coefficients of the polynomial of the small lattice spacing $a$ via $\alpha_{\text{small}}=\alpha_{\text{small}}(a) \lesssim  (a/L)$. 
It is common in condensed matter systems that the non-relativistic fermion $\hat{f}_{q}$ at a UV Hamiltonian is \emph{dressed} up to be a relativistic fermionic spinor $\psi_q$ at IR
under RG flow, see \Fig{fig:flow} and \Fig{kx_chiral}.

We comment that only the leading term in the mapping is important,
the full account for the exact mapping from the fermion operator $f_q$ to $\psi_q$ is immaterial
to our model, because of two main reasons:\\

\noindent
$\bullet$(i) Our lattice construction is based on 
several layers of Chern insulators, and the chirality of each layer's edge states are protected by a topological number - the first Chern number $C_1 \in \mathbb{Z}$. 
Such an integer Chern number cannot be deformed by small perturbation,
thus it is {\bf non-perturbative topologically robust}, hence the chirality of edge states will be protected and will not be eliminated by small perturbations.
The origin of our \emph{fermion chirality} (breaking parity and time reversal symmetries) is an emergent phenomena
due to the \emph{complex hopping} amplitude of some hopping constant $t_{ij}'$ or $t_{ij}  \in \mathbb{C}$.
Beside, it is well-known that Chern insulator can produce the gapless fermion energy spectrum 
at low energy. 
More details and the energy spectrum are explicitly presented in Sec.\ref{numeric}.
\\

\noindent
$\bullet$(ii) The 
properly-designed interaction 
effect (from boundary fully gapping rules) is a {\bf non-perturbative topological effect} (as we will show in Sec.\ref{anomaly-gap} and Appendix \ref{AppendixE}). 

\color{black}

\subsubsection{Numerical simulation for the free 
fermion theory with a nontrivial Chern number $C_1$\label{numeric}}

\begin{figure}[tb] 
{\includegraphics[width=.45\textwidth]{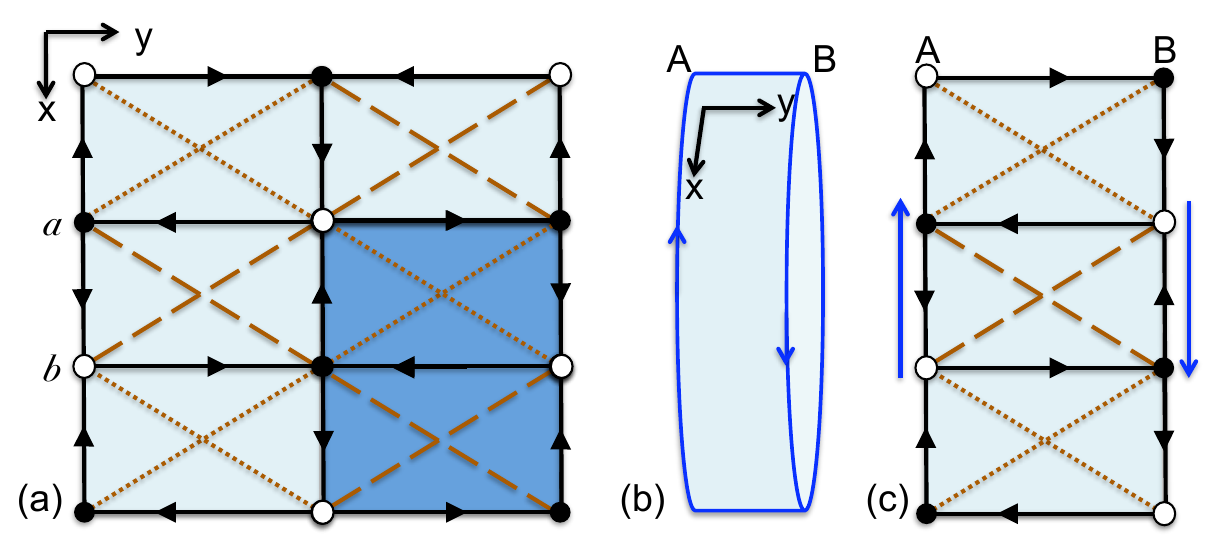}}
\caption{Chiral $\pi$-flux square lattice: (a) A unit cell is indicated as the shaded darker region,
containing two sublattice as a black dot $a$ and a white dot $b$. The lattice Hamiltonian has
hopping constants, $t_1 e^{\ii\pi/4}$ along the black arrow direction,
 $t_2$ along dashed brown links,
 $-t_2$ along dotted brown links.
(b) Put the lattice on a cylinder.
(c) The ladder: the lattice on a cylinder with a square lattice width.
The chirality of edge state is along the direction of blue arrows.}
\label{chiral-pi-flux}
\end{figure}

 Follow from Sec.\ref{sec3-5-4-0} and \ref{subsubsec:kinetic}, here we provide a concrete lattice realization for free fermions part of Eq.(\ref{H3540}) (with $G_1=G_2=0$),
and show that the Chern insulator provides the desired gapless fermion energy spectrum 
(say, a left-moving Weyl fermion on the edge A and a right-moving Weyl fermion on the edge B, and totally a Dirac fermion for the combined).
We adopt the chiral $\pi$-flux square lattice model\cite{Wen:1990fv} 
in Fig.\ref{chiral-pi-flux} as an example.
This lattice model can be regarded as a free theory of 3-5-4-0 fermions of Eq.(\ref{cf}) with its mirror conjugate.
We will explicitly show filling the first Chern number\cite{Thouless:1982zz} $C_1=-1$ band of the lattice on a cylinder would give the edge states of a free fermion with U(1) charge $3$,
similar four copies of model together render 3-5-4-0 free fermions theory of Eq.(\ref{H3540}).

\begin{figure}[!h] 
{\includegraphics[width=.35\textwidth]{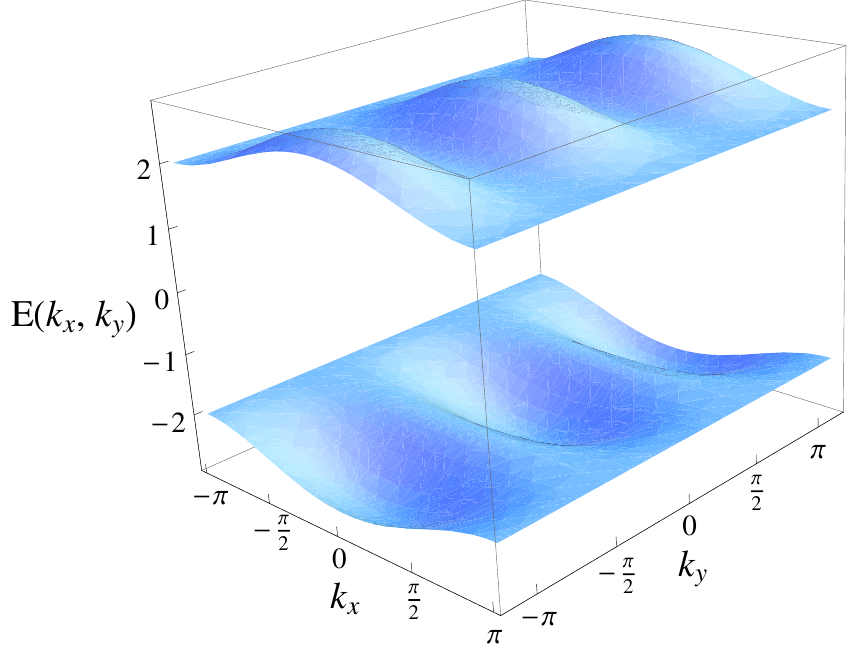}}
\caption{Two nearly-flat energy bands $\E_{\pm}$ in Brillouin zone
for the kinetic hopping terms of our model Eq.(\ref{H3540}).}
\label{BZflatband}
\end{figure}

We design hopping constants $t_{ij,3}=t_1 e^{\ti \pi/4}$ along the black arrow direction in Fig.\ref{chiral-pi-flux}, and its hermitian conjugate determines $t_{ij,3}=t_1 e^{-\ti \pi/4}$ along the opposite hopping direction;
$t'_{ij,3}=t_2$ along dashed brown links, $t'_{ij,3}=-t_2$ along dotted brown links.
The shaded blue region in Fig.\ref{chiral-pi-flux} indicates a unit cell, containing two sublattice as a black dot $a$ and a white dot $b$.
If we put the lattice model on a torus with periodic boundary conditions for both $x,y$ directions,
then we can write the Hamiltonian in $\mathbf{k}=(k_x,k_y)$ space in Brillouin zone (BZ), as
$H=\sum_\mathbf{k} f^\dagger_\mathbf{k} H(\mathbf{k}) f_\mathbf{k}$,
where $f_\mathbf{k}=(f_{a,\mathbf{k}},f_{b,\mathbf{k}})$.
For two sublattice $a$ and $b$, we have a generic pseudospin form of Hamiltonian $H(\mathbf{k})$,
\be
H(\mathbf{k})=B_0(\mathbf{k}) + \vec{B}(\mathbf{k}) \cdot \vec{\sigma}. 
\ee
$\vec{\sigma}$ are Pauli matrices $(\sigma_x,\sigma_y,\sigma_z)$. In this model $B_0(\mathbf{k})=0$ and
$\vec{B}=(B_x(\mathbf{k}),B_y(\mathbf{k}),B_z(\mathbf{k}))$ have three components
in terms of $\mathbf{k}$ and lattice constants $a_x,a_y$. The eigenenergy $\E_{\pm}$ of $H(\mathbf{k})$ provide two nearly-flat energy bands, shown in Fig.\ref{BZflatband}, from
$H(\mathbf{k}) | \psi_{\pm}(\mathbf{k}) \rangle = \E_{\pm} | \psi_{\pm}(\mathbf{k}) \rangle $.

For the later purpose to have the least mixing between edge states on the left edge A and right edge B on a cylinder in Fig.\ref{chiral-pi-flux}(b), here we fine tune 
$t_2/t_1=1/2$.
For convenience, we simply set $t_1=1$ as the order magnitude of $\E_{\pm}$. We set lattice constants $a_x=1/2,a_y=1$ such that BZ has $-\pi \leq k_x <\pi,-\pi \leq k_y <\pi$.
The first Chern number\cite{Thouless:1982zz} of the 
energy band $|\psi_{\pm}(\mathbf{k}) \rangle$ is
\be
C_1=\frac{1}{2\pi}\int_{\mathbf{k} \in \text{BZ}} \dd^2\mathbf{k}\;  \epsilon^{\mu \nu } \partial_{k_\mu} \langle \psi(\mathbf{k}) | -\ii \partial_{k_\nu} |  \psi(\mathbf{k}) \rangle.
\ee
We find $C_{1,\pm}=\pm 1$ for two bands.
The $C_{1,-}=-1$ lower energy band indicates the clockwise chirality of edge states when we put the lattice on a cylinder as in Fig.\ref{chiral-pi-flux}(b).
Overall it implies the chirality of the edge state on the left edge A moving along $-\hat{x}$ direction, and on the right edge B moving along $+\hat{x}$ direction 
- the clockwise chirality as in Fig.\ref{chiral-pi-flux}(b), consistent with the earlier result
$C_{1,-}=-1$ of Chern number.
This edge chirality is demonstrated in Fig.\ref{kx_chiral}.
Details are explained in its captions and in Appendix \ref{sec:App-Chern-band}.

The above construction is for free fermion edge states with U(1) charge $3$ of the 3$_L$-5$_R$-4$_L$-0$_R$ fermion model. 
Add the same copy with $C_{1,-}=-1$ lower band gives another layer of U(1) charge $4$ free fermion.
For another layers of U(1) charge $5$ and $0$, we simply adjust hopping constant $t_{ij}$ to $t_1 e^{-\ti \pi/4}$ along the black arrow direction and $t_1 e^{\ti\pi/4}$ along the opposite direction in Fig.\ref{chiral-pi-flux}, which makes $C_{1,-}=+1$.
Stack four copies of chiral $\pi$-flux ladders with $C_{1,-}=-1,+1,-1,+1$ provides the lattice model of 3-5-4-0 free fermions with its mirror conjugate.

The lattice model so far is an effective 1+1D non-chiral theory. 
We claim the interaction terms ($G_1,G_2\neq 0$) can gap out the mirror edge states on the edge B.
The simulation including interactions can be numerically expansive, even so on a simple ladder model.
Because of higher power interactions, one can no longer diagonalize the model in $\mathbf{k}$ space as the case of the quadratic free-fermion Hamiltonian.
For 
interacting case, one may need to apply exact diagonalization in real space, or density matrix renormalization group (DMRG\cite{White:1992zz}), which is powerful in 1+1D.
%
We must acknowledge that the numerical simulation of the full interacting systems
is far beyond the scope of this present article. 
However, after the completion of our work,
Ref.~\onlinecite{ZengZhuWangYou2202.12355} performs this demanding numerical task on our model,
and successfully achieves the lattice chiral fermion.
\onecolumngrid
\begin{widetext}
\begin{figure*}[!h] 
{(a)\includegraphics[height=.198\textwidth]{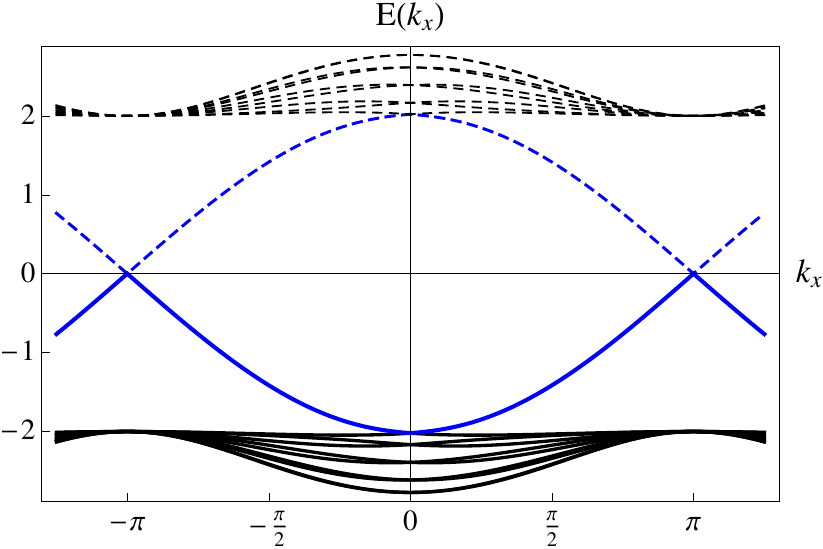} 
~~ (b)\includegraphics[height=.198\textwidth]{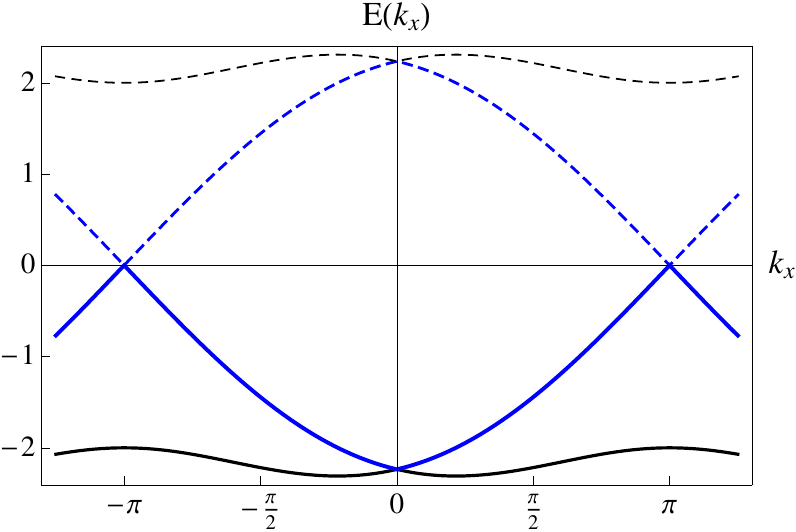}
~~ (c)\includegraphics[height=.197\textwidth]{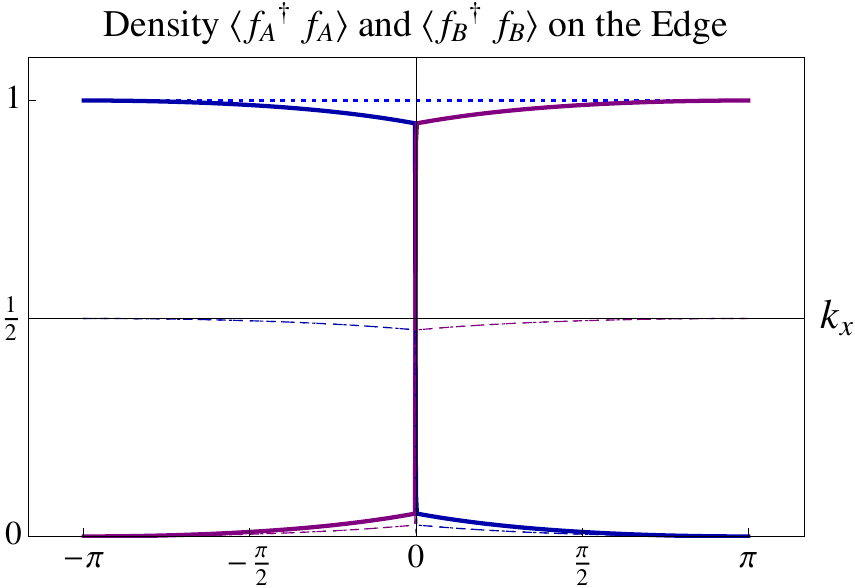}}
\caption{ The energy spectrum $\E(k_x)$ and the density matrix $\langle f^\dagger f \rangle$ of the chiral $\pi$-flux model 
on a cylinder:
{\bf (a)} On a 10-sites width ($9a_y$-width) cylinder: The blue curves are edge states spectrum. The black curves are for states extending in the bulk.
The chemical potential at zero energy fills eigenstates in solid curves, and leaves eigenstates in dashed curves unfilled.
{\bf (b)} On the ladder, a 2-sites width ($1a_y$-width) cylinder: the same as the (a)'s convention.
{\bf (c)} The density $\langle f^\dagger f \rangle$ of the edge eigenstates (the solid blue curve in (b)) on the ladder lattice.
The dotted blue curve shows the total density sums to 1,
the darker purple curve shows $\langle f_{\A}^\dagger f_{\A} \rangle$ on the left edge A, and
the lighter purple curve shows $\langle f_{\B}^\dagger f_{\B} \rangle$ on the right edge B.
The dotted darker (or lighter) purple curve shows density $\langle f_{\A,a}^\dagger f_{\A,a} \rangle$ (or $\langle f_{\B,a}^\dagger f_{\B,a} \rangle$) on sublattice $a$, while
the dashed darker (or lighter) purple curve shows density $\langle f_{\A,b}^\dagger f_{\A,b} \rangle$ (or $\langle f_{\B,b}^\dagger f_{\B,b} \rangle$) on sublattice $b$.
This edge eigenstate has the left edge A density with majority quantum number $k_x<0$, and has the right edge B density with majority quantum number $k_x>0$.
Densities on two sublattice $a,b$ are equally distributed as we desire.\\
{\bf Note}:
Here we do \emph{not} use the domain wall fermion approach with a large extra dimension, 
and we do not require a 1D domain wall in an infinite large 2D lattice system.
We cannot over-emphasize that our 1D spatial lattice model (effectively 1D ladder, or a 2D cylinder with a finite width along $y$, 
here we focus on the quadratic free part of Hamiltonian in Eq.~(\ref{H3540}))
with a finite Hilbert space can already effectively simulate the relativistic 1+1D Weyl fermion doubling theory at low energy.
}
\label{kx_chiral}
\end{figure*}
\end{widetext}
\twocolumngrid

\subsection{ Interaction gapping terms and the intermediate/strong coupling scale} \label{sec:mapping-strong-gapping}

Similar to Sec.\ref{subsubsec:kinetic}, for the {\bf interaction gapping terms} of the Hamiltonian, we can do the mapping based on
Eq.(\ref{eq:subf}), where the leading terms on the lattice is 
\bea \label{subsec:gap_map}
&& g_{a}  \cos(\ell_{a,I}^{} \cdot\Phi_{I}) \\
&& = U_{\text{interaction}}\big( \tilde{\psi}_{q}, \dots,  \nabla^n_x \tilde{\psi}_{q},\dots  \big)  \nonumber \\
&& \to U_{\text{point.split.}}
\bigg(\hat{f}_{q}(j), \dots  \big( \hat{f}^n_{q}(j)\big)_{pt.s.}, \dots
\bigg) 
+ \alpha_{\text{small}} \dots \nonumber
\eea
Again, potentially there may contain subleading pieces, such as
further higher order derivatives $\alpha_{\text{small}} \nabla^n_x \psi_q$ with a small coefficient $\alpha_{\text{small}}$, 
or tiny mixing of the different U(1)-charge flavors $\alpha_{\text{small}}' {\psi_{q_1}\psi_{q_2}\dots}$. 
However, using the same RG analysis in Sec.\ref{subsubsec:kinetic},
at both the weak coupling and the strong coupling fixed points,
we learn that those $\alpha_{\text{small}}$ terms are only {\bf subleading-perturbative effects}
which are {\emph{ further irrelevant perturbation}} at the infrared comparing
to the dominant piece (which is the kinetic term for the weak $g$ coupling, but is replaced by the cosine term for the strong $g$ coupling).


One more question to ask is: what is {\bf the scale of coupling $G$} such that the gapping term becomes dominant and 
the {\bf B edge states form the energy gaps, but 
maintaining} (without interfering with) 
{\bf the gapless A edge states?}

To answer this question, we first know 
the absolute value of energy magnitude 
for each term in the desired Hamiltonian for our chiral fermion model:
\begin{widetext}
\be  \label{eq:energyscale}
|G \text{ gapping term}| \gtrsim |t_{ij},t_{ij}' \text{ kinetic term}|  \gg  |G \text{ higher order $\nabla^n_x$ and mixing terms}| \gg |t_{ij},t_{ij}' \text{ higher order } \psi_q \dots \nabla^n_x \psi_q|. 
\ee
\end{widetext}

For {\bf field theory}, the gapping terms (the cosine potential term or the multi-fermion interactions) are irrelevant for a weak $g$ coupling,
this implies that $g$ needs to be large enough. Here the $g \equiv (g_a)/a^2$ really means the dimensionless quantity $g_a$.

For {\bf lattice model}, however, the dimensional analysis is very different. 
Since the $G$ coupling of gapping terms and the hopping amplitude $t_{ij}$ both have the dimension of energy $[G] =[t_{ij}]=1$,
this means that the scale of the dimensionless quantity of $|G|/|t_{ij}|$ is important.  
(The $|t_{ij}|$ and $|t_{ij}'|$ are about the same order of magnitude.)


Presumably we can design the lattice model under Eq.(\ref{eq:lengthscale}), $a < \xi <  w <  L$,
such that their ratios between each length scale are about the same.
We expect the ratio of couplings of ${|G|}$ to ${|t_{ij}|}$ is about the ratio of energy gap ${\Delta_m}$ to kinetic energy fluctuation ${\delta E_k}$ caused by $t_{ij}$ hopping,
thus \emph{very roughly} 
\be
\frac{|G|}{|t_{ij}|} \sim \frac{\Delta_m}{\delta E_k} \sim \frac{(\xi)^{-1}}{(w)^{-1}} \sim \frac{w}{\xi} \sim \frac{L}{w}  \sim \frac{\xi}{a}. 
\ee
We expect that the scales at strong coupling $G$ is about
\be
|G| \gtrsim  |{t_{ij}}|  \cdot  \frac{\xi}{a}  
\ee 
this magnitude can support our lattice chiral fermion model with mirror-fermion decoupling. 
If $G$ is too much smaller than $ |{t_{ij}}|  \cdot  \frac{\xi}{a} $, then 
mirror sector stays gapless.
On the other hand, if $|G|/|{t_{ij}}|$ is too much stronger 
or simply $|G|/|{t_{ij}}| \to \infty $ 
may cause either of two disastrous cases: \\
(i) Both edges would be gapped and the whole 2D plane becomes \emph{dead without kinetic hopping}.
\\
(ii) The B edge (say at site $n \hat{y}$) becomes completely gapped, but forms a dead layer ---
an overly-high-energy 1D line decoupled from the remain lattice. The neighbored line (along $(n-1) \hat{y}$) next to edge B 
experiences no interaction thus may still form mirror gapless states near B. 
(This may be another reason why CGP fails in Ref.\onlinecite{CGP1247} due to implementing overlarge strong coupling.)\\
So either the two cases caused by too much strong $|G|/|{t_{ij}}|$ is not favorable.
Only $|G| \gtrsim  |{t_{ij}}|  \cdot    \frac{\xi}{a}$, we can have the mirror sector at edge $B$ gapped,
meanwhile keep the chiral sector at edge $A$ gapless. 
$\frac{|G|}{|{t_{ij}}| }$ is somehow larger than order 1 is what we referred as the {\bf intermediate(-strong) coupling}.
\be
\frac{|G|}{|{t_{ij}}| } \gtrsim   O(1).
\ee
(Our $O(1)$ means some finite values, possibly as large as $10^1,10^2$, etc, but still finite. And the kinetic term is \emph{not} negligible.)
The sign of $G$ coupling shall not matter, since in the cosine potential language, either $g_1,g_2$ greater or smaller than zero
are related by sifting the minimum energy vaccua of the cosine potential.\\

To summarize, the two key messages in Sec.\ref{sec:mapping-field-lattice} are:\\ 
\noindent
$\bullet$ First, the free-kinetic hopping part of lattice model has been simulated and there gapless energy spectra have been computed shown in Figures.
The energy spectra indeed show the gapless Weyl fermions on each edge.
So, 
the continuum field theory to a lattice model mapping is immaterial to the subleading terms of Eq.(\ref{eq:subf}), 
the physics is as good or as exact as we expect for the free kinetic part.
We comment that this lattice realization of quantum hall-like states with chiral edges have been implemented for long in condensed matter, dated back as early such as Haldane's work.\cite{Haldane:1988zza}\\
\noindent
$\bullet$ Second, by adding the interaction gapping terms, the spectra will be modified from the mirror gapless edge to the mirror gapped edge. 
The continuum field theory to a lattice model mapping based on
Eq.(\ref{eq:subf}) for the \emph{gapping terms} in Eq.(\ref{subsec:gap_map}) is as good or as exact as the \emph{free kinetic part} Eq.(\ref{subsec:free_map}),
because the mapping is the same procedure as in Eq.(\ref{eq:subf}). 
Since the subleading correction for the free and for the interacting parts are {\emph{ further irrelevant perturbation}} at the infrared,
the {\bf non-perturbative topological effect} of the gapped edge contributed from the leading terms remains. 

\color{black}

In the next section, 
we will provide a
{\bf topological non-perturbative proof}
to justify that the $G_1,G_2$ interaction terms can gap out mirror edge states, without employing numerical methods,
but purely based on an analytical derivation.

\section{Topological Non-Perturbative Proof of Anomaly Matching Conditions = Boundary Fully Gapping Rules} \label{anomaly-gap-proof}

As Sec.~\ref{sec3-5-4-0} and \ref{sec:mapping-field-lattice} prelude, we now show that
Eq.(\ref{H3540}) indeed gaps out the mirror edge states on the edge B in
Fig.\ref{3540}. This proof will support the evidence that Eq.(\ref{H3540})
gives the non-perturbative lattice definition of the 1+1D chiral fermion theory
of Eq.(\ref{cf}).

In Sec.\ref{SPT-CS}, we first provide a generic way to formulate our model, with an insulating bulk but with gapless edge states.
This can be done via the {\bf bulk-edge correspondence},
namely the Chern-Simons theory in the bulk and the Wess-Zumino-Witten (WZW) model on the boundary.
More specifically, for our case with U(1) symmetry chiral matter theory, 
we only need a U(1)$^N$ rank-$N$ Abelian $K$ matrix Chern-Simons theory\cite{Wen:1992uk} 
in the bulk and the multiplet chiral boson theory on the boundary.
We can further fermionize the multiplet chiral boson theory to the multiplet chiral fermion theory.

In Sec.\ref{anomaly-hall}, we provide a physical understanding between
the anomaly matching conditions and the effective Hall conductance.
This intuition will be helpful to understand the relation between the anomaly matching conditions and Boundary Fully Gapping Rules, to be discussed in Sec.\ref{anomaly-gap}. 

\noindent
\subsection{Bulk-Edge Correspondence - 2+1D Bulk Abelian SPTs by Chern-Simons theory \label{SPT-CS}}

With our 3$_L$-5$_R$-4$_L$-0$_R$ chiral fermion model in mind, below we will trace back to fill in the
background how we obtain this model from the understanding of symmetry-protected topological states (SPTs).
This understanding in the end leads to a more general construction.

We first notice that the bosonized action of the free part of chiral fermions in Eq.(\ref{b3540}), can be
regarded as the edge action $S_{\partial}$ of a bulk U(1)$^N$ Abelian $K$ matrix Chern-Simons theory $S_{bulk}$ 
(on a 2+1D manifold ${\cM}$ with the 1+1D boundary ${\partial \cM}$):\cite{Wen:1992uk}
\be
S_{bulk}=
\frac{K_{IJ}}{4\pi}\int_{\cM}   a_I \wedge \dd  a_J = \frac{K_{IJ}}{4\pi}\int_{\cM}  \dd t\, \dd^2x \varepsilon^{\mu\nu\rho} a^I_{\mu} \partial_\nu a^J_{\rho},\;\;\;\;\;\;
\label{CSbulk} 
\ee
\be
S_{\partial}= \frac{1}{4\pi} \int_{\partial \cM} \dd t \; \dd x \; K_{IJ} \partial_t \Phi_{I} \partial_x \Phi_{J} -V_{IJ}\partial_x \Phi_{I}   \partial_x \Phi_{J}. \;\;\;\;\;\;\;
\label{CSboundary}
\ee
Here $a_\mu$ is 
an intrinsic dynamical 1-form gauge field from a low energy 
viewpoint.
Both indices $I,J$ run from $1$ to $N$.
Given a symmetric integer-valued $K_{IJ}$ matrix, it is known the ground state degeneracy (GSD, counting zero energy modes) of this theory on the $\mathbb{T}^2$ torus is 
$$\GSD=|\det K|.\cite{Wen:1992uk}$$
$V_{IJ}$ is the symmetric `velocity' matrix, we can simply choose $V_{IJ}=\mathbb{I}$, without losing generality of our argument.
The U(1)$^N$ gauge transformation is $a_I \to a_I + \dd \lambda_I$ and $\Phi_I \to \Phi_I+ \lambda_I$.
The bulk-edge correspondence is meant to have
the gauge non-invariances of the bulk-only and the edge-only cancel with each other, so that the total gauge invariances is achieved 
from the full bulk and edge as a whole.

We will consider only an even integer $N \in 2 \Z^+$. The reason is that only such even number of edge modes, we can potentially gap out the edge states.
(For odd integer $N$, such a set of gapping interaction terms generically \emph{do not} exist, so the mirror edge states remain gapless.)

To formulate 3$_L$-5$_R$-4$_L$-0$_R$ fermion model, as shown in Eq.(\ref{b3540}), we need a rank-4 $K$ matrix $\bigl( {\begin{smallmatrix}
1 &0 \\
0 & -1
\end{smallmatrix}}  \bigl) \oplus \bigl( {\begin{smallmatrix}
1 &0 \\
0 & -1
\end{smallmatrix}}  \bigl)$. Generically, for a general U(1) chiral fermion model, we can use a canonical fermionic matrix
\be \label{eq:Kf}
K^f_{N\times N} =\bigl( {\begin{smallmatrix}
1 &0 \\
0 & -1
\end{smallmatrix}}  \bigl) \oplus \bigl( {\begin{smallmatrix}
1 &0 \\
0 & -1
\end{smallmatrix}}  \bigl) \oplus \bigl( {\begin{smallmatrix}
1 &0 \\
0 & -1
\end{smallmatrix}}  \bigl) \oplus\dots
\ee
Such a matrix is special, because it describes a more-restricted Abelian Chern-Simons theory with GSD$=|\det K^f_{N\times N}|=1$ on the $\mathbb{T}^2$ torus.
In the condensed matter language, the uniques GSD implies it has no long-range entanglement, and it has no intrinsic topological order. 
Such a state may be wronged to be regarded as only a trivial insulator, but actually this is recently-known to be potentially nontrivial as
the symmetry-protected topological states (SPTs).\footnote{This paragraph 
is for readers with interests in SPTs: SPTs are short-range entangled states with onsite symmetry in the bulk.\cite{Chen:2011pg}
For SPTs, there is no long-range entanglement, no fractionalized quasiparticles (fractional anyons) and no fractional statistics in the bulk.
The bulk onsite symmetry becomes a non-onsite symmetry on the boundary.
The boundary non-onsite symmetry cannot be dynamically gauged in its own dimension ---
the obstruction of gauging a global symmetry is indeed the 't Hooft anomaly.\cite{'tHooft:1979bh}
However, this understanding indicates that if the boundary theory
happens to be 't Hooft anomaly-free under the $G$-symmetry, then the bulk SPTs is in fact a trivial gapped bulk insulator/vacuum that can be removed.
So this anomaly-free boundary theory
can be defined non-perturbatively on the same dimensional lattice with an onsite $G$-symmetry.}

The bulk \emph{spin Chern-Simons theory} with
a $K^f_{N\times N}$ matrix of $|\det K|=1$
describes {\bf fermionic SPT states}.
A spin Chern-Simons theory only exists on the spin manifold, which has spin structure and can further define spinor bundles.\cite{Belov:2005ze}
However, there is another simpler class of SPT states, the {\bf bosonic SPT states}, which
 is described by 
the canonical form  $K^{b\pm}_{N\times N}$\cite{Wang:2012am,canonical,Ye:2013upa} with blocks of
$ \bigl( {\begin{smallmatrix}
0 &1 \\
1 & 0
\end{smallmatrix}} \bigl)$ and a set of all positive 
(or negative) coefficients $\E_8$ lattices $K_{\E_8}$,\cite{Wang:2012am,canonical,Lu:2012dt,{Plamadeala:2013zva}} 
namely,
\bea \label{eq:Kb0}
K^{b0}_{N\times N} &=&\bigl( {\begin{smallmatrix}
0 &1 \\
1 & 0
\end{smallmatrix}}  \bigl) \oplus \bigl( {\begin{smallmatrix}
0 &1 \\
1 & 0
\end{smallmatrix}}  \bigl) \oplus \dots. \\
K^{b \pm}_{N\times N} &=&K^{b0}_{}   \oplus  (\pm K_{\E_8}) \oplus (\pm K_{\E_8}) \oplus  \dots \nonumber
\eea
The $K_{\E_8}$ matrix Chern-Simons theory 
describes the 8-multiplet chiral bosons moving in the same direction, thus it cannot be gapped by adding multi-field interaction among themselves.
We will neglect $\E_8$ chiral boson states but only focus on $K^{b0}_{N\times N}$, 
for the reason to consider \emph{only the gappable states}.
The $K$-matrix form of Eq.(\ref{eq:Kf}),(\ref{eq:Kb0}) is called
the \emph{unimodular indefinite symmetric integral matrix}.

After fermionizing the boundary action Eq.(\ref{CSboundary}) with $K^f_{N\times N}$ matrix, we obtain multiplet chiral fermions 
(with several pairs, each pair contains the left-right moving Weyl fermions forming a Dirac fermion).
\bea \label{CSferboundary}
&S_\Psi&=\int_{\partial \cM}  \dd t \; \dd x \; ( \ti\bar{\Psi}_{\A} \Gamma^\mu  \partial_\mu \Psi_{\A} ),   
\eea
with
$\Gamma^0=\underset{j=1}{\overset{N/2}{\bigoplus}} \gamma^0$, $\Gamma^1 =\underset{j=1}{\overset{N/2}{\bigoplus}}  \gamma^1$, $\Gamma^5\equiv\Gamma^0\Gamma^1$, $\bar{\Psi}_i \equiv \Psi_i \Gamma^0$
and $\gamma^0=\sigma_x$, $\gamma^1=\ti\sigma_y$, $\gamma^5\equiv\gamma^0\gamma^1=-\sigma_z $.\\

\noindent
{\bf Symmetry transformation for the edge states-}

The edge states of $K^f_{N\times N}$ and $K^{b0}_{N\times N}$ Chern-Simons theory are non-chiral, 
in the sense there are equal number of left and right moving modes.
However, we can make them with a charged `chirality' respect to a global (or external probed, or dynamical gauge) symmetry group.
 For the purpose to build up our `chiral fermions and chiral bosons' model with `charge chirality,' we consider the simplest possibility to couple it to a global U(1) symmetry with
 a charge vector $\mathbf{t}$. (This is the same as the symmetry charge vector of SPT states.\cite{Lu:2012dt,Ye:2013upa,Hung:2013nla})

\noindent
{\bf Chiral Bosons}:
For the case of a multiplet chiral boson theory of Eq.(\ref{CSboundary}), the group element $g_{\theta}$ of U(1) symmetry acts on chiral fields as
\bea \label{eq:U(1)}
&& g_{\theta}: W^{\U(1)_{\theta}}=\mathbb{I}_{N \times N}, \;\; \delta \phi^{\U(1)_{\theta}} =\theta \mathbf{t},
\eea
with the following symmetry transformation,
\bea \label{chiralbosonsym}
&&\phi \to   W^{\U(1)_{\theta}}  \phi+\delta \phi^{\U(1)_{\theta}}=\phi+\theta \mathbf{t}.
\eea

To derive this boundary symmetry transformation from the bulk Chern-Simons theory via bulk-edge correspondence,
we first write down the charge coupling bulk Lagrangian term, namely $\frac{\mathbf{q}^I}{2\pi} \; \epsilon^{\mu\nu\rho} A_{\mu} \partial_\nu a^I_{\rho}$, 
where the global symmetry current ${\mathbf{q}^I} J^{I \mu}= \frac{\mathbf{q}^I}{2\pi} \; \epsilon^{\mu\nu\rho} \partial_\nu a^I_{\rho}$ is coupled to an external gauge field $A_\mu$.   
The bulk U(1)-symmetry current ${\mathbf{q}^I} J^{I \mu}$ induces a boundary
U(1)-symmetry current ${\mathbf{q}^I} j^{I \mu}=   \frac{\mathbf{q}^I}{2\pi}   \; \epsilon^{\mu\nu} \partial_\nu \phi_{I}$. 
This implies the boundary symmetry operator  
is $S_{sym}=\exp(\ti \,\theta\, \frac{\mathbf{q}^I}{2\pi}   \int  \partial_x \phi_{I})$, 
with an arbitrary U(1) angle $\theta$.
The induced symmetry transformation 
on $\phi_I $ is: 
\bea \label{chiralbosonsym2}
&&(S_{sym}) \phi_I (S_{sym})^{-1}=\phi_I- \ti \theta\int \dd x \frac{\mathbf{q}^l}{2\pi}[\phi_I,\partial_x \phi_l] \nonumber \\
&&=\phi_I + \theta  (K^{-1})_{I l}  {\mathbf{q}^l}  \equiv \phi_I +\theta \mathbf{t}_I,
\eea
here we have used the canonical commutation relation $[\phi_I,\partial_x \phi_l]=\ti 2 \pi \,(K^{-1})_{I l}  $.
Compare the two Eq.(\ref{chiralbosonsym}),(\ref{chiralbosonsym2}), 
we learn that 
$$\mathbf{t}_I \equiv (K^{-1})_{I l}   {\mathbf{q}^l}.$$
The charge vectors $\mathbf{t}_I$ and ${\mathbf{q}^l}$ are related by an inverse of the $K$ matrix.
The generic interacting or gapping terms\cite{Wang:2012am,Levin:2013gaa,Lu:2012dt} 
for the multiplet chiral boson theory are the sine-Gordon or the cosine term
\be
S_{\partial,\text{gap}}= \int \dd t \; \dd x\;  \sum_{a} g_{a}  \cos(\ell_{a,I}^{} \cdot\Phi_{I}).  
\label{eq:Sgap}
\ee
If we insist that $S_{\partial,\text{gap}}$ obeys U(1) symmetry, to make Eq.(\ref{eq:Sgap}) invariant under Eq.(\ref{chiralbosonsym2}), we have to impose 
\bea
&&\ell_{a,I}^{} \cdot\Phi_{I} \to \ell_{a,I}^{} \cdot(\Phi_{I}+ \delta \phi^{\U(1)_{\theta}}) \text{mod}\; 2\pi \nonumber \\
&&\text{so}\;\;\; 
{\ell_{a,I}^{} \cdot\mathbf{t}_I =0}  \label{eq:t_symmetry}\\
&&\Rightarrow 
{\ell_{a,I}^{} \cdot (K^{-1})_{I l} \cdot  {\mathbf{q}^l} =0}.
 \label{eq:gapsym}
\eea

The above generic U(1) symmetry transformation works for bosonic $K^{b0}_{N\times N}$ as well as fermionic $K^f_{N\times N}$. 

\noindent
{\bf Chiral Fermions}:
In the case of fermionic $K^f_{N\times N}$, we will do one more step to fermionize the multiplet chiral boson theory. Fermionize the free kinetic part from Eq.(\ref{CSboundary}) to Eq.(\ref{CSferboundary}),
as well as the interacting cosine term:
\bea \label{eq:cosine-multif}
&&g_{a}  \cos(\ell_{a,I}^{} \cdot\Phi_{I}) \nonumber \\
&&\to  \prod_{I=1}^{N} \tilde{g}_{a} 
\Big( \big( ({\psi}_{q_I})    (\nabla_x {\psi}_{q_I}) \dots   (\nabla_x^{|\ell_{a,I}|-1} {\psi}_{q_I}) \big)^{\epsilon} +\text{h.c.}\Big)\nonumber\\ 
&&\equiv U_{\text{interaction}}\big( {\psi}_{q}, \dots,  \nabla^n_x  {\psi}_{q},\dots  \big)
\eea
to a multi-fermion interaction. 
The ${\epsilon}$ is defined as the complex conjugation operator which depends on ${\text{sgn}(\ell_{a,I})}$, the sign of $\ell_{a,I}$. 
When ${\text{sgn}(\ell_{a,I})}=-1$, we define ${\psi}^\epsilon \equiv {\psi}^\dagger$ and also for the higher power polynomial terms.
Again, we absorb the normalization factor and the Klein factors through normal ordering of bosonization into the factor $\tilde{g}_{a}$.
The precise factor is not of our concern, 
since our goal is a non-perturbative lattice model. Obviously, the U(1) symmetry transformation for fermions is 
\be \label{chiralfermionsym}
{\psi}_{q_I} \to {\psi}_{q_I} e^{\ti \mathbf{t}_I \theta} ={\psi}_{q_I} e^{\ti (K^{-1})_{I l} \cdot  {\mathbf{q}^l}.  \theta}.
\ee
In summary, we have shown a framework to describe U(1) symmetry chiral fermion/boson model using the bulk-edge correspondence, 
the explicit Chern-Simons/WZW actions are given in Eq.(\ref{CSbulk}),(\ref{CSboundary}),(\ref{CSferboundary}),(\ref{eq:Sgap}),(\ref{eq:cosine-multif}), and their symmetry realization 
Eq.(\ref{chiralbosonsym2}),(\ref{chiralfermionsym}) and constraints are given in Eq.(\ref{eq:t_symmetry}),(\ref{eq:gapsym}). 
Their physical properties are tightly associated with 
the fermionic/bosonic SPT states.

\color{black}

\noindent
\subsection{Anomaly Matching Conditions and Effective Hall Conductance \label{anomaly-hall} }

The bulk-edge correspondence is meant,
not only to achieve the gauge invariance by canceling the non-invariance of bulk-only and boundary-only,
but also to have the boundary anomalous current flow can be transported into the extra dimensional bulk.
This is known as Callan-Harvey effect\cite{Callan_Harvey} in high energy physics. It is also known as
 Laughlin thought experiment,\cite{Laughlin:1981jd} or simply the quantum-hall-like state bulk-edge correspondence in condensed matter theory.

The goal of this subsection is to provide a concrete physical understanding of the anomaly matching conditions and effective Hall conductance : \\

\noindent
$\bullet$ (i) The anomalous current inflowing from the boundary is transported into the bulk. We now show that this thinking 
can easily derive the 1+1D U(1) Adler-Bell-Jackiw (ABJ) anomaly,
or Schwinger's 1+1D quantum electrodynamics (QED) anomaly.

\begin{figure}[h!] 
{(a)\includegraphics[width=.15\textwidth]{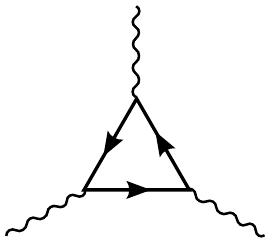}
~~ (b)\includegraphics[width=.20\textwidth]{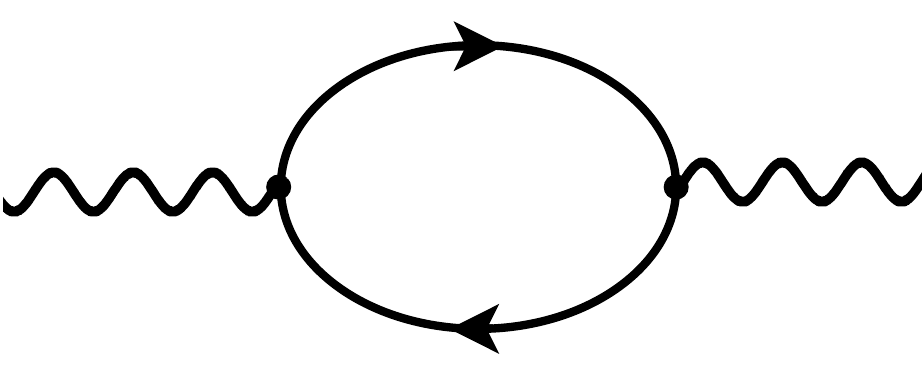}}
\caption{Feynman diagrams with solid lines representing chiral fermions and wavy lines representing U(1) gauge bosons: (a) 3+1D chiral fermionic anomaly shows $\mathcal{A}=\sum_q (q_L^3-q_R^3)$
(b) 1+1D chiral fermionic anomaly shows $\mathcal{A}=\sum_{q} (q_L^2-q_R^2)$.}
\label{anomaly}
\end{figure}

We will focus on the U(1) chiral anomaly, which is ABJ anomaly\cite{Adler:1969gk,Bell:1969ts} type.
It is well-known that ABJ anomaly can be captured by the anomaly factor $\mathcal{A}$ of the 1-loop polygon Feynman diagrams (see Fig.\ref{anomaly}).
The anomaly matching condition requires
\be
\mathcal{A}=\tr[T^aT^bT^c\dots]=0.
\ee
Here $T^a$ is the matrix representation of the Lie algebra generator of the global or gauge symmetry, 
which corresponds to one of the vertices of 1-loop polygon Feynman diagrams. 

For example, the 3+1D chiral anomaly 1-loop triangle diagram of U(1) symmetry in Fig.\ref{anomaly}(a) with chiral fermions on the loop gives $\mathcal{A}=\sum (q_L^3-q_R^3)$.
Similarly, the 1+1D chiral anomaly 1-loop diagram of U(1) symmetry in Fig.\ref{anomaly}(b) with chiral fermions on the loop gives $\mathcal{A}=\sum (q_L^2-q_R^2)$.
Here $L$ and $R$ stand for the left-moving and right-moving modes.

\begin{figure}[h!] 
{\includegraphics[width=.24\textwidth]{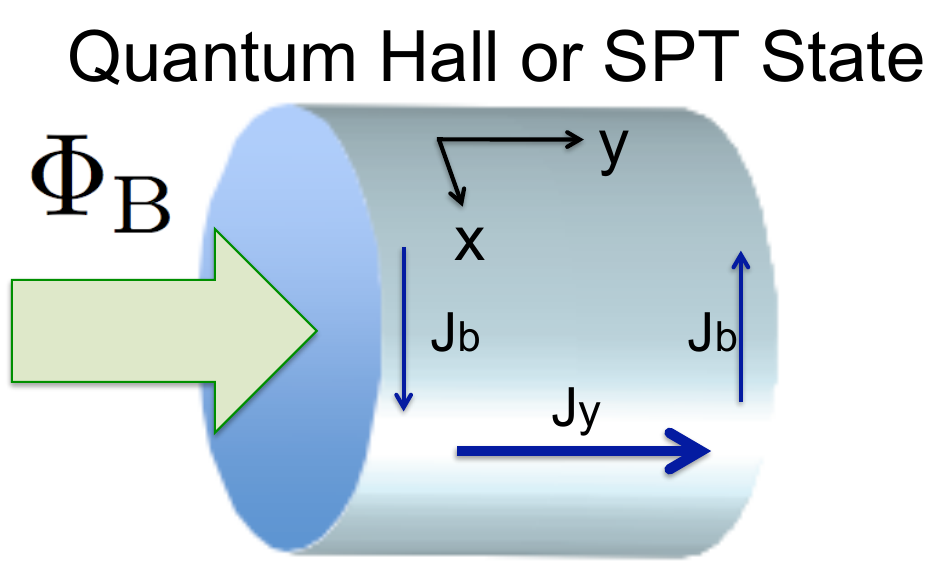}}
\caption{A physical picture illustrates how the anomalous current $J$ of the boundary theory along $x$ direction leaks to the extended bulk system along $y$ direction. 
Laughlin flux insertion $\dd \Phi_B/ \dd t=-\oint E \cdot \dd L$ induces the electric $E_x$ field along the $x$ direction.
The effective Hall effect shows $J_y=\sigma_{xy}E_x=\sigma_{xy}\varepsilon^{\mu\nu}\,\partial_{\mu} A_{\nu}$, with the effective Hall conductance $\sigma_{xy}$
probed by an external U(1) gauge field $A$. 
The anomaly-free condition implies no anomalous bulk current, so $J_y=0$ for any flux $\Phi_B$ or any $E_x$, thus we derive the anomaly-free condition must be $\sigma_{xy}=0$.}
\label{cylinder}
\end{figure}

How to derive this anomaly matching condition from 
a condensed matter theory viewpoint? 
Conceptually, we understand that
``{A $d$-dimensional anomaly free theory (which satisfies the anomaly matching condition) 
means that there is no anomalous current leaking from its $d$-dimensional spacetime (as the boundary)
to an extended bulk theory of $d+1$-dimension.}'' 

More precisely, for a 1+1D U(1) anomalous theory realization of the above statement,
we can formulate it as the boundary of a 2+1D bulk as in Fig.\ref{cylinder} with a Chern-Simons action ($S=\int\big(\frac{K}{4\pi} \;a\wedge \dd a+ \frac{q}{2\pi}A \wedge \dd a$\big)). 
Here the field strength $F=\dd A$ is equivalent to the external U(1) flux in the Laughlin's flux-insertion thought experiment\cite{Laughlin:1981jd} threading through the cylinder
(see a precise derivation in Appendix of Ref.\onlinecite{Santos:2013uda}). 
Without losing generality, let us first focus on the boundary action of Eq.(\ref{CSboundary}) as a chiral boson theory with only one edge mode. We derive its equations of motion as 
\bea 
\partial_{\mu}\,j_{\textrm{b} }^{\mu}
&=&
\frac{\sigma_{xy}}{2}\,
\varepsilon^{\mu\nu}\,F_{\mu\nu}
={\sigma_{xy}}\,
\varepsilon^{\mu\nu}\,\partial_{\mu} A_{\nu}
=
J_{y}, \label{eq:J=sy}\\
\partial_{\mu}\,  j_{\textrm{L}}&=&\partial_{\mu} (\frac{q}{2\pi}\epsilon^{\mu\nu} \partial_\nu \Phi)=\partial_{\mu} (q\bar{\psi} \gamma^\mu  P_L  \psi)=+J_{y},\;\;\; \\
\partial_{\mu}\,  j_{\textrm{R}}&=&-\partial_{\mu} (\frac{q}{2\pi}\epsilon^{\mu\nu} \partial_\nu \Phi)=\partial_{\mu} (q\bar{\psi}  \gamma^\mu P_R \psi)=-J_{y}.\;\;\;
\eea
Here we derive the Hall conductance, easily obtained from its definitive relation $J_y ={\sigma_{xy}} E_x $ in Eq.(\ref{eq:J=sy}), as\cite{W}
$$\sigma_{xy}=qK^{-1}q/(2\pi).$$

Here $j_{\textrm{b}}$ stands for the edge current, with a left-moving current $j_L=j_{\textrm{b}}$ on one edge and a right-moving current $j_R=-j_{\textrm{b}}$ on the other edge, as in Fig.\ref{cylinder}. 
We convert a compact bosonic phase $\Phi$ to the fermion field $\psi$ by bosonization.
We can combine currents $ j_{\textrm{L}}+j_{\textrm{R}}$ as the vector current $j_{\textrm{V}}$, then find its U(1)$_V$ current conserved.
We combine currents $ j_{\textrm{L}}-j_{\textrm{R}}$ as the axial current $j_{\textrm{A}}$, then we obtain the famous ABJ U(1)$_A$ anomalous current in 1+1D (or Schwinger 1+1D QED anomaly).
\bea
\partial_{\mu}\,  j_{\textrm{V}}^{\mu}&=&\partial_{\mu}\, ( j_{\textrm{L}}^{\mu}+j_{\textrm{R}}^{\mu} )= 0,\\ 
\partial_{\mu}\,  j_{\textrm{A}}^{\mu}&=&\partial_{\mu}\, ( j_{\textrm{L}}^{\mu}-j_{\textrm{R}}^{\mu} )=\sigma_{xy} \varepsilon^{\mu\nu}\,F_{\mu\nu}. 
\eea
This simple physical derivation shows that the left and right edges' boundary theories (living on the edge of a 2+1D U(1) Chern-Simons theory) can
combine to be a 1+1D anomalous world of Schwinger's 1+1D QED.

In other words, when the anomaly-matching condition holds ($\mathcal{A}=0$), then
there is no anomalous leaking current into the extended bulk theory,\cite{Callan_Harvey} as in Fig.\ref{cylinder}, so no `effective Hall conductance' for this anomaly-free theory.\cite{Kao:1996ey}

It is straightforward to generalize the above discussion to a rank-$N$ $K$ matrix Chern-Simons theory. It is easy to show that
the Hall conductance in a 2+1D system for a generic $K$ matrix is (via $ {\mathbf{q}_l} =K^{}_{I l} \,\mathbf{t}_I$)
\bea
 {\sigma_{xy}=\frac{1}{2 \pi} \mathbf{q} \cdot {K}^{-1} \cdot \mathbf{q} =\frac{1}{2 \pi} \mathbf{t} \cdot  {K}^{} \cdot  \mathbf{t}.} \;\;\;
\eea
For a 2+1D fermionic system for $K^f$ matrix of Eq.(\ref{eq:Kf}), 
\bea
 \sigma_{xy} =\frac{q^2}{2 \pi} \mathbf{t} {(K^{f}_{N \times N})}^{}\mathbf{t} =\frac{1}{2 \pi}\sum_q (q_{L}^2-q_{R}^2)=\frac{1}{2 \pi}\mathcal{A}.\;\;\;
\eea
%
Remarkably, this physical picture demonstrates that we can reverse the logic, starting from the `{\bf effective Hall conductance} of the bulk system'
to derive the {\bf anomaly factor} from the relation 
\be
{
{ \mathcal{A}\; (\text{anomaly factor})= 2\pi \sigma_{xy} \;(\text{effective Hall conductance})  }  }
\label{eq:A=hall}
\ee
And from the ``no anomalous current in the bulk'' means that ``$\sigma_{xy}=0$'', we can further understand ``the anomaly matching condition $\mathcal{A}=2\pi \sigma_{xy}=0$.'' 

For the U(1) symmetry case, we can explicitly derive the anomaly matching condition for fermions and bosons:\\

\noindent
{\bf Anomaly Matching Conditions for 1+1D chiral fermions with U(1) symmetry}
\be
\mathcal{A}= 2\pi \sigma_{xy}=q^2 \mathbf{t} {(K^{f}_{N \times N})}^{}\mathbf{t} =\sum^{N/2}_{j=1} (q_{L,j}^2-q_{R,j}^2)=0. \label{Anomalyf}
\ee
\noindent
{\bf Anomaly Matching Conditions for 1+1D chiral bosons with U(1) symmetry}
\be
\mathcal{A}= 2\pi \sigma_{xy}=q^2 \mathbf{t} {(K^{b0}_{N \times N})}^{}\mathbf{t} =\sum^{N/2}_{j=1} 2q_{L,j}q_{R,j}=0. \label{Anomalyb}
\ee
Here $q\mathbf{t}\equiv(q_{L,1},q_{R,1}, q_{L,2},q_{R,2},\dots,, q_{L,N/2},q_{R2,N/2})$. 
(For a bosonic theory, we note that the bosonic charge for this theory is described by non-chiral Luttinger liquids. 
One should identify the left and right moving charge as $q_L' \propto q_L+q_R$ and $q_R'\propto q_L-q_R$.)


\noindent
\subsection{Anomaly Matching Conditions and Boundary Fully Gapping Rules  } \label{anomaly-gap}

This subsection is the main emphasis of our work, and we encourage the readers paying extra 
attentions on the result presented here. 
We will first present a heuristic physical argument on the rules that under what situations the boundary states can be gapped,
named as the {\bf Boundary Fully Gapping Rules}. We will then provide a \emph{topological non-perturbative} proof using the notion of Lagrangian subgroup and the exact sequence, following our
previous work Ref.\onlinecite{Wang:2012am} and the work in Ref.\onlinecite{{Kapustin:2013nva},{Levin:2013gaa}}.

\subsubsection{Physical picture}

\begin{figure}[h!] 
{\includegraphics[width=.25\textwidth]{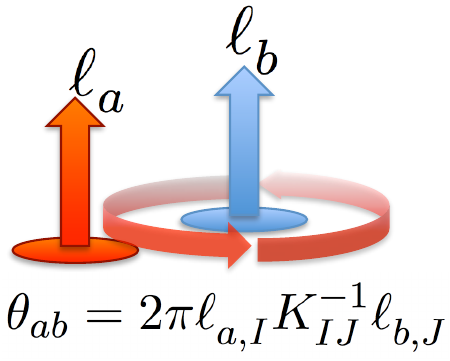} }
\caption{The braiding statistical angle $\theta_{ab}$ of two quasiparticles labeled by $\ell_a,\ell_b$, obtained from
the phase gain $e^{\ii \theta_{ab}}$ in the wavefunction by winding $\ell_a$ around $\ell_b$. 
Here the effective 2+1D Chern-Simons action with the internal 1-form gauge field $a_I$ is $\int\big( \frac{1}{4\pi} K_{IJ} a_I \wedge \dd a_J + a \wedge * j(\ell_a)+ a \wedge * j(\ell_b)\big)$. 
One can integrate out $a$ field to obtain the Hopf term, which coefficient as a
self-statistical angle $\ell_a$ is $\theta_{aa}/2 \equiv \pi\ell_{a,I}^{} K^{-1}_{IJ} \ell_{a,J}^{}$
and the mutual-statistical angle between $\ell_a,\ell_b$ is 
$\theta_{ab} \equiv2 \pi\ell_{a,I}^{} K^{-1}_{IJ} \ell_{b,J}^{}$.\cite{W}
}
\label{braiding}
\end{figure}

Here is the physical intuition: To define a {\bf topological gapped boundary conditions}, it means that the energy spectrum of the edge states are gapped.
We require the gapped boundary to be stable against quantum fluctuations in order to prevent it from flowing back to the gapless states.
Such a gapped boundary must take a stable classical values at the partition function of edge states. From the bosonization techniques,
we can map the multi-fermion interactions to the cosine potential term $g_a \cos(\ell_a \cdot \Phi)$. 
From the bulk-edge correspondence, we learn to regard the 1+1D chiral fermion/boson theory as the edge states of a $K$ matrix Chern-Simons theory,
and further learn that 
the $\ell_a$ vector indeed corresponds to a Wilson line operator $\exp(\ti \int \ell_{a,I} a_I )$
of anyons [integer anyons (fermions or bosons) for $\det(K)=1$ matrix (e.g. SPT states), fractional anyons for $\det(K)>1$ (e.g. Topological Orders).]
However, the nontrivial \emph{braiding statistics}  of anyons of $\ell_a$ vectors will cause quantum fluctuations
to the partition function (or the path integral)
\be
\mathbf{Z}_{\text{statistics}} \sim  \exp[\ti \theta_{ab}]= \exp[ \ti \, 2\pi \, \ell_{a,I}^{} K^{-1}_{IJ} \ell_{b,J}^{}].
\ee
Here the Abelian braiding statistics angle can be derived from 
the effective action between anyon vectors $\ell_a, \ell_b$
by integrating out the internal gauge field $a$ of the Chern-Simons action
$\int\big( \frac{1}{4\pi} K_{IJ} a_I \wedge \dd a_J + a \wedge * j(\ell_a)+ a \wedge * j(\ell_b)\big)$. (See Fig.\ref{braiding}).
In order to define a \emph{classically-stable} topological gapped boundary, we need to stabilize the unwanted quantum fluctuations.
We are forced to choose the trivial statistics for the Wilson lines from the set of interaction terms $g_a \cos(\ell_a \cdot \Phi)$. This requires 
the \emph{trivial statistics} rule
\be
{\text{\bf{Rule}}\;\bf{(1)}} \;\;\;\;\; \ell_{a,I}^{} K^{-1}_{IJ} \ell_{b,J}^{}=0,
\ee
known as the Haldane null condition.\cite{h95}


What else rules do we require? For a total $N$ edge modes, $N_L=N_R=N/2$ number of left/right moving free Weyl fermion modes,
we need to have \emph{at least} $N/2$ interaction terms to open the energy gap. This can be intuitively understood as a pair of modes
can be gapped together if it is a pair of one left-moving to one right-moving mode. It turns out that
if we include \emph{more} linear-independent interactions of $\ell_a$ than $N/2$ terms, such $\ell_a$ cannot be compatible with the
previous set of $N/2$ terms for a compatible trivial mutual or self statistics $\theta_{ab}=0$. 
So we arrive the {\text{\bf{Rule}}\;\bf{(2)}}, ``\emph{no more or no less than the exact $N/2$ interaction terms}.''
And implicitly, we must have  the {\text{\bf{Rule}}\;\bf{(3)}}, ``\emph{$N_L=N_R=N/2$ number of left/right moving modes}.''

So from this physical picture, we have the following rules in order to gap out the edge states of Abelian $K$-matrix Chern-Simons theory:\\


\noindent
\frm{
{\bf Boundary Fully Gapping Rules}\cite{h95,Wang:2012am,Levin:2013gaa,Barkeshli:2013jaa,Lu:2012dt,Hung:2013nla} -
There exists a {\it Lagrangian subgroup}\cite{Levin:2013gaa,Barkeshli:2013jaa,Kapustin:2010hk} $\Gamma^\partial \equiv \{ \sum_a c_a \ell_{a,I}^{} |  c_a \in \Z,  \ell_{a,I} \in \Z\}$
(or named as the {\it boundary gapping lattice}\cite{Wang:2012am} in the $K_{N \times N}$ Abelian Chern-Simons theory),
such that giving a set of interaction terms as the cosine potential terms $g_a \cos(\ell_a \cdot \Phi)$:

\noindent
{\bf{(1)}} $\forall  \ell_a^{}, \ell_b^{} \in \Gamma^\partial_{}$, the self and mutual statistical angles $\theta_{ab}$ 
are zeros among quasiparticles. 
Namely,  
\be
\theta_{ab} \equiv2 \pi\ell_{a,I}^{} K^{-1}_{IJ} \ell_{b,J}^{} =0.
\ee
(For $a=b$, the self-statistical angle  $\theta_{aa}/2 =0$ is called the self-null condition.  And for $a\neq b$, the mutual-statistical angle $\theta_{ab}=0$ is called the mutual-null conditions.\cite{h95})

\noindent
{\bf{(2)}} The dimension of the lattice $\Gamma^\partial$ is $N/2$, where $N$ must be an even integer.
This means the Chern-Simons lattice $\Gamma^\partial$ is spanned by $N/2$ linear independent vectors of $\ell_a^{}$. 

\noindent
{\bf{(3)}} The signature of $K$ matrix (the number of left moving modes $-$ the number of left moving modes) is zero.
Namely $N_L=N_R=N/2$.

\noindent
{\bf{(4)}} ${\ell}_a^{} \in \Gamma_e$, where $\Gamma_e$ is composed by column vectors of $K$ matrix, namely $\Gamma_e=\{ \sum_J c_J K_{IJ} \mid  c_J \in \mathbb{Z} \}$.
$\Gamma_e$ is names as 
the non-fractionalized Chern-Simons lattice.\cite{{Wang:2012am,Wen:1992uk},{particle lattice}}
}


The {\bf{Rule (4)}} is an extra rule, which is not of our main concern here.
This extra rule is for the ground state degeneracy (GSD) matching between the bulk GSD and the boundary GSD while applying the cutting-glueing(or sewing) relations, studied in Ref.\onlinecite{Wang:2012am}.
(Note that the bulk GSD is the topological ground state degeneracy for a bulk closed manifold without boundary,
the boundary GSD is the topological GSD for a compact manifold with gapped boundaries.)
Since we have the unimodular indefinite symmetric integral $K$ matrix of Eq.(\ref{eq:Kf}),(\ref{eq:Kb0}), so {\bf{Rule (4)}} is always true,
for our chiral fermion/boson models.

\subsubsection{{Topological non-perturbative} proof}  \label{sec:topo-nonp-proof}

The above physical picture is suggestive, but not yet rigorous enough mathematically.
Here we will formulate some topological non-perturbative proofs for {\bf Boundary Fully Gapping Rules}, and its equivalence to the {\bf anomaly-matching conditions} for the case of
U(1) symmetry. The first approach is using the topological quantum field theory (TQFT) along the logic of Ref.\onlinecite{Kapustin:2010hk}.
The new ingredient for us is to find \emph{the equivalence of the gapped boundary to the anomaly-matching conditions}. 
We intentionally save the details in 
Appendix \ref{AppendixE}, especially in \ref{sec:TQFT-proof}.

For a field theory, the boundary condition is defined by a Lagrangian submanifold in the
space of Cauchy boundary condition data on the boundary. 
For a topological gapped boundary condition of a TQFT with a gauge group, 
we must choose a Lagrangian subspace in the Lie algebra of the gauge group. 
A subspace is {\bf Lagrangian} \emph{if and only if} it is both {\bf isotropic and coisotropic}. 

Specifically, for $\mathbf{W}$ be a linear subspace of a finite-dimensional vector space $\mathbf{V}$. Define the symplectic complement of $\bfW$ to be the subspace $\bfW^{\perp}$ as
\be
\bfW^{\perp} = \{v\in \bfV \mid \omega(v,w) = 0, \;\;\; \forall w\in \bfW\}
\ee
Here $\omega$ is the symplectic form, in the matrix form
$
\omega=\begin{pmatrix} 0 & \mathbf{1} \\ -\mathbf{1} & 0\end{pmatrix}
$
with $0$ and $\mathbf{1}$ are the block matrix of the zero and the identity.
The symplectic complement $\bfW^{\perp}$ satisfies:
$(\bfW^{\perp})^{\perp} = \bfW$, $\dim \bfW + \dim \bfW^\perp = \dim \bfV$. 
We have:\\

\noindent
$\bullet$  $\bfW$ is Lagrangian if and only if it is both isotropic and coisotropic, namely, 
 if and only if $\bfW = \bfW_{\perp}$. 
In a finite-dimensional $\bfV$, a Lagrangian subspace $\bfW$ is an isotropic one whose dimension is half that of $\bfV$. 

Now let us focus on the $K$-matrix $\text{U(1)}^N$ Chern-Simons theory,
the symplectic form $\omega$  is given by (with the restricted $a_{\parallel,I}$ on ${\partial \cM}$ ) 
\be \label{eq:CSsymp}
\omega=\frac{K_{IJ}}{4\pi} \int_{ \cM}  (\delta a_{\parallel,I}) \wedge \dd (\delta a_{\parallel,J}).
\ee
The bulk gauge group $\text{U(1)}^N  \cong \mathbb{T}_\Lambda$ as the torus, is the quotient space of $N$-dimensional vector space $\bfV$ by a subgroup $\Lambda \cong \Z^N$.
Locally the gauge field $a$ is a 1-form,
which has values in the Lie algebra of $\mathbb{T}_\Lambda$, we can denote this Lie algebra $\mathbf{t}_\Lambda$ as the vector space $\mathbf{t}_\Lambda =\Lambda \otimes \mathbb{R}$.

Importantly, for \emph{topological gapped boundary},
$a_{\parallel,I}$ lies in a Lagrangian subspace of $\mathbf{t}_\Lambda$ implies that the {\bf boundary gauge group} ($ \equiv \mathbb{T}_{\Lambda_0}$)
is a {\bf Lagrangian subgroup}. We can rephrase it in terms of the exact sequence
for the vector space of Abelian group $\Lambda \cong \Z^N$ and its subgroup $\Lambda_0$:
\bea
0 \to \Lambda_0 \overset{\mathbf{h}}{\to} \Lambda \to \Lambda/\Lambda_0 \to 0.
\eea
Here $0$ means the trivial zero-dimensional vector space and 
$\mathbf{h}$ is an injective map from $\Lambda_0$ to $\Lambda$. 
A self-consistent {\bf boundary condition} must %
define a Lagrangian submanifold with
respect to this symplectic form $\omega$.

The generic Lagrangian subgroup condition applies to $K$-matrix with the above symplectic form Eq.(\ref{eq:CSsymp}) renders three conditions on $\bfW $:\\
\frm{
\noindent
$\bullet(i)$  The subspace $\bfW$ is isotropic with respect to the symmetric bilinear form $K$. \\

\noindent 
$\bullet(ii)$  The subspace dimension is a half of the dimension of $\mathbf{t}_\Lambda$. \\

\noindent 
$\bullet(iii)$ The signature of $K$ is zero. This means that $K$ has the same number of positive and negative eigenvalues.\\
}
Now we can examine the if and only if conditions $\bullet(i)$,$\bullet(ii)$,$\bullet(iii)$ listed above. 

For $\bullet(i)$ ``The subspace is isotropic with respect to the symmetric bilinear form $K$'' to be true, we have an extra condition on the injective ${\mathbf{h}}$ matrix (${\mathbf{h}}$ with
$N \times (N/2)$ components) for the $K$ matrix:
\bea \label{eq:TQFThKh}
\boxed{ {\mathbf{h}^T} K {\mathbf{h}}=0}.
\eea
Since $K$ is invertible ($\det(K) \neq 0$), by defining a $N \times (N/2)$-component $\mathbf{L} \equiv K {\mathbf{h}}$, we have an equivalent condition:
\be  \label{eq:TQFTLKL}
\boxed{\mathbf{L}^T K^{-1} \mathbf{L}=0}.
\ee

For $\bullet(ii)$, ``the subspace dimension is a half of the dimension of $\mathbf{t}_\Lambda$'' is true if
$\Lambda_0$ is a rank-$N/2$ integer matrix. 

For $\bullet(iii)$, ``the signature of $K$ is zero'' 
is true, because our $K_{b0}$ and fermionic $K_{f}$ matrices implies that we have same number of left moving modes ($N/2$) and right moving modes ($N/2$), with $N\in 2 \Z^+$ an even number. 

Lo and behold, these above conditions $\bullet(i)$,$\bullet(ii)$,$\bullet(iii)$ are equivalent to the {\bf boundary fully gapping rules} listed earlier. 
We can interpret $\bullet(i)$ as trivial statistics by
either writing in the column vector of ${\mathbf{h}}$ matrix (${\mathbf{h}} \equiv \Big(\eta_{1}, \eta_{2}, \dots,  \eta_{N/2}  \Big)$ with $N \times (N/2)$-components):
\be 
\boxed{\eta_{a,I'} K_{I'J'} \eta_{b,J'}=0}.
\ee
or writing in the column vector of ${\mathbf{L}}$ matrix ($\mathbf{L} \equiv \Big(\ell_{1}, \ell_{2}, \dots,  \ell_{N/2}  \Big)$ with $N \times (N/2)$-components): 
\be 
\boxed{\ell_{a,I} K^{-1}_{IJ} \ell_{b,J}=0}.
\ee
for any $\ell_a, \ell_b \in \Gamma^\partial \equiv \{ \sum_\alpha c_\alpha \ell_{\alpha,I}^{} |  c_\alpha,\ell_{\alpha,I} \in \Z\}$ of boundary gapping lattice (Lagrangian subgroup).
Namely,

\frm{The \emph{boundary gapping lattice} $\Gamma^\partial$ is basically the $N/2$-dimensional 
integer-valued vector space of a Chern-Simons lattice spanned by the $N/2$-independent 
integer-valued column vectors of $\mathbf{L}$ matrix, $\mathbf{L} \equiv \Big(\ell_{1}, \ell_{2}, \dots,  \ell_{N/2}  \Big)$.}

Moreover, we can go a step further 
to relate the above rules equivalent to the {\bf anomaly-matching conditions}. 
By adding the corresponding cosine potential $g_a \cos(\ell_a \cdot \Phi)$ 
to the edge states of U(1)$^N$ Chern-Simons theory, we break the symmetry down to
$$
\U(1)^N \to \U(1)^{N/2}.
$$
What are the remained $\U(1)^{N/2}$ symmetry? By Eq.(\ref{eq:t_symmetry}),
this remained $\U(1)^{N/2}$ symmetry is generated by a number of $N/2$ of $\mathbf{t}_{b,I}$ vectors satisfying ${\ell_{a,I}^{} \cdot\mathbf{t}_{b,I} =0}$. 
We can easily construct 
\be \label{eq:t=KinvL}
\mathbf{t}_{b,I} \equiv K^{-1}_{IJ} \ell_{b,J}, \;\;\; \mathbf{t}  \equiv K^{-1} \mathbf{L}
\ee 
with $N/2$ number of them (or define $\mathbf{t}$ as
the linear-combination of $\mathbf{t}_{b,I} \equiv \sum_{I'} c_{I I'} (K^{-1}_{I'J} \ell_{b,J})$). 
It turns out that
$\U(1)^{N/2}$ symmetry is exactly generated by $\mathbf{t}_{b,I}$ with $b=1,\dots, N/2$, and
these remained unbroken symmetry with $N/2$ of U(1) generators are {\bf anomaly-free} and {\bf mixed anomaly-free}, due to
\be
\boxed{\mathbf{t}_{a,I'} K_{I'J'} \mathbf{t}_{b,J'}={\ell_{a,I'} K^{-1}_{I'J'}   \ell_{b,J'}}=0}.
\ee
Indeed, $\mathbf{t}_a$ must be anomaly-free, because that by defining an
$N \times N/2$ matrix
$\mathbf{t} \equiv \Big(\mathbf{t}_{1}, \mathbf{t}_{2}, \dots,  \mathbf{t}_{N/2}  \Big) =\Big(\eta_{1}, \eta_{2}, \dots,  \eta_{N/2}  \Big)$ of
Eq.(\ref{eq:TQFThKh}), 
we must have:
\bea \label{eq:TQFTtKt}
\boxed{{\mathbf{t}^T} K {\mathbf{t}}=0}, \;\;\; \text{where }\mathbf{t} = \mathbf{h}. 
\eea
This is exactly the anomaly factor and the effective Hall conductance discussed in Sec.\ref{anomaly-hall}.

In summary of the above, we have provided a topological non-perturbative proof that
the {\bf Boundary Fully Gapping Rules} (following Ref.\onlinecite{Kapustin:2010hk}), and its extension
to the equivalence relation to the {\bf anomaly-matching conditions}. 
We emphasize that
{\bf Boundary Fully Gapping Rules} provide a topological statement on the gapped boundary conditions, which is non-perturbative,
while the {\bf anomaly-matching conditions} are also non-perturbative in the sense that the conditions hold at any energy scale,
from low energy IR to high energy UV.
Thus, the equivalence between the twos is remarkable, especially that both are \emph{non-perturbative statements} (namely the proof we provide is
as exact as integer number values without allowing any small perturbative expansion).
Our proof apply to
a bulk U(1)$^N$ $K$ matrix Chern-Simons theory (describing bulk Abelian topological orders or Abelian SPT states)
with boundary multiplet chiral boson/fermion theories. More discussions can be found in Appendix \ref{appendixC}, \ref{appendixD}, \ref{AppendixE}.

\subsubsection{Perturbative arguments} \label{sec:III-perturb}

Apart from the non-perturbative proof using TQFT, we can use other well-known techniques to show the boundary is gapped when the {\bf Boundary Fully Gapping Rules} are satisfied.
Using the techniques systematically studied in Ref.\onlinecite{Wang:2013vna} and detailed in Appendix \ref{app:approachII}, 
it is convenient to map the $K_{N\times N}$-matrix multiplet chiral boson theory to $N/2$ copies of non-chiral Luttinger liquids, each copy with an action
\bea \label{eq:non-chiral-Lutt}
&&\int \dd t \, \dd x \; \Big( \frac{1}{4\pi}( (\partial_t \bar{\phi}_{a} \partial_x \bar{\theta}_{a} +\partial_x \bar{\phi}_{a} \partial_t \bar{\theta}_{a}) -V_{IJ} \partial_x \Phi_{I}   \partial_x \Phi_{J}) \nonumber \\
&&+  g_{}  \cos(\beta \; \bar{\theta}_{a}) \Big)
\eea
at large coupling $g$ at the low energy ground state. Notice that the mapping sends 
$\Phi \to \Phi''=( \bar{\phi}_1, \bar{\phi}_2, \dots, \bar{\phi}_{N/2}, \bar{\theta}_1, \bar{\theta}_2, \dots,  \bar{\theta}_{N/2})$ in a new basis, such that
the cosine potential only takes one field $\bar{\theta}_{a}$ decoupled from the full multiplet.
However, this mapping has been shown 
to be possible \emph{if} $\mathbf{L}^T K^{-1} \mathbf{L}=0$ is satisfied.

When the mapping is done (in Appendix \ref{app:approachII}), we can simply study a single copy of non-chiral Luttinger liquids,
and which, by changing of variables, 
is indeed equivalent to the action of Klein-Gordon fields with a sine-Gordon cosine potential studied by S.\,Coleman.\cite{Coleman:1978ae}
There are various ways to show the existence of energy gap of this sine-Gordon action.
For example,
there is a duality between the quantum sine-Gordon action of bosons and the massive Thirring model of fermions in 1+1D.
In the sense, it is an integrable model, and the Zamolodchikov formula is known \cite{Zamolodchikov:1995xk,Lukyanov:1996jj} and Bethe ansatz can be applicable.
The energy gap is known unambiguously at the large $g$.
 
In short, from the mapping to decoupled $N/2$-copies of non-chiral Luttinger liquids with gapped spectra
together with the anomaly-matching conditions proved in Appendix \ref{appendixC}, \ref{appendixD}, we obtain the relations:
\frm{\center{the U(1)$^{N/2}$ anomaly-free theory ($\mathbf{q}^T \cdot {K}^{-1} \cdot \mathbf{q} = \mathbf{t}^T \cdot  {K}^{} \cdot  \mathbf{t}=0$) with gapping terms $\mathbf{L}^T K^{-1} \mathbf{L}=0$ satisfied. \\
$\updownarrow$\\ 
the $K$ matrix multiplet-chirla boson theories with gapping terms $\mathbf{L}^T K^{-1} \mathbf{L}=0$ satisfied.\\
$\downarrow$ \\ 
 $N/2$-decoupled-copies of non-chiral Luttinger liquid actions with gapped energy spectra.}}

\noindent
$\bullet$ We can also answer other questions using \emph{perturbative analysis}: (Please see Appendix \ref{sec-app-gap-zero} for the details of calculation.)\\ 
(Q1) How can we see explicitly the formation of energy gap necessarily requiring trivial braiding statistics among Wilson line operators (the $\ell_a$ vectors)?\\
(A1) To evaluate the mass gap, we need to know the energy gap of the lowest energy state, namely the \emph{zero mode}. The mode expansion of chiral boson $\Phi$ field on
a compact circular $S^1$ boundary of size $0\leq x<L$ is
\be
\Phi_I(x) ={\phi_{0}}_{I}+K^{-1}_{IJ} P_{\phi_J} \frac{2\pi}{L}x+\ti \sum_{n\neq 0} \frac{1}{n} \alpha_{I,n} e^{-\ii n x \frac{2\pi}{L}},
\ee
where zero modes ${\phi_{0}}_{I}$ and winding modes $P_{\phi_J}$ satisfy the commutator $[{\phi_{0}}_{I},  P_{\phi_J}]=\ti\delta_{IJ}$; and the Fourier modes satisfy generalized Kac-Moody algebra: 
$[\alpha_{I,n} , \alpha_{J,m} ]= n K^{-1}_{IJ}\delta_{n,-m}$.
A \emph{perturbative} way to figure the zero mode's mass is to learn when the zero mode ${\phi_{0}}_{I}$ can be pinned down at the minimum of cosine potential,
with only quadratic fluctuations. In that case, we can evaluate the mass by solving the simple harmonic oscillator problem. 
This requires the following approximation to hold
\bea
&& g_{a} \int_0^{L}\dd x\; \cos(\ell_{a,I}^{} \cdot\Phi_{I}) \nonumber \\
&& \to g_{a} \int_0^{L}\dd x\; \cos(\ell_{a,I}^{} \cdot ({\phi_{0}}_{I}+K^{-1}_{IJ} P_{\phi_J} \frac{2\pi}{L}x) ) \nonumber\\
&& \to g_{a} L \; \cos(\ell_{a,I}^{} \cdot {\phi_{0}}_{I}) \delta_{(\ell_{a,I}^{} \cdot K^{-1}_{IJ} P_{\phi_J} ,0)}.
\eea
In the second line, one neglects the higher energetic Fourier modes; while to have the third line to be true,
it demands a commutator, $[\ell_{a,I}^{}  {\phi_{0}}_{I}, \;\ell_{a,I'}^{}  K^{-1}_{I'J} P_{\phi_J}] = 0$. 
Remarkably, this demands  the null-condition $\ell_{a,J}^{} K^{-1}_{I'J} \ell_{a,I'}^{}=0$,
and the Kronecker delta function restricts the Hilbert space of winding modes $P_{\phi_J}$ residing on the \emph{boundary gapping lattice} $\Gamma^\partial$
due to $\ell_{a,I}^{} \cdot K^{-1}_{IJ} P_{\phi_J}=0$.
Thus, we see that, even at the perturbative level, the formation of energy gap requires trivial braiding statistics among the $\ell_a$ vectors of interaction terms.\\

\noindent
(Q2) What is the scale of the mass gap?\\
(A2) At the \emph{perturbative} level, we compute from a quantum simple harmonic oscillator solution and find the mass gap $\Delta_{m}$ 
above the zero mode: 
$$
\Delta_{m} \simeq \sqrt{2\pi\, g_a \ell_{a,l1} \ell_{a,l2} V_{IJ} K^{-1}_{I l1} K^{-1}_{J l2} }. 
$$

\noindent
(Q3) What happens to the mass gap if we include \emph{more} (\emph{incompatible}) \emph{interaction terms or less interaction terms} with respect to the set of interactions dictated by {\bf Boundary Fully Gapping Rules}
(adding $\ell' \notin \Gamma^\partial$, namely $\ell'$ is not a linear combination of column vectors of $\mathbf{L}$)?\\
(A3) Let us check the \emph{stability} of the mass gap against any \emph{incompatible} interaction term $\ell'$ (which has nontrivial braiding statistics respect to at least one of $\ell_a \in \Gamma^\partial$),
by adding an extra interaction $g' \cos(\ell'_I \cdot \Phi_I)$ to the original set of interactions $\sum_{a} g_{a}  \cos(\ell_{a,I}^{} \cdot\Phi_{I})$.
 We find that as $\ell_{a,I}^{} K^{-1}_{IJ} \ell'_{J} \neq 0$ for the newly added $\ell'$, then the energy spectra for zero modes as well as the higher Fourier modes have the \emph{unstable} form: 
\be \label{eq:Enunstable}
E_n= \big( \sqrt{ \Delta_m^2 + \# (\frac{2\pi n}{L})^2+ \sum_{a} \# g_{a}  \,g' (\frac{L}{n})^2 \dots+\dots } + \dots \big).
\ee
Here $\#$ are denoted as some numerical factors.
Comparing to the case for $g'=0$ (without $\ell'$ term), the energy changes from the \emph{stable} form
$E_n= \big( \sqrt{ \Delta_m^2 + \# (\frac{2\pi n}{L})^2 } + \dots \big)$
to the \emph{unstable} form Eq.(\ref{eq:Enunstable}) at long-wave length low energy ($L \to \infty$) , due to the disastrous term $g_{a}  \,g' (\frac{L}{n})^2$.
The energy has an infinite jump, either from $n=0$ (zero mode) to $n\neq 0$ (Fourier modes), or at $L \to \infty$.

With any incompatible interaction term of $\ell'$, the pre-formed mass gap shows an instability. 
This indicates the \emph{perturbative} analysis may not hold, and the zero modes cannot be pinned down at the minimum.
The consideration of instanton tunneling and talking between different minimum may be important when $\ell_{a,I}^{} K^{-1}_{IJ} \ell'_{J} \neq 0$.
In this case, we expect the massive gapped phase is not stable, and the phase could be gapless.
{\bf Importantly, this can be one of the reasons why the numerical attempts of Chen-Giedt-Poppitz model finds gapless phases instead of gapped phases.}
\emph{The immediate reason is that their Higgs terms induce many extra interaction terms, not compatible 
with the terms dictated by Boundary Fully Gapping Rules.
As we checked explicitly, many of their induced terms break the U(1)$_{\text{2nd}}$ symmetry 0-4-5-3, which is not compatible to the set inside $\Gamma^\partial$ or $\mathbf{L}$ matrix.}
See further discussions in Sec.\ref{sec:Interaction-terms}.
\\

\subsubsection{Preserved U(1)$^{N/2}$ symmetry and a unique ground state} 

We would like to discuss the symmetry of the system further. 
As we mention in Sec.\ref{sec:topo-nonp-proof}, the symmetry is broken down from 
$\U(1)^N \to \U(1)^{N/2}$ by adding $N/2$ gapping terms with $N=4$.  
In the case of gapping terms $\ell_1=(1,1,-2,2)$ and $\ell_2=(2,-2,1,1)$,
we can find the unbroken symmetry by Eq.(\ref{eq:t=KinvL}), 
where the symmetry charge vectors are $\mathbf{t}_1=(1,-1,-2,-2)$ and $\mathbf{t}_2=(2,2,1,-1)$.
The symmetry vector can have another familiar linear combination $\mathbf{t}_1=(3,5,4,0)$ and $\mathbf{t}_2=(0, 4, 5, 3)$, which indeed matches to our original 
U(1)$_{\text{1st}}$ 3-5-4-0 and U(1)$_{\text{2nd}}$ 0-4-5-3 symmetries.
Similarly, the two gapping terms can have another linear combinations: $\ell_1=(3,-5,4,0)$ and $\ell_2=(0,4,-5,3)$.
We can freely choose any linear-independent combination set of the following,
%
%
%
\bea \label{eq:Lt3540dual}
&&\mathbf{L}=\left(
\begin{array}{cc}
 3 & 0 \\
 -5 & 4 \\
 4 & -5 \\
 0 & 3
\end{array}
\right),\left(
\begin{array}{cc}
 1 & 2 \\
 1 & -2 \\
 -2 & 1 \\
 2 & 1
\end{array} 
\right), \dots \\
&&\Longleftrightarrow
\mathbf{t}=\left(
\begin{array}{cc}
 3 & 0 \\
 5 & 4 \\
 4 & 5 \\
 0 & 3
\end{array}
\right),\left(
\begin{array}{cc}
 1 & 2 \\
 -1 & 2 \\
 -2 & 1 \\
 -2 & -1
\end{array} 
\right), \dots. \nonumber
\eea
and we emphasize the vector space spanned by the column vectors of $\mathbf{L}$ 
and $\mathbf{t}$ (the complement space of $\mathbf{L}$'s) 
will be the entire 4-dimensional vector space $\mathbb{Z}^4$.
In Appendix \ref{sec-app-3540}, we will provide the lattice construction for the alternative $\mathbf{L}$, see Eq.(\ref{H3540App}).

Now we like to answer: \\
(Q4) Whether the $\U(1)^{N/2}$ symmetry stays unbroken when the mirror sector becomes gapped by the strong interactions?\\
(A4) The answer is Yes. We can check: There are two possibilities that $\U(1)^{N/2}$ symmetry is broken. 
\noindent
One is that it is \emph{explicitly broken} by the interaction term.
This is not true. 
\noindent
The second possibility is that the ground state (of our chiral fermions with the gapped mirror sector) \emph{spontaneously or explicitly break}
the $\U(1)^{N/2}$ symmetry. This possibility can be checked by calculating its {\bf ground state degeneracy (GSD) on the cylinder with gapped boundary}.
Using the method developing in our previous work Ref.\onlinecite{Wang:2012am}, also in Ref.\onlinecite{{Kapustin:2013nva},{Wang:2013vna},{Lan2014uaaLWW1408.6514}}, 
we find GSD=1,
there is only a unique ground state. Because there is only one lowest energy state, it cannot \emph{spontaneously or explicitly break} the remained symmetry.
The GSD is 1 as long as the $\ell_a$ vectors are chosen to be the minimal vector, namely the greatest common divisor(gcd) among each component of any $\ell_a$ is 1,
${|\gcd(\ell_{a,1}, \ell_{a,2}, \dots,  \ell_{a,N/2} \Big) |}=1$, such
that 
$$
\ell_a \equiv \frac{(\ell_{a,1}, \ell_{a,2}, \dots,  \ell_{a,N/2}  )}{|\gcd(\ell_{a,1}, \ell_{a,2}, \dots,  \ell_{a,N/2}) |}.
$$

In addition, 
thanks to Coleman-Mermin-Wagner theorem, there is \emph{no spontaneous symmetry breaking for any continuous symmetry in 1+1D, due to no Goldstone modes 
in 1+1D}, we can safely conclude that $\U(1)^{N/2}$ symmetry stays unbroken. 
\color{black}

 
 To summarize the whole Sec.\ref{anomaly-gap-proof}, we provide both non-perturbative and perturbative analysis on {\bf Boundary Fully Gapping Rules}. 
 This applies to a generic $K$-matrix U(1)$^{N}$ Abelian Chern-Simons theory with a boundary multiplet chiral boson theory.
 (This generic $K$ matrix theory describes general Abelian topological orders including all Abelian SPT states.)

In addition, in the case when K is \emph{unimodular indefinite symmetric integral matrix}, for both fermions $K=K^{f}$ and bosons $K=K^{b0}$, we have further proved:
\frm{{\bf Theorem:} The boundary fully gapping rules of 1+1D boundary/2+1D bulk with unbroken 
U(1)$^{N/2}$ symmetry preserving $\leftrightarrow$ ABJ's U(1)$^{N/2}$ anomaly matching conditions in 1+1D.}
Similar to our non-perturbative algebraic result on topological gapped boundaries, 
the {\bf 't Hooft anomaly matching} here is a non-perturbative statement, being exact from IR to UV, insensitive to the energy scale.

\noindent
\section{General Construction of Non-Perturbative Anomaly-Free chiral matter model from SPTs \label{model}}

As we already had an explicit example of 3$_L$-5$_R$-4$_L$-0$_R$ 
chiral fermion model introduced in Sec.\ref{sec3-5-4-0},\ref{numeric}, 
and we had paved the way building up tools and notions in Sec.\ref{anomaly-gap-proof}, 
now we are finally here to present our general model construction.
Our construction of non-perturbative anomaly-free chiral fermions and bosons model with onsite U(1) symmetry is the following.\\

\noindent
{\bf Step 1}: We start with a $K$ matrix Chern-Simons theory as in Eq.(\ref{CSbulk}),(\ref{CSboundary})
for \emph{unimodular indefinite symmetric integral $K$ matrices}, both fermions $K=K^{f}$ of Eq.(\ref{eq:Kf})
and bosons $K=K^{b0}$ of Eq.(\ref{eq:Kb0}) (describing generic Abelian SPT states with GSD on torus is $|\det(K)|=1$.)\\

\noindent
{\bf Step 2}: We assign charge vectors $\mathbf{t}_a$ of U(1) symmetry 
as in Eq.(\ref{eq:U(1)}), which satisfies the anomaly matching condition Eq.(\ref{Anomalyf}) for fermionic model,
or satisfies Eq.(\ref{Anomalyb}) for bosonic model. 
We can assign up to $N/2$ charge vector
$\mathbf{t} \equiv \Big(\mathbf{t}_{1}, \mathbf{t}_{2}, \dots,  \mathbf{t}_{N/2} \Big)$ with a total U(1)$^{N/2}$ symmetry with the matching
$\mathcal{A}= {{\mathbf{t}^T} K {\mathbf{t}}=0}$
such that the model is anomaly and mixed-anomaly free. \\

\noindent
{\bf Step 3}: In order to be a \emph{chiral} theory, it needs to \emph{violate the parity symmetry}.  
In our model construction, assigning $q_{L,j} \neq q_{R,j}$ generally fulfills our aims by breaking both parity and time reversal symmetry. (See Appendix \ref{appendixA} for details.)\\

\noindent
{\bf Step 4}: By the equivalence of the anomaly matching condition and boundary fully gapping rules
(proved in Sec.\ref{sec:topo-nonp-proof} and Appendix \ref{appendixC},\ref{appendixD}), our anomaly-free theory guarantees that
a proper choice of gapping terms of Eq.(\ref{eq:Sgap}) can fully gap out the edge states. 
For $N_L=N_R=N/2$ left/right Weyl fermions, there are $N/2$ gapping terms ($\mathbf{L} \equiv \Big(\ell_{1}, \ell_{2}, \dots,  \ell_{N/2}  \Big)$), and
the U(1) symmetry can be extended to U(1)$^{N/2}$ symmetry by finding the corresponding $N/2$ charge vectors ($\mathbf{t} \equiv \Big(\mathbf{t}_{1}, \mathbf{t}_{2}, \dots,  \mathbf{t}_{N/2}  \Big)$). 
The topological non-perturbative proof found in Sec.\ref{sec:topo-nonp-proof} guarantees the duality relation:
\bea  \label{eq:TQFTLKLTKT}
&&\boxed{\mathbf{L}^T \cdot K^{-1} \cdot \mathbf{L}=0 \underset{ \mathbf{L}=K^{} \mathbf{t}  }{\overset{ \mathbf{t}=K^{-1} \mathbf{L} }{\longleftrightarrow}} \mathbf{t}^T \cdot  {K}^{} \cdot  \mathbf{t}=0}.
\eea
Given $K$ as a $N \times N$-component matrix of $K^{f}$ or $K^{b0}$, we have $\mathbf{L}$ and $\mathbf{t}$ are both $N \times (N/2)$-component matrices.

So our strategy is that constructing the bulk SPTs on a 2D spatial lattice with two edges
(for example, 
a cylinder in Fig.\ref{3540},Fig.\ref{cylinder}). 
The low energy edge property of the 2D lattice model has the same continuum field theory\cite{fermionization1} as we had in Eq.(\ref{CSboundary}),
and selectively only fully gapping out states on one mirror edge with a large energy gap by adding symmetry-allowed gapping terms Eq.(\ref{eq:Sgap}),
while leaving the other side gapless edge states untouched.\cite{Wen:2013ppa}

In summary, 
we start with a chiral edge theory of SPT states with $ \cos(\ell_{I}^{} \cdot\Phi^B_{I})$ gapping terms on the edge B, which action is
\bea 
S_{\Phi} 
&=&\frac{1}{4\pi}  \int \dd t \dd x  \big(K^{\A}_{IJ}  \partial_t \Phi^{\A}_I   \partial_x \Phi^{\A}_{J} -V_{IJ}  \partial_x \Phi^{\A}_I   \partial_x \Phi^{\A}_{J}\big)\;\;\;\nonumber \\
&+&\frac{1}{4\pi}  \int \dd t \dd x \big(K^{\B}_{IJ}  \partial_t \Phi^{\B}_I   \partial_x \Phi^{\B}_{J} -V_{IJ}  \partial_x \Phi^{\B}_I   \partial_x \Phi^{\B}_{J} \big) \label{eq:bosonize-all}        \;\;\;\nonumber\\
&+&\int \dd t \dd x  \; \sum_{a} g_{a}  \cos(\ell_{a,I}^{} \cdot \Phi^{\B}_{I}).  \;\;\;\;\;\;\;
 \eea
We fermionize the action to:
\bea 
&S_\Psi&=\int  \dd t \; \dd x \; ( \ii\bar{\Psi}_{\A} \Gamma^\mu  \partial_\mu \Psi_{\A}+ \ii\bar{\Psi}_{\B} \Gamma^\mu  \partial_\mu \Psi_{\B}  \nonumber   \\
 &\;&+U_{\text{interaction}}\big( \tilde{\psi}_{q}, \dots,  \nabla^n_x \tilde{\psi}_{q},\dots  \big) ).    \label{eq:fermionize-all}
\eea
with
$\Gamma^0$, $\Gamma^1$, $\Gamma^5$ follow the notations of Eq.(\ref{CSferboundary}).

The gapping terms on the field theory side need to be irrelevant operators or marginally irrelevant operators 
with appropriate strength (to be order 1 intermediate-strength for the dimensionless lattice coupling $|G|/|t_{ij}| \gtrsim O(1)$), 
so it can gap the mirror sector, but it is weak enough to keep the original light sector gapless.

Use several copies of Chern bands to simulate the free kinetic part of Weyl fermions, 
and convert the higher-derivative fermion interactions $U_{\text{interaction}}$ to the point-splitting
$U_{\text{point.split.}}$ term on the lattice, we propose its corresponding lattice Hamiltonian 
\bea 
\label{eq:H-int-all}
H&=&\sum_{q} 
\bigg(  \sum_{\langle i, j \rangle}
\big(t_{ij,q}\; \hat{f}^\dagger_{q}(i)
\hat{f}_{q}(j)+\text{h.c.}\big) \\
&+& \sum_{\langle\langle i, j
\rangle\rangle} \big(  t'_{ij,q}
\;\hat{f}^\dagger_{q}(i) \hat{f}_{q}(j)+\text{h.c.}\big) \bigg) \nonumber \\ 
&+& \sum_{j \in \B} U_{\text{point.split.}}
\bigg(\hat{f}_{q}(j), \dots  \big( \hat{f}^n_{q}(j)\big)_{pt.s.}, \dots
\bigg).\;\; 
\nonumber 
\eea
The key to avoid Nielsen-Ninomiya challenge\cite{Nielsen:1980rz,Nielsen:1981xu,Nielsen:1981hk} is that our model has the \emph{properly-desgined} interactions. 
\frm{ We have obtained {\bf a 1+1D non-perturbative lattice Hamiltonian construction (and realization) of anomaly-free massless chiral fermions (and chiral bosons)} on one gapless edge. Since the extra-dimensional trivially gapped bulk (that separates two 1+1D edges) has a finite width, 
the whole system can be effectively regarded as a 1+1D system.
}

For readers with interests, In Appendix \ref{app:more_models}, we will demonstrate 
a step-by-step construction on several lattice Hamiltonian models of chiral fermions
and chiral bosons, based on our general prescription above. 
In short, our approach is generic for constructing many anomaly-free
lattice chiral matter models in 1+1D.
  
\section{Conclusion: Compare ours with {Eichten-Preskill} and {Chen-Giedt-Poppitz} models \label{summary}} 

We have proposed a 1+1D lattice Hamiltonian definition of non-perturbative anomaly-free chiral matter models with U(1) symmetry. 
Our 3$_L$-5$_R$-4$_L$-0$_R$ fermion model is under the framework of the mirror fermion decoupling approach. 
However, some importance essences make our model distinct from the lattice models of {Eichten-Preskill}\cite{EP8679} and {Chen-Giedt-Poppitz} 3-4-5 model.\cite{CGP1247}
The differences between our and theirs are listed below.

\subsection{Onsite v.s. non-onsite symmetry}
\noindent
\underline{\bf Onsite or non-onsite symmetry}.
Our model only implements onsite symmetry, which can be easily to be gauged. 
While {Chen-Giedt-Poppitz} model implements Ginsparg-Wilson (GW) fermion approach with non-onsite symmetry (details explained in Appendix \ref{appendixB}).
To have 
GW relation $\{D,\gamma^5\}=2aD\gamma^5D$ to be true ($a$ is the lattice constant), the
Dirac operator is non-onsite (not strictly local) as $D(x_1,x_2) \sim e^{-|x_1-x_2|/{\xi} }$ but with a distribution range $\xi$. The axial U(1)$_A$ symmetry is modified
$$\delta \psi(y) = \sum_w \ti\, \theta_A \hat{\gamma}_5 (y,w) \psi(w),\;\;\; \delta \bar{\psi}(x) = \ti\, \theta_A \bar{\psi}(x)  {\gamma}_5$$
with the operator $\hat{\gamma}_5(x,y) \equiv \gamma_5 - 2 a \gamma_5 D(x,y)$. 
Since its axial U(1)$_A$ symmetry transformation contains $D$ and the Dirac operator $D$ is non-onsite, the GW approach necessarily implements non-onsite symmetry.
GW fermion has non-onsite symmetry in the way that it cannot be written as the tensor product structure on each site:
$U(\theta_A)_{\text{non-onsite}} \neq \otimes _j U_j(\theta_A)$, for $e^{\ti \theta_A} \in \U(1)_A$. 
The Neuberger-Dirac operator also contains such a non-onsite symmetry feature.
The non-onsite symmetry is the signature property of the boundary theory of SPT states.
The non-onsite symmetry causes GW fermion difficult (or impossible for certain groups) to be gauged to a chiral gauge theory,
because the gauge theory is originally defined by gauging the local (on-site) degrees of freedom.

\subsection{Interaction terms}
\label{sec:Interaction-terms}

\noindent
\underline{\bf Interaction terms}.
Our model has properly chosen a particular set of interactions satisfying the Eq.(\ref{eq:TQFTLKLTKT}), from the Lagrangian subgroup algebra
to define a topological gapped boundary conditions. On the other hand, {Chen-Giedt-Poppitz} model proposed different kinds of interactions - 
all Higgs terms obeying U(1)$_{\text{1st}}$ 3-5-4-0 symmetry (Eq.(2.4) of Ref.\onlinecite{CGP1247}), 
including the 
Yukawa-Dirac terms: 
\bea \label{eq:yukawa-d}
&&\int \dd t \dd x\Big(\mathrm{g}_{30} \psi_{L,3}^\dagger \psi_{R,0} \phi_h^{{\dagger}3}+
\mathrm{g}_{40} \psi_{L,4}^\dagger \psi_{R,0}  \phi_h^{{\dagger}4} \nonumber\\
&&+\mathrm{g}_{35} \psi_{L,3}^\dagger \psi_{R,5}  \phi_h^{2}+\mathrm{g}_{45} \psi_{L,4}^\dagger \psi_{R,5}  \phi_h^{1}+ \text{h.c.}\Big),
\eea
with Higgs field $ \phi_h(x,t)$ carrying charge $(-1)$. There are also Yukawa-Majorana terms:
\bea \label{eq:yukawa-m}
&&\int \dd t \dd x\Big(\ti \mathrm{g}_{30}^{M} \psi_{L,3} \psi_{R,0} \phi_h^{3}+
\ti \mathrm{g}_{40}^{M} \psi_{L,4} \psi_{R,0}  \phi_h^{4} \nonumber\\
&&+\ti \mathrm{g}_{35}^{M} \psi_{L,3} \psi_{R,5}  \phi_h^{8}+ \ti \mathrm{g}_{45}^{M} \psi_{L,4} \psi_{R,5}  \phi_h^{9}+ \text{h.c.}\Big),
\eea 
{Notice that the Yukawa-Majorana coupling has an extra imaginary number $\ti$ in the front, and implicitly there is also a
Pauli matrix $\sigma_y$ if we write the Yukawa-Majorana term in the two-component Weyl basis.}

The question is: {\bf How can we compare between interactions of ours and Ref.\onlinecite{CGP1247}'s?}
If integrating out the Higgs field $\phi_h$, we find that:\\
$(\star 1)$ Yukawa-Dirac terms of Eq.(\ref{eq:yukawa-d}) 
\emph{cannot} generate any of our multi-fermion interactions of $\mathbf{L}$ in Eq.(\ref{eq:Lt3540dual}) for our 3$_L$-5$_R$-4$_L$-0$_R$ model.\\
$(\star 2)$ Yukawa-Majorana terms of Eq.(\ref{eq:yukawa-m}) 
\emph{cannot} generate any of our multi-fermion interactions of $\mathbf{L}$ in Eq.(\ref{eq:Lt3540dual}) for our 3$_L$-5$_R$-4$_L$-0$_R$ model.\\
$(\star 3)$ Combine Yukawa-Dirac and Yukawa-Majorana terms of Eq.(\ref{eq:yukawa-d}),(\ref{eq:yukawa-m}), one can indeed generate 
the multi-fermion interactions of $\mathbf{L}$ in Eq.(\ref{eq:Lt3540dual}); however, many more multi-fermion interactions 
outside of the Lagrangian subgroup (not being spanned by $\mathbf{L}$) are generated. Those extra unwanted multi-fermion interactions \emph{do not} obey the boundary fully gapping rules.
As we have shown in Sec.\ref{sec:III-perturb} and Appendix \ref{sec-app-gap-zero}, those extra unwanted interactions induced by the Yukawa term will cause the pre-formed mass gap unstable
due to the nontrivial braiding statistics between the interaction terms. {This explains 
why the massless mirror sector is observed in Ref.\onlinecite{CGP1247}.
In short, we know that  Ref.\onlinecite{CGP1247}'s interaction terms are different from us, and know that the properly-designed interactions are crucial, and
our proposal 
will succeed the mirror-sector-decoupling even if Ref.\onlinecite{CGP1247} fails.}\\

\noindent
\underline{\bf{Symmetry breaking: $\U(1)^{N}\to \U(1)^{N/2} \to \U(1)$}}.
We have shown that for a given $N_L=N_R=N/2$ equal-number-left-right moving mode theory, 
the $N/2$ gapping terms break the symmetry from $\U(1)^{N}\to \U(1)^{N/2}$. 
Its remained $ \U(1)^{N/2}$ symmetry is unbroken and mixed-anomaly free.
{\bf Is it possible to further add interactions to break $ \U(1)^{N/2}$ to a smaller symmetry, such as a single U(1)? }
For example, breaking the U(1)$_{\text{2nd}}$ 0-4-5-3 of 3$_L$-5$_R$-4$_L$-0$_R$ model to only a single U(1)$_{\text{\text{1st}}}$ 3-5-4-0 symmetry remained.
We argue that it is doable. %
Adding any extra explicit-symmetry-breaking term may be incompatible to the original Lagrangian subgroup and thus potentially ruins the stability of the energy gap. 
Nonetheless,
{\bf as long as we add an extra interaction term (breaking the U(1)$_{\text{2nd}}$ symmetry), which is irrelevant operator with 
a tiny coupling},
it can be weak enough not driving the system to gapless states.
Thus, our setting to obtain 3-5-4-0 symmetry is still quite different from {Chen-Giedt-Poppitz} where the {\bf universal intermediate/strong couplings} are applied.



\begin{figure}[!h]
{\includegraphics[width=.40\textwidth]{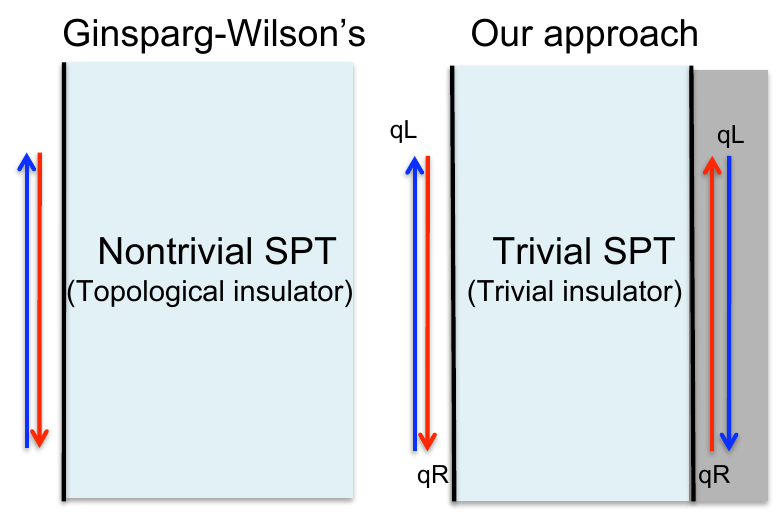}}
\caption{Ginsparg-Wilson fermions can be viewed as putting gapless states on the edge of a nontrivial SPT state (e.g. topological insulator).
Our approach can be viewed as putting gapless states on the edge of a trivial SPT state (trivial insulator).}
\label{G-WandOurs}
\end{figure}

We show that GW fermion approach implements the {\it non-onsite symmetry} (more in Appendix \ref{appendixB}), 
thus GW can avoid the fermion-doubling no-go theorem (limited to an {\it onsite symmetry}) to
obtain chiral fermion states. 
Remarkably, this also suggests that 
\frm{The nontrivial edge states of SPT order,\cite{Chen:2011pg} such as topological insulators\cite{TI4,TI5,TI6}  alike,
can be obtained in its own dimension (without the need of an extra dimension to the bulk)
by implementing the {\it non-onsite symmetry} as Ginsparg-Wilson fermion approach.}
To summarize, 
so far we have learned (see Fig.\ref{G-WandOurs}),
\begin{itemize}
\item {\bf Nielsen-Ninomiya theorem} claims that local free chiral fermions on the lattice with onsite (U(1) or chiral\cite{U(1)sym}) symmetry have fermion-doubling problem in an even-dimensional spacetime.
\item {\bf  Ginsparg-Wilson (G-W) fermions}: quasi-local free chiral fermions on the lattice with non-onsite U(1) symmetry\cite{U(1)sym} have no fermion doublers. 
G-W fermions correspond to gapless edge states of a nontrivial SPT state. 
\item {\bf Our 3-5-4-0 chiral fermion and general model constructions}: local interacting chiral fermions on the lattice with onsite U(1) symmetry\cite{U(1)sym} have no fermion-doublers.
Our model corresponds to unprotected gapless edge states of a trivial SPT state (i.e. a trivial insulator).
\end{itemize}
%


We should also clarify that, from SPT classification viewpoint, 
all our chiral fermion models are in the same class of 
$K^f=({\begin{smallmatrix}
1 &0 \\
0 & -1
\end{smallmatrix}} )$ with $\mathbf{t}=(1,-1)$, a {\it trivial class} in the fermionic SPT with U(1) symmetry.\cite{Lu:2012dt,Ye:2013upa,JWunpublished} %
All our chiral boson models are in the same class of 
$K^b=({\begin{smallmatrix}
0 &1 \\
1 & 0
\end{smallmatrix}} )$ with $\mathbf{t}=(1,0)$, a {\it trivial class} in the bosonic SPT with U(1) symmetry.\cite{Lu:2012dt,Ye:2013upa,JWunpublished} %
In short, we understand that 
\frm{ From the 2+1D bulk theory viewpoint, all our chiral matter models are {\it equivalent} to the {\it trivial class} of SPT (trivial bulk insulator),
thus the 2+1D bulk can be entirely removed.
However,
the 1+1D boundary theories with different U(1) charge vectors $\mathbf{t}$ 
can be regarded as {\it different} chiral matter theories on its own 1+1D.}

\noindent
{\bf{Proof of a Special Case and some Conjectures}}

At this stage, we already fulfill proposing our models.
On the other hand, the outcome of our proposal becomes fruitful with deeper implications.
We prove that,  at least for 1+1D boundary/2+1D bulk SPT states with U(1) symmetry, 
\frm{ There are equivalence relations between \\
(a) `` 't Hooft anomaly matching conditions satisfied'', \\
(b) ``the boundary fully gapping rules satisfied'',  \\
(c) ``the effective Hall conductance is zero,'' and\\
(d) ``a bulk trivial SPT (i.e.\,trivial insulator), 
with unprotected boundary edge states (realizing an onsite symmetry) which can be decoupled from the bulk.''} 



Rigorously speaking, what we actually prove in Sec.\ref{sec:topo-nonp-proof} and Appendix \ref{appendixC},\ref{appendixD} is the equivalence of:
\frm{{\bf Theorem:} ABJ's perturbative local 
U(1) anomaly matching condition in 1+1D $\leftrightarrow$ the boundary fully gapping rules of 1+1D boundary/2+1D bulk with unbroken U(1) symmetry
for an equal number of left-right moving Weyl-fermion modes ($N_L=N_R$,  $c_L=c_R$) of 1+1D theory.}  

We show  
that the U(1)-anomaly free condition with zero chiral central charge $c_L-c_R=0$ (thus no gravitational anomaly)
is equivalent to the U(1) symmetric interaction gapping rule.
Under the Narain-Chern-Simons lattice level quantization \cite{Narain:1986am} in the context of chiral boson and Chern-Simons
theories, 
when the numbers of left and right 1+1D Weyl fermions equal to $N_L=N_R=N/2 \in \Z^+$,
we have constructed the symmetry group $(\U(1)^{N/2})_{\text{anomaly free}}^{\text{'t Hooft}}$ and 
the dual group $(\U(1)^{N/2})^{\text{gapping}}_{\text{term}}$ sectors via a short exact sequence:
\be
1\to {(\U(1)^{N/2})_{\text{anomaly free}}^{\text{'t Hooft}} \to \U(1)^{N } \to (\U(1)^{N/2})^{\text{gapping}}_{\text{term}}} \to 1.
\ee
The $(\U(1)^{N/2})_{\text{anomaly free}}^{\text{'t Hooft}}$ as a normal subgroup 
is the maximal group carrying the anomaly-free chiral U(1)$^{N/2}$ symmetry.
The $(\U(1)^{N/2})_{\text{gapping term}}$ as a quotient group is the symmetry-breaking group as the \emph{Poincar\'e dual}
space in the quantized lattice space in the sense of Narain lattice of Chern-Simons coupling.\cite{Narain:1986am} 

Note that 
possible modifications are needed for more generic symmetry cases:\\
(i)  For unbalanced left-right moving modes, the unbalanced chirality also implies the additional \emph{gravitational anomaly}.\\
(ii) For a bulk with \emph{topological order} (instead of pure SPT states), even if the boundary is gappable without breaking the symmetry,
there still can be nontrivial signature on the boundary, such as degenerate ground states (with gapped boundaries) or surface topological order.
This modifies the above specific Theorem to a more general Conjecture on the equivalence relation:
\frm{{\bf A Possible Conjecture:} The anomaly matching condition in an even-dimensional $(d+1)$D $\leftrightarrow$ 
the boundary fully gapping rules of $(d+1)$D boundary/$(d+2)$D bulk with unbroken $G$ symmetry
for an equal number of left-right moving degrees of freedom 
of $(d+1)$D theory, such that the system with arbitrary gapped boundaries has 
\emph{a unique non-degenerate ground state} (GSD=1),\cite{Wang:2012am,{Kapustin:2013nva},Lan2014uaaLWW1408.6514} \emph{no surface topological order},\cite{Vishwanath:2012tq} \emph{no symmetry/quantum number fractionalization}\cite{Wang:2014tia}
and \emph{without any nontrivial} (\emph{anomalous}) \emph{boundary signature}.
}

However, for an arbitrary given theory,
we {\it do not} know ``all kinds of anomalies,''  and thus in principle we {\it do not} know ``all anomaly matching conditions.''  
However, our work reveals some deep connection between the ``anomaly matching conditions'' and the ``boundary fully gapping rules.'' 
Alternatively, if we take the following statement as a definition instead,
\frm{{\bf A Proposed Definition:}  The {\bf anomaly matching conditions} in $(d+1)$D (all anomalies need to be cancelled) for symmetry $G$ $\leftrightarrow$  
the {\bf boundary fully gapping rules} without breaking symmetry $G$ and without anomalous boundary signatures for a $(d+1)$D boundary.} 
then the Theorem and the Proposed Definition together reveal that 
{the only anomaly type of \emph{a theory with an equal number of left/right-hand Weyl fermion modes}
and only with a U(1) symmetry in 1+1D is the perturbative local or ABJ's U(1) anomaly.}

Arguably the most interesting future direction is to test our above conjecture for more general cases, 
such as other dimensions or other symmetry groups.
One may test the above statements via the modular invariance\cite{{Levin:2013gaa},Sule:2013qla} of boundary theory.
%
It will also be profound
to address, the boundary fully gapping rules for non-Abelian symmetry,
and for
nonperturbative global or 
non-ABJ anomalies,\cite{Wen:2013oza,Wen:2013ppa,Witten:1982fp, Wang:2018qoy} including the
old and the new SU(2) anomalies.\cite{Witten:1982fp, Wang:2018qoy}\\

Though being numerically challenging, it will be interesting to test our models on the lattice (see a recent successful attempt
Ref.~\onlinecite{ZengZhuWangYou2202.12355}). 
Our {local spatial-lattice Hamiltonian with a finite Hilbert space, 
onsite symmetry and short-range hopping/interaction terms} is exactly a 
{condensed matter system we can realize in the lab}. It may be possible in the future we can simulate the lattice chiral model
in the physical instant time using the condensed matter set-up in the lab (such as in cold atoms system). 
Such a real-quantum-world simulation may be much faster than any classical computer or quantum computer.\\

\color{black}

\noindent
{\bf Note added}: After the completion of this present work in 2013, the authors have learned the potential relations between the
``$G$-symmetric anomaly-free condition'' to the ``trivial cobordism class in the cobordism group given by the classifying space B$G$ of symmetry group $G$'' 
stated in \Refe{Wang:2018cai}. The cobordism theory is relevant for the classifications of topological phases, studied recently in Ref.~\onlinecite{Kapustin2014tfa1403.1467,
Kapustin2014dxa1406.7329,
Freed2016rqq1604.06527,
Wan2018bns1812.11967}.
Our topological non-perturbative statement on 
the equivalence relation between the 't
Hooft anomaly matching conditions and the boundary fully gapping rules
is in fact a special case of the \emph{deformation classes of QFTs} advocated by Seiberg,\cite{Seiberg2019}
determined by two inputs: (1) Global symmetries and (2) Their 't Hooft anomalies.
Our major claim is a special case as the \emph{trivial deformation class of QFTs} 
that:

\frm{{\bf Any $G$-symmetry anomaly-free theory can be deformed from
a $G$-symmetry-preserving gapless phase to 
a $G$-symmetry-preserving gapped phase via the ``Symmetric Mass Generation\cite{WangYou2204.14271}''
by preserving the full $G$-symmetry along the deformation.}}

In addition, recently Ref.~\onlinecite{Wang:2018cai}
checks the classifications of all 't Hooft anomalies (including perturbative local and non-perturbative global anomalies) 
for the weakly-gauged standard models from the 16n-number of chiral Weyl-fermions in 3+1D, 
 the $so(10)$ grand unification (more precisely, Spin(10) chiral gauge theory). 
Ref.~\onlinecite{Wang:2018cai} shows that the only possible $\Z_2$ anomaly class for the  
Spin(10) symmetry for chiral Weyl-fermions in 3+1D.
Ref.~\onlinecite{Wang:2018cai, Wang:2018qoy} also find that the new anomaly is absent in the $so(10)$ grand unification,
therefore the $so(10)$ grand unification is all anomaly-free.
%
%
Ref.~\onlinecite{Wang:2018cai} also finds that the same conclusion holds for the $so(18)$ grand unification (more precisely, the Spin(18) chiral gauge theory).
This analysis supports the non-perturbatively lattice regularization of these ``standard models'' via a 3+1D local lattice model of Ref.~\onlinecite{Wen:2013ppa, Wang:2018cai}.
Recently Ref.~\onlinecite{Tong2104.03997} also provides alternative arguments to support our claim that
the 1+1D interaction we designed in Section \ref{sec3-5-4-0} can gap out the mirror chiral fermions.

\begin{acknowledgments}
We are grateful to Erich Poppitz, John Preskill, and Edward Witten for very helpful feedback and generous comments on our work.
We thank Michael Levin for important conversations
at the initial stage 
and for his comments on the manuscript. 
JW also thanks Jordan Cotler for very helpful feedbacks and for his interests. 
JW thanks Roman Jackiw, Anton Kapustin, Thierry Giamarchi, Alexander Altland, Sung-Sik Lee, Yanwen Shang, Duncan Haldane,  
Shinsei Ryu, David Senechal, Eduardo Fradkin, Subir Sachdev, Ting-Wai Chiu, Jiunn-Wei Chen, 
Chenjie Wang, and  Luiz Santos 
for comments. 
JW thanks H.\,He, L.\,Cincio, R.\,Melko and G.\,Vidal for comments on DMRG.

This work is supported by NSF Grant No. DMR-1005541, NSFC 11074140, and
NSFC 11274192. It is also supported by the BMO Financial
Group and the John Templeton Foundation. 
Research at Perimeter Institute is supported by the
Government of Canada through Industry Canada and by the Province of Ontario
through the Ministry of Research.
JW gratefully acknowledges the support from Institute for Advanced Study, 
the Corning Glass Works Foundation Fellowship and NSF Grant PHY-1314311 and PHY-1606531.
XGW is partially supported by NSF grant
DMR-1506475 and DMS-1664412.
This work is also supported by NSF Grant DMS-1607871 ``Analysis, Geometry and Mathematical Physics'' and Center for Mathematical Sciences and
Applications at Harvard University.

\end{acknowledgments}

\begin{widetext}
\end{widetext}



\appendix
\begin{center} 
\bf{Appendix}
\end{center} 

In Appendix \ref{appendixA},
we discuss the $C,P,T$ symmetry in a 1+1 D fermion theory.
In Appendix \ref{appendixB}, we show that Ginsparg-Wilson fermions realizing its axial U(1) symmetry by a non-onsite symmetry transformation.
In Appendix \ref{appendixC}  and \ref{appendixD} , 
under the specific assumption for a $2+1$D bulk Abelian symmetric protected topological (SPT) states\cite{Wen:2013oza,Wen:2013ppa,Chen:2011pg} with U(1) symmetry, 
we prove that 
\frm{
{\bf Boundary fully gapping rules} (in Sec.\ref{anomaly-gap})\cite{h95,Wang:2012am,Levin:2013gaa,Barkeshli:2013jaa,Lu:2012dt} are sufficient and necessary conditions of the {\bf 't Hooft anomaly matching condition} (in Sec.\ref{anomaly-hall}).\cite{'tHooft:1979bh} 
}
The SPT order (explained in Sec.\ref{SPT-CS}) are short-range entangled states with some onsite symmetry $G$ in the bulk.
For the nontrivial SPT order, the symmetry $G$ is realized as a non-onsite symmetry on the boundary.\cite{Chen:2011pg,Chen:2012hc,Santos:2013uda} 
The 1+1D edge states are protected to be gapless as long as the symmetry $G$ is unbroken on the boundary.\cite{Chen:2011pg,Lu:2012dt}
Importantly, SPT has no long-range entanglement, so no gravitational anomalies.\cite{Wen:2013oza,Wen:2013ppa}
The only anomaly here is the ABJ's U(1) anomaly\cite{Adler:1969gk,Bell:1969ts,Donoghue:1992dd} for chiral matters. 

Appendix \ref{AppendixE} includes several approaches for proving boundary fully gapping rules.
In Appendix \ref{AppendixF}, we discuss the property of our Chern insulator in details, and provide
additional models of lattice chiral fermions and chiral bosons.

\section{$C$, $P$, $T$ symmetry in the 1+1D fermion theory \label{appendixA}}

Here we show the charge conjugate $C$, parity $P$, time reversal $T$ symmetry transformation for the 1+1D Dirac fermion theory.
Recall that the massless Dirac fermion Lagrangian is $\mathcal{L}=\bar{\Psi} \ii \gamma^\mu \partial_\mu \Psi$. Here the Dirac fermion field 
$\Psi= \Psi (t,\vec{x}) = \Psi (x)$ 
can be written as a two-component spinor.
For convenience, but without losing the generality, we choose the Weyl basis, so $\Psi=(\psi_L,\psi_R)$, where 
each component of $\psi_L,\psi_R$ is a chiral Weyl fermion with left and right chirality respectively. Specifically, gamma matrices in the Weyl basis are
\bea
&&\gamma^0 = \sigma_x 
= \bigl( {\begin{smallmatrix}
0 &1 \\
1 & 0
\end{smallmatrix}}  \bigl),\;\;\;
\gamma^1 = \ii \sigma_y
= \bigl( {\begin{smallmatrix}
0 &1 \\
-1 & 0
\end{smallmatrix}}  \bigl),\;\;\; \cr
&&\gamma^5 =\gamma^0\gamma^1= -\sigma_z
=\bigl( {\begin{smallmatrix}
-1 &0 \\
0 & 1
\end{smallmatrix}}  \bigl).
\eea
satisfies Clifford algebra $\{\gamma^\mu,\gamma^\nu\}=2 \eta^{\mu\nu}$, here the signature of the Minkowski metric is $(+,-)$.
The projection operators are
\be
P_L=\frac{1-\gamma^5}{2}=\bigl( {\begin{smallmatrix}
1 &0 \\
0 &0
\end{smallmatrix}}  \bigl),\;\; P_R=\frac{1+\gamma^5}{2}=\bigl( {\begin{smallmatrix}
0 &0 \\
0 & 1
\end{smallmatrix}}  \bigl), 
\ee
mapping a massless Dirac fermion to two Weyl fermions, i.e. $\mathcal{L}=\ii \psi^\dagger_{L} (\partial_t-\partial_x) \psi_{L}+\ii \psi^\dagger_{R} (\partial_t+\partial_x) \psi_{R}$.
We derive the $P, T, C$ transformation on the fermion field operator $\hat{\Psi}$ in $1+1$D, up to some overall complex phases $\eta_P, \eta_T$
degree of freedom, 
\bea \label{eq:Psym}
&&P \hat{\Psi}(t,\vec{x}) P^{-1}=\eta_P\, \gamma^0 \hat{\Psi}(t,-\vec{x}) = 
\eta_P\,
\bigl( {\begin{smallmatrix}
0 &1 \\
1 & 0
\end{smallmatrix}}  \bigl) \hat{\Psi}(t,-\vec{x}). \cr 
&&T \hat{\Psi}(t,\vec{x}) T^{-1}=\eta_T\, \gamma^0 \hat{\Psi}(-t,\vec{x})
= \eta_T\,
\bigl( {\begin{smallmatrix}
0 &1 \\
1 & 0
\end{smallmatrix}}  \bigl) \hat{\Psi}(-t,\vec{x}). 
\cr  
&&C \hat{\Psi}(t,\vec{x}) C^{-1}=\gamma^0\gamma^1  \hat{\Psi}^*(t,\vec{x})
= \bigl( {\begin{smallmatrix}
-1 &0 \\
0 & 1
\end{smallmatrix}}  \bigl)
 \hat{\Psi}^*(t,\vec{x}).\cr
 &&CPT \hat{\Psi}(t,\vec{x}) T^{-1} P^{-1} C^{-1}  =\gamma^0\gamma^1   \hat{\Psi}^*(-t,-\vec{x}).
\eea 
We can quickly derive these symmetry transformations (which are preserved by a massive Dirac fermion theory).

\paragraph{$P$ symmetry:}

Recall for the \emph{passive $P$ transformations} on the coordinates, we require that
sending $(t,\vec{x})$ to $(t,-\vec{x})$, 
which is a $\Z_2$ symmetry on the vector coordinates. 
But we shall, instead, consider the \emph{active transformation} viewpoint on the fields,
in terms of operators,
such that
$P \hat{\Psi}({x}) P^{-1} = 
P \hat{\Psi}(t,\vec{x}) P^{-1} =  
\hat{\Psi'}(t,\vec{x}) 
= {\rm M}_P \hat{\Psi}(t,-\vec{x})
= {\rm M}_P \hat{\Psi}(x')$,
where the ${\rm M}_P$ is a linear transformation matrix on the $\hat{\Psi}(t,-\vec{x}) \equiv \hat{\Psi}(x')$
and now we also define $(t,-\vec{x})\equiv x'$ ($x'^\mu$ or $\partial'_\mu=\partial/\partial x'^\mu$ for each transformed coordinate, $\mu=0,1$).

We want the $P$ transformation on the Dirac equation
$(\ii \gamma^\mu \partial_\mu- m) \Psi({x})$ becomes
$(\ii \gamma^\mu \partial'_\mu- m) \Psi'({x'}) = 
(\ii \gamma^\mu \partial'_\mu- m) ( {\rm M}_P \Psi(t,\vec{x}) )= 0$
where the prime notation implies under the $P$ transformation.
Then we can identify $\Psi'({x'})={\rm M}_P \Psi(t,\vec{x}) $,
or conversely $\Psi'({x})={\rm M}_P \Psi(t,-\vec{x}) $.
To achieve this goal, we multiply Dirac equation by $\gamma^0$, we obtain 
$\gamma^0(\ii \gamma^\mu \partial_\mu- m) \Psi(t,\vec{x})=(\ii \gamma^\mu \partial'_\mu- m) (\gamma^0\Psi(t,\vec{x}))=0$.
In comparison, this means we should identify $\Psi'({x'}) ={\rm M}_P \Psi(t,\vec{x})= \gamma^0\Psi(t,\vec{x})$ up to a phase.
In other words, renaming the coordinates, we get
$\Psi'({x}) ={\rm M}_P \Psi(t,-\vec{x})= \gamma^0\Psi(t,-\vec{x})$.
In the operator form, we derive: 
$P \hat{\Psi}(t,\vec{x}) P^{-1} =  
\hat{\Psi'}(t,\vec{x}) 
= {\rm M}_P \hat{\Psi}(t,-\vec{x})
=  \gamma^0 \hat{\Psi}(t,-\vec{x})$  up to a $\eta_P$ phase, hence Eq.(\ref{eq:Psym}). 
The parity $P$ transforms the left ($L$) and the right ($R$) to each other, which is also a $\Z_2$ symmetry on the spinor.

\paragraph{$T$ symmetry:}

Recall a \emph{passive $T$ symmetry transformation} on the vector coordinates,
sending $(t,\vec{x})$ to $(-t,\vec{x})$ as a $\Z_2$ symmetry transformation; 
for convenience we also redefine the transformed coordinates $(-t,\vec{x})\equiv x'^\mu$.
It is much easier to derive the fact that $T \ii T^{-1}= - \ii$
in the \emph{passive} $T$ transformation, we see that the generic Schr\"odinger equation
$\ii \partial_t \Psi(t,\vec{x})=H\Psi(t,\vec{x})$ acted by $T$
is $(T\ii T^{-1})(T\partial_t T^{-1})(T \Psi(t,\vec{x}))=(THT^{-1} )(T \Psi(t,\vec{x})) = H(T \Psi(t,\vec{x}))$, 
where the last line assumes that $THT^{-1} =H$ is time-reversal invariant.
The \emph{passive} transformation acts on the coordinates so $(T\partial_t T^{-1})=-\partial_t$.
Then $(T \Psi(t,\vec{x}))$ is a time-reversed solution of the original Schr\"odinger equation if and only if $T\ii T^{-1} = - \ii$.

Below we shall, instead, consider the \emph{active transformation} viewpoint on the fields, in terms of operators,
so that we can derive
$T \hat{\Psi}({x}) T^{-1} = 
T \hat{\Psi}(t,\vec{x}) T^{-1} =  
\hat{\Psi'}(t,\vec{x}) 
= {\rm M}_T \hat{\Psi}(-t,\vec{x})
= {\rm M}_T \hat{\Psi}(x')$,
where the ${\rm M}_T$ is a linear transformation matrix on the $\hat{\Psi}(-t,\vec{x}) \equiv \hat{\Psi}(x')$.
We want the $T$ transformation on the Dirac equation
$(\ii \gamma^\mu \partial_\mu- m) \Psi({x})$ becomes
$(\ii \gamma^\mu \partial'_\mu- m) \Psi'({x'}) = 
(\ii \gamma^\mu \partial'_\mu- m) ( {\rm M}_T \Psi(t,\vec{x}) )= 0$
where the prime notation implies under the $T$ transformation.
Then we can identify $\Psi'({x'})={\rm M}_T \Psi(t,\vec{x}) $,
or conversely $\Psi'({x})={\rm M}_T \Psi(-t,\vec{x}) $.
To this end,
we massage the Dirac equation in terms of Schr\"odinger equation form, $\ii \partial_t \Psi(t,\vec{x})=H\Psi(t,\vec{x})=
(-\ii \gamma^0\gamma^j \partial_j+m \gamma^0)\Psi(t,\vec{x})$,
here $j$ are spatial coordinates (we have $j=1$ only for 1+1D).
Then we wanted to identify the $T$ transformed equation to be
$\ii \partial_{t'} \Psi'(t',\vec{x})=H\Psi'(t',\vec{x})=
(-\ii \gamma^0\gamma^1 \partial_1+m \gamma^0)\Psi'(t',\vec{x})$.
Since $T$ is anti-unitary, $T$ can be written as $T=UK$ with a unitary transformation part $U$ and an extra $K$ does the complex conjugate so $T^{-1} \ii T = -\ii$. Then $T^{-1} H T^{}=H$ imposes
the constraints $U^{-1} \gamma^0 U=\gamma^{0*}$ and $U^{-1} \gamma^j U=-\gamma^{j*}$. 
In 1+1D Weyl basis, since $\gamma^{0}$ and $\gamma^{1}$ both are reals, 
we conclude that $U=\gamma^0$ up to a complex phase. 
This means we should identify 
$\Psi'(-t,\vec{x}) = \gamma^0 K \Psi(t,\vec{x})$ up to a phase to make a time-reversal symmetric solution. 
So in the operator form, $T\hat{\Psi}(t,\vec{x})T^{-1} 
={\eta_T\, \gamma^0 \hat{\Psi}(-t,\vec{x})}$, also
$Tz \hat{\Psi}(t,\vec{x})T^{-1} 
={\eta_T\, z^* \gamma^0 \hat{\Psi}(-t,\vec{x})}$ since the antiunitary $T$ sends a complex number $z \in \C$ to its conjugate $z^*$.

\paragraph{$C$ symmetry:}
For the $C$ transformation, we transform a particle to its anti-particle. This means that we flip the charge $q$ (in the term coupled to a background gauge field $A$), 
which can be done by taking the complex conjugate on the Dirac equation,
$\big[-\ii \gamma^{\mu*} (\partial_\mu+i q A_\mu) -m\big] \Psi^*(t,\vec{x})=0$, where
$-\gamma^{\mu*}$ satisfies Clifford algebra. 
We can rewrite the equation as 
$\big[\ii \gamma^{\mu} (\partial_\mu+i q A_\mu) -m\big] \Psi_c(t,\vec{x})=0$, 
by identifying the charge conjugated state as $\Psi_c={\rm M}_C \gamma^0 \Psi^*$
and imposing the constraint $- {\rm M}_C  \gamma^0 \gamma^{\mu*} \gamma^0 {\rm M}_C ^{-1}= \gamma^{\mu}$.
Additionally, we already have $\gamma^0 \gamma^{\mu} \gamma^0=\gamma^{\mu \dagger}$. 
So the constraint reduces to $- {\rm M}_C  \gamma^{\mu T}  {\rm M}_C^{-1}= \gamma^{\mu}$.
In the 1+1D Weyl basis, we obtain $-  {\rm M}_C  \gamma^{0}   {\rm M}_C^{-1}= \gamma^{0}$ and $ {\rm M}_C  \gamma^{1}   {\rm M}_C^{-1}= \gamma^{1}$. Thus,
$ {\rm M}_C=\eta_C \, \gamma^{1}$  up to a phase,
and we derive $\Psi_c= \gamma^0 \gamma^{1}\Psi^*$ as a state. 
In the operator form, we obtain $C\hat{\Psi}(t,\vec{x})C^{-1}=\hat{\Psi}_c(t,\vec{x})=\gamma^0 \gamma^{1}\hat{\Psi}^*(t,\vec{x})$.

The important feature is that our chiral matter theory has the parity $P$ and time reversal $T$ symmetry explicitly broken.
Because both $P$ and $T$ exchange between left-handness and right-handness particles, i.e. $\psi_L,\psi_R$ becomes $\psi_R,\psi_L$.
Thus both $P$ and $T$ transformations switch left/right moving charges by switching its charge carrier. 
If $q_L \neq q_R$ and if no field redefinition can restore the charge assignment, 
then our chiral matter theory breaks $P$ and $T$.

Our chiral matter theory, however, does not break charge conjugate symmetry $C$.
Because the symmetry transformation acting on the state induces 
$C \Psi C^{-1} = -\sigma_z \Psi^*=\bigl( {\begin{smallmatrix}
-1 &0 \\
0 &1
\end{smallmatrix}}  \bigl)  \Psi^*$, while $\psi_L$/$\psi_R$ maintains its left-handness/right-handness as $\psi_L$/$\psi_R$.

Of course, if the theory is a vector-like Dirac theory with $q_L = q_R$, then it preserves $C$, $P$, and $T$.

\section{Ginsparg-Wilson fermions with a non-onsite U(1) symmetry as SPT edge states\label{appendixB}} 

We firstly review the meaning of onsite symmetry and non-onsite symmetry transformation,\cite{Chen:2011pg,2011PhRvB..84w5141C} 
and then we will demonstrate that Ginsparg-Wilson fermions realize the U(1) symmetry in the non-onsite symmetry manner.

\subsection{On-site symmetry and non-onsite symmetry}

The onsite symmetry transformation as an operator $U(g)$, with $g \in G$ of the symmetry group, transforms the state $| v \rangle$ globally, by  $U(g) | v \rangle$.
The onsite symmetry transformation $U(g)$ must 
be written in the tensor product form acting on each site $i$,\cite{Chen:2011pg,2011PhRvB..84w5141C}
\be
U(g)= \otimes _i U_i(g), \ \ \ g \in G. 
\ee

For example, consider a system with only two sites. Each site with a qubit degree of freedom (i.e. with $| 0 \rangle$ and $| 1 \rangle$ eigenstates on each site). 
The state vector $| v \rangle$ for the two-sites system is $| v \rangle=\sum_{j_1, j_2 }  c_{j_1, j_2 } | j_1\rangle \otimes | j_2 \rangle=\sum_{j_1, j_2 }  c_{j_1, j_2 } | j_1, j_2 \rangle$ with $1,2$ site indices and 
$|j_1 \rangle, |j_2 \rangle$ are eigenstates chosen among $|0 \rangle, |1 \rangle$.

An example for the onsite symmetry transformation can be,
\bea
U_{\text{onsite}}&=&| 00\rangle \langle 00| +| 01 \rangle \langle 01 |-| 10 \rangle \langle 10 |-| 11\rangle \langle 11 | \nonumber\\
&=&(| 0\rangle \langle 0| -| 1\rangle \langle 1| )_1 \otimes (| 0\rangle \langle 0| +| 1\rangle \langle 1| )_2 \nonumber \\
&=& \otimes _i U_i(g).
\eea
Here $U_{\text{onsite}}$ is in the tensor product form, where $U_1(g)=(| 0\rangle \langle 0| -| 1\rangle \langle 1| )_1$ and $U_2(g)=(| 0\rangle \langle 0| +| 1\rangle \langle 1| )_2$, again with $1,2$ subindices are site indices.
Importantly, this operator does not contain non-local information between the neighbored sites.

A non-onsite symmetry transformation $U(g)_{\text{non-onsite}}$ cannot be expressed as a tensor product form: 
\be
U(g)_{\text{non-onsite}} \neq \otimes _i U_i(g), \ \ \ g \in G. 
\ee
An example for the non-onsite symmetry transformation can be the $CZ$ operator,\cite{2011PhRvB..84w5141C} 
\bea
CZ=| 00\rangle \langle 00| +| 01 \rangle \langle 01 |+| 10 \rangle \langle 10 |-| 11\rangle \langle 11 |. \nonumber
\label{eq:CZ}
\eea
$CZ$ operator contains non-local information between the neighbored sites, which flips the sign of the state vector if both sites $1,2$ are in the eigenstate $|1 \rangle$. 
One cannot achieve writing $CZ$ as a tensor product structure. 

Now let us discuss how to gauge the symmetry. Gauging an onsite symmetry simply requires replacing the group element $g$ in the symmetry group 
to $g_i$ with a site dependence, i.e. replacing a global symmetry to a local (gauge) symmetry. All we need to do is,
\be
U(g)= \otimes _i U_i(g)  \stackrel{\rm Gauge}{\Longrightarrow} U(g_i)= \otimes _i U_i(g_i), 
\label{eq:gauge}
\ee
with $g_i \in G$.
Following Eq.(\ref{eq:gauge}),
it is easy to gauge such an onsite symmetry 
to obtain a chiral fermion theory coupled to a 
gauge field. 
Since our chiral matter theory is implemented with an onsite U(1) symmetry, it is easy to gauge our chiral matter theory to be a U(1) chiral gauge theory.

On the other hand, a non-onsite symmetry transformation cannot be written as a tensor product form. 
So, it is difficult (or unconventional) to gauge a non-onsite symmetry.
As we will show below Ginsparg-Wilson fermions realizing a non-onsite symmetry, so that is why it is difficult to gauge it.

\subsection{Ginsparg-Wilson relation, Wilson fermions and non-onsite symmetry}

Below we attempt to show 
that Wilson fermions implemented with Ginsparg-Wilson (G-W) relation realizing the symmetry transformation by the non-onsite manner.
Follow the notation of Ref.\onlinecite{Fujikawa:2004cx}, the generic form of the Dirac fermion $\psi$ path integral on the lattice (with the lattice constant $a$) is
\be
\int \mathcal{D}{\bar{\psi}} \mathcal{D}{\psi} \exp[a^{\text{d}_m} \sum_{x_1,x_2} \bar{\psi}(x_1) D(x_1,x_2) \psi(x_2)].
\ee 
Here 
the exponent $\text{d}_m$ is the dimension of the spacetime.
For example, the action of Wilson fermions with Wilson term (the term with the front coefficient $r$) can be written as:\\
\fontsize{10}{10pt} \selectfont

\begin{widetext}
\bea
S_{\Psi}&=&a^{\text{d}_m}  \Big( \sum_{x,\mu} \frac{\ti}{2a} (\bar{\psi}(x) \gamma^\mu U_\mu(x) \psi(x+a^\mu) - \bar{\psi}(x+a^\mu) \gamma^\mu U^\dagger_\mu(x) \psi(x) ) \nonumber 
- \sum_{x} \frac{1}{a}  m_0 \bar{\psi}(x) {\psi}(x)  \nonumber \\
&+&\frac{r}{2a}\sum_{x,\mu} (- \bar{\psi}(x)  U_\mu(x) \psi(x+a^\mu)- \bar{\psi}(x+a^\mu)  U^\dagger_\mu(x) \psi(x) +2 \bar{\psi}(x) \psi(x) )\Big). 
\eea
\end{widetext}
Here $U_\mu(x) \equiv \exp(\ii a g A_\mu)$  
are the gauge field connection.
At the weak $g$ coupling, it is also fine for us simply consider $U_\mu(x) \simeq 1$.
One can find its Fermion propagator:
\be
(\sum_\mu \frac{1}{a} \gamma^\mu \sin(a k_\mu) - m_0 -\sum_\mu \frac{r}{a}(1-\cos(a k_\mu) ))^{-1}. \nonumber
\ee

The Wilson fermions with $r \neq 0$ kills the doubler (at $k_\mu=\pi/a$) by giving a mass of order $r/a$ to it. As $a \to 0$, the doubler disappears from the spectrum with an infinite large mass.

This Dirac operator $D(x_1,x_2)$ is not strictly local, but decreases exponentially as
\be
D(x_1,x_2) \sim e^{-|x_1-x_2|/{\xi} }
\ee
with $\xi=\text{(local range)} \cdot a$ as some localized length scale of the Dirac operator. We call $D(x_1,x_2)$ as a quasi-local operator, which is strictly \emph{non-local}.

One successful way to treat the lattice Dirac operator is imposing the Ginsparg-Wilson (G-W) relation:\cite{Wilson:1974sk} 
\be
\{D,\gamma^5\}=2aD\gamma^5D.
\ee
Thus in the continuum limit $a \to 0$, this relation becomes $\{ {\not}D,\gamma^5\}=0$. One can choose a Hermitian $\gamma^5$,
and ask for the Hermitian property on $\gamma^5 D$, which is $(\gamma^5 D)^\dagger=D^\dagger \gamma^5=\gamma^5 D$.

It can be shown that the action (in the exponent of the path integral) is invariant under the axial U(1) chiral transformation with a $\theta_A$ rotation:
\be
\delta \psi(y) = \sum_w \ti \theta_A \hat{\gamma}^5 (y,w) \psi(w), \;\;\; \delta \bar{\psi}(x) = \ti \theta_A \bar{\psi}(x)  {\gamma}^5 \;\;\; 
\label{eq:GWU(1)A}
\ee
where
\be
\hat{\gamma}^5(x,y) \equiv \gamma^5 - 2 a \gamma^5 D(x,y).  
\ee
The chiral anomaly on the lattice can be reproduced from the Jacobian $J$ of the path integral measure:
\be
J=\exp[-\ti \theta_A \tr(\hat{\gamma}^5+{\gamma}^5)]=\exp[-2 \ti \theta_A \tr({\Gamma}^5)]
\ee
here ${\Gamma}^5(x,y) \equiv \gamma^5 -  a \gamma^5 D(x,y)$. The chiral anomaly follows the index theorem
$\tr({\Gamma}^5)=n_+-n_-$, with $n_\pm$ counts the number of zero mode eigenstates $\psi_j$, with zero eigenvalues, i.e. 
$\gamma^5 D \psi_j=0$, where the projection is $\gamma^5 \psi_j=\pm \psi_j$ for $n_\pm$ respectively.

Note that G-W relation can be rewritten as
\be
\gamma^5 D+ D\hat{\gamma}^5=0.
\ee

Importantly, now axial U(1)$_A$ transformation in Eq.(\ref{eq:GWU(1)A}) involves with $\hat{\gamma}_5(x,y)$ which contains the piece of quasi-local operators 
$D(x,y) \sim e^{-|x_1-x_2|/{\xi} }$.
Thus, it becomes apparent that U(1)$_A$ transformation Eq.(\ref{eq:GWU(1)A}) is a non-onsite symmetry which carries nonlocal information between different sites $x_1$ and $x_2$. It is analogous to the 
$CZ$ symmetry transformation in Eq.(\ref{eq:CZ}), which contains the entangled information between neighbored sites $j_1$ and $j_2$.

Thus we have shown G-W fermions realizing axial U(1) symmetry (U(1)$_A$ symmetry) with a non-onsite symmetry transformation.
While the left and right chiral symmetries, U(1)$_L$ and U(1)$_R$, mix between the linear combination of vector U(1)$_V$ symmetry and axial U(1)$_A$ symmetry. 
So U(1)$_L$ and U(1)$_R$ have non-onsite symmetry transformations, too. In short,

\frm{The axial U(1)$_A$ symmetry in G-W fermion is a non-onsite symmetry. Also the left and right chiral symmetry U(1)$_L$ and U(1)$_R$ in G-W fermion are non-onsite symmetry.}

The non-onsite symmetry here indicates the nontrivial edge states of bulk SPTs,\cite{Chen:2011pg,Chen:2012hc,Santos:2013uda}
thus Ginsparg-Wilson fermions can be regarded as gapless edge states of some bulk fermionic SPT order. 
With the above analysis, we emphasize again that our approach in the main text is different from Ginsparg-Wilson fermions -  
while our approach implements only onsite symmetry, Ginsparg-Wilson fermion implements non-onsite symmetry.
In Chen-Giedt-Poppitz model,\cite{CGP1247} the Ginsparg-Wilson fermion is implemented.
Thus this is one of the major differences between Chen-Giedt-Poppitz and our approaches.

\section{Proof: Boundary Fully Gapping Rules $\to $ Anomaly Matching Conditions \label{appendixC} }
\begin{figure}[!h]
{\includegraphics[width=.20\textwidth]{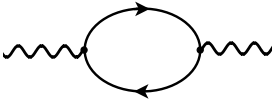}}
\caption{Feynman diagrams with solid lines representing chiral fermions and wavy lines representing U(1) gauge bosons: 1+1D chiral fermionic anomaly shows $\mathcal{A}=\sum (q_L^2-q_R^2)$.
For a generic 1+1D theory with U(1) symmetry, $\mathcal{A}=q^2 \mathbf{t} K^{-1}\mathbf{t}$.}
\label{anomaly1+1D}
\end{figure}

Here we show that if boundary states can be fully gapped (there exists a boundary gapping lattice $\Gamma^\partial$ satisfies boundary fully gapping rules ({\bf{1}})({\bf{2}})({\bf{3}}) in Sec.\ref{anomaly-gap}\cite{h95,Wang:2012am,Levin:2013gaa,Barkeshli:2013jaa,Lu:2012dt}) with U(1) symmetry unbroken,
then 
the boundary theory is an anomaly-free theory free from ABJ's U(1) anomaly. 
This theory satisfies the effective Hall conductance $\sigma_{xy}=0$, so the anomaly factor $\mathcal{A}=0$ by Eq.(\ref{eq:A=hall}) in Sec.\ref{anomaly-hall},
and illustrated in Fig.\ref{anomaly1+1D}.\\

Importantly, for $N$ numbers of  1+1D Weyl fermions, 
in order to gap out the mirror sector, our model enforces $N \in 2 \mathbb{Z}^+$ is an even positive integer,
and requires equal numbers of left/right moving modes $N_L=N_R=N/2$. When there is no interaction, we have a total U(1)$^N$ symmetry for the free theory.
We will then introduce the properly-designed gapping terms, and (if and only if) there are $N/2$ allowed gapping terms. 
The total symmetry is further broken from U(1)$^N$ down to U(1)$^{N/2}$ due to $N/2$ gapping terms.
 
{ 
The remained U(1)$^{N/2}$ symmetry stays unbroken for the following reasons:\\
(i) The gapping terms obey the U(1)$^{N/2}$ symmetry. The symmetry is thus {\bf not explicitly broken}.\\
(ii) In 1+1D, there is {\bf no spontaneous symmetry breaking} of a continuous symmetry (such as our U(1) symmetry) due to Coleman-Mermin-Wagner-Hohenberg theorem.\\
(iii) We explicitly check the ground degeneracy of our model with a gapped boundary has {\bf a unique ground state}, following the procedure of Ref.\onlinecite{{Wang:2012am},{Kapustin:2013nva}}. 
Thus, a unique ground state implies that there is no spontaneous symmetry breaking degeneracy nor topological ground state degeneracy.
}

{
Below we will prove that all the remained U(1)$^{N/2}$ symmetry is anomaly-free and mixed-anomaly-free. We will prove for both fermionic and bosonic cases together, 
under Chern-Simons symmetric-bilinear $K$ matrix notation,
with fermions $K=K^{f}$ and bosons $K=K^{b0}$,  where $K=K^{-1}$.
}\\

\noindent
{\bf{Proof}}:
{
There are $N/2$ linear-independent terms of $\ell_a$ for $\cos(\ell_a \cdot \Phi)$ in the boundary gapping terms $\Gamma^\partial$, for 
$\{\ell_a \}=\{\ell_1,\ell_2,\dots, \ell_{N/2} \} \in \Gamma^\partial$.
To find the remained unbroken U(1)$^{N/2}$ symmetry, we notice that we can define charge vectors 
\be
\mathbf{t}_a \equiv K^{-1} \ell_a^{}
\ee
where any $\ell_a \in \Gamma^\partial$ is allowed, and $a=1,\dots, N/2$. So there are totally $N/2$ charge vectors.
These $\mathbf{t}_a$ charge vectors are linear-independent because all $\ell_a^{}$ are linear-independent to each other.}

Now we show that these ${N/2}$ charge vectors $\mathbf{t}_a$ span the whole unbroken U(1)$^{N/2}$-symmetry.
Indeed, follow the condition Eq.(\ref{eq:t_symmetry}), this is true:
\be
 \ell_{c,I}^{} \cdot\mathbf{t}_a= \ell_{c}^{} K^{-1} \ell_a^{}=0
\ee
for all $\ell_c \in \Gamma^\partial$.
This proves that ${N/2}$ charge vectors $\mathbf{t}_a$ are exactly the U(1)$^{N/2}$-symmetry generators.
We end the proof by showing our construction is indeed an anomaly-free theory among all U(1)$^{N/2}$-symmetries or all U(1) charge vectors $\mathbf{t}_a$,
thus we check that they satisfy the anomaly matching conditions:
\be
\mathcal{A}_{(a,b)}=2\pi \sigma_{xy,(a,b)}=  q^2 \mathbf{t}_a K^{}\mathbf{t}_b= q^2  \ell_a^{} K^{-1} \ell_b^{}=0.
\ee
Here $\ell_a,\ell_b \in \{\ell_1,\ell_2,\dots, \ell_{N/2} \}$, where we use $K=K^{-1}$.
Therefore, our U(1)$^{N/2}$-symmetry theory is {\bf fully anomaly-free} ($\mathcal{A}_{(a,a)}=0$) and {\bf mixed anomaly-free} ($\mathcal{A}_{(a,b)}=0$ for $a\neq b$).
We thus proved
\frm{{\bf Theorem:} The boundary fully gapping rules of 1+1D boundary/2+1D bulk with unbroken U(1) symmetry $\rightarrow$ ABJ's U(1) anomaly matching condition in 1+1D.}
  for both fermions $K=K^{f}$ and bosons $K=K^{b0}$. (Q.E.D.)

\section{Proof: Anomaly Matching Conditions $\to$ Boundary Fully Gapping Rules \label{appendixD} }
Here we show that if 
the boundary theory is an anomaly-free theory (free from ABJ's U(1) anomaly), 
which satisfies the anomaly factor $\mathcal{A}=0$ (i.e. the effective Hall conductance $\sigma_{xy}=0$ in the bulk, in Sec.\ref{anomaly-hall}), then
boundary states can be fully gapped with U(1) symmetry unbroken.
Given a charge vector $\mathbf{t}$, we will prove in the specific case of U(1) symmetry, by finding the set of 
boundary gapping lattice $\Gamma^\partial$ satisfies boundary fully gapping rules ({\bf{1}})({\bf{2}})({\bf{3}})  in Sec.\ref{anomaly-gap}.\cite{h95,Wang:2012am,Levin:2013gaa,Barkeshli:2013jaa,Lu:2012dt}
We denote the charge vector as $\mathbf{t}=(t_1,t_2,t_3,\dots,t_N)$.
We will prove this for fermions $K=K^{f}$ and bosons $K=K^{b0}$ separately. Note the fact that $K=K^{-1}$ for both $K^{f}$ and $K^{b0}$.

\subsection{Proof for fermions $K=K^{f}$}
Given a $N$-component charge vector 
\bea
\mathbf{t}=(t_1,t_2, \dots, t_N)
\eea
of a U(1) charged anomaly-free theory satisfying $\mathcal{A}=0$, which means $\mathbf{t} {(K^{f})^{-1}} \mathbf{t}=0$. Here the fermionic $K^{f}$ matrix is written in this canonical form,
\be
K^f_{N\times N} =\bigl( {\begin{smallmatrix}
1 &0 \\
0 & -1
\end{smallmatrix}}  \bigl) \oplus \bigl( {\begin{smallmatrix}
1 &0 \\
0 & -1
\end{smallmatrix}}  \bigl) \oplus \dots
\ee

We now construct $\Gamma^\partial$ obeying boundary fully gapping rules.
We choose
\be
\ell_{1}=(K^{f}) \mathbf{t}
\ee
which satisfies self-null condition
$\ell_{1} (K^{f})^{-1} \ell_{1}=0$. To complete the proof, we continue to find out a total set of $\ell_{1}, \ell_{2}, \dots, \ell_{N/2}$, so $\Gamma^\partial$ is a dimension $N/2$ Chern-Simons-charge lattice (Lagrangian subgroup).

For $\ell_{2}$, we choose its form as
\be
\ell_{2}=(\ell_{2,1},\ell_{2,1},\ell_{2,3},\ell_{2,3},0,\dots,0)
\ee
where even component of $\ell_{2}$ duplicates its odd component value,
to satisfy
$\ell_{2}  (K^{f})^{-1} \ell_{1} =\ell_{2} (K^{f})^{-1} \ell_{2}=0$. The second constraint is automatically true for our choice of $\ell_{2}$.
The first constraint is achieved by solving $\ell_{2,1}( t_1-t_2)+\ell_{2,3}( t_3-t_4)=0$. We can properly choose $\ell_{2}$ to satisfy this constraint. 

For $\ell_{n}$, by mathematical induction, we choose its form as
\be
\ell_{n}=(\ell_{n,1},\ell_{n,1},\ell_{n,3},\ell_{n,3},\dots,\ell_{n,2n-1},\ell_{n,2n-1} 0,\dots,0)
\ee
where even component of $\ell_{n}$ duplicates its odd component value,
to satisfy
\be
\ell_{n}  (K^{f})^{-1} \ell_{j}, \;\;\;j=1,\dots,n,
\ee
for any $n$.
For $2\leq j \leq n$, the constraint is automatically true for our choice of $\ell_{n}$ and $\ell_{j}$.
For $\ell_{n}  (K^{f})^{-1} \ell_{1}=0$, it leads to the constraint:
$\ell_{n,1}( t_1-t_2)+\ell_{n,3}( t_3-t_4)+\dots +\ell_{n,2n-1}( t_{2n-1}-t_{2n}) =0$, we can generically choose
$\ell_{n,2n-1}\neq 0$ to have a new $\ell_{n}$ independent from other $\ell_{j}$ with $1\leq j \leq n-1$.

Notice the gapping term obeys U(1) symmetry, because $ \ell_{n} \cdot \mathbf{t}=\ell_{n} (K^{f})^{-1} \ell_{1}=0$ is always true for all $\ell_{n}$.
Thus we have constructed a dimension $N/2$ Lagrangian subgroup $\Gamma^\partial=\{\ell_{1}, \ell_{2}, \dots, \ell_{N/2}\}$ which obeys 
the boundary fully gapping rules ({\bf{1}})({\bf{2}})({\bf{3}}) in Sec.\ref{anomaly-gap}. (Q.E.D.)

\subsection{Proof for bosons $K=K^{b0}$}
Similar to the proof of fermion, we start with a given $N$-component charge vector $\mathbf{t}$, 
\bea
\mathbf{t}=(t_1,t_2, \dots, t_N),
\eea
of a U(1) charged anomaly-free theory satisfying $\mathcal{A}=0$, which means $\mathbf{t} {(K^{b0})^{-1}} \mathbf{t}=0$.

Here the bosonic $K^{b0}$ matrix is written in this canonical form,
\be
K^{b0}_{N\times N} =\bigl( {\begin{smallmatrix}
0 &1 \\
1 & 0
\end{smallmatrix}}  \bigl) \oplus \bigl( {\begin{smallmatrix}
0 &1 \\
1 & 0
\end{smallmatrix}}  \bigl) \oplus \dots
\ee

We now construct $\Gamma^\partial$ obeying boundary fully gapping rules.
We choose
\be
\ell_{1}=(K^{b0}) \mathbf{t}
\ee
which satisfies self-null condition
$\ell_{1} (K^{b0})^{-1} \ell_{1}=0$. To complete the proof, we continue to find out a total set of $\ell_{1}, \ell_{2}, \dots, \ell_{N/2}$, so $\Gamma^\partial$ is a dimension $N/2$ Chern-Simons-charge lattice (Lagrangian subgroup).

For $\ell_{2}$, we choose its form as
\be
\ell_{2}=(\ell_{2,1},0,\ell_{2,3},0,\dots,0)
\ee
where even components of $\ell_{2}$ are zeros,
to satisfy
$\ell_{2}  (K^{b0})^{-1} \ell_{1} =\ell_{2} (K^{b0})^{-1} \ell_{2}=0$. The second constraint is automatically true for our choice of $\ell_{2}$.
The first constraint is achieved by $\ell_{2,1}( t_1)+\ell_{2,3}( t_3)=0$. We can properly choose $\ell_{2}$ to satisfy this constraint. 

For $\ell_{n}$, by mathematical induction, we choose its form as
\be
\ell_{n}=(\ell_{n,1},0,\ell_{n,3},0,\dots,\ell_{n,2n-1}, 0,\dots,0)
\ee
where even components of $\ell_{n}$ are zeros,
to satisfy
\be
\ell_{n}  (K^{b0})^{-1} \ell_{j}, \;\;\;j=1,\dots,n,
\ee
for any $n$.
For $2\leq j \leq n$, the constraint is automatically true for our choice of $\ell_{n}$ and $\ell_{j}$.
For $\ell_{n}  (K^{b0})^{-1} \ell_{1}=0$, it leads to the constraint:
$\ell_{n,1}( t_1)+\ell_{n,3}( t_3)+\dots \ell_{n,2n-1}( t_{2n-1}) =0$, we can generically choose
$\ell_{n,2n-1}\neq 0$ to have a new $\ell_{n}$ independent from other $\ell_{j}$ with $1\leq j \leq n-1$.

Notice the gapping term obeys U(1) symmetry, because $ \ell_{n} \cdot \mathbf{t}=\ell_{n} (K^{b0})^{-1} \ell_{1}=0$ is always true for all $\ell_{n}$.
Thus we have constructed a dimension $N/2$ Lagrangian subgroup $\Gamma^\partial=\{\ell_{1}, \ell_{2}, \dots, \ell_{N/2}\}$ which obeys
the boundary fully gapping rules ({\bf{1}})({\bf{2}})({\bf{3}})  in Sec.\ref{anomaly-gap}. (Q.E.D.)
\frm{{\bf Theorem:} ABJ's U(1) anomaly matching condition in 1+1D $\rightarrow$ the boundary fully gapping rules of 1+1D boundary/2+1D bulk with unbroken U(1) symmetry.}
We emphasize again that although we start with a single-U(1)-anomaly-free theory (aiming for a single U(1)-symmetry), it turns out that the full symmetry
after adding interacting gapping terms will result in a theory with an enhanced total U(1)$^{N/2}$ symmetry. 
The $N/2$ number of gapping terms break a total U(1)$^{N}$ symmetry (for $N$ free Weyl fermions) down to U(1)$^{N/2}$ symmetry.
The derivation follows directly from the statement in Appendix \ref{appendixC}, which we shall not repeat it.

We comment that our proofs in Appendix \ref{appendixC} and \ref{appendixD} are algebraic and topological, thus it is a non-perturbative result 
(instead of a perturbative result in the sense of doing weak or strong coupling expansions).

\section{More about the Proof of ``Boundary Fully Gapping Rules''} \label{AppendixE}

This section aims to demonstrate that the {\bf Boundary Fully Gapping Rules} 
used throughout our work, indeed can gap the edge states.
We discuss this proof here 
to make our work self-contained and to further convince the readers.

\subsection{Canonical quantization} \label{app:canonical-quant}
Here we set up the canonical quantization of the bosonic field $\phi_{I}$ for a multiplet chiral boson theory of Eq.(\ref{CSboundary}) on a 1+1D spacetime, with a spatial $S^1$ compact circle. 
The canonical quantization means that imposing a commutation relation between $\phi_{I}$ and its conjugate momentum field 
{$\Pi_{I}(x)=\frac{\delta {L}}{\delta (\partial_t \phi_{I} )}=\frac{1}{4\pi} K_{IJ} \partial_x \phi_{J}$.} 
Since $\phi_{I}$ is a compact phase of a matter field,
its bosonization contains both zero mode ${\phi_{0}}_{I}$ and winding momentum $P_{\phi_J}$, in addition to Fourier modes $\alpha_{I,n}$:\cite{Wang:2012am} 
\be \label{eq:mode}
\Phi_I(x) ={\phi_{0}}_{I}+K^{-1}_{IJ} P_{\phi_J} \frac{2\pi}{L}x+\ti \sum_{n\neq 0} \frac{1}{n} \alpha_{I,n} e^{-\ii n x \frac{2\pi}{L}}.
\ee
The periodic boundary has a size of length $0\leq x<L$, with $x$ identified with $x+L$. 
We impose the commutation relation for zero modes and winding modes, and generalized Kac-Moody algebra for Fourier modes:
\be
[{\phi_{0}}_{I},  P_{\phi_J}]=\ti\delta_{IJ},\;\; [\alpha_{I,n} , \alpha_{J,m} ]= n K^{-1}_{IJ}\delta_{n,-m}.
\ee
Consequently, the commutation relations for the canonical quantized fields are:
\bea
[\phi_I(x_1), K_{I'J} \partial_x \phi_{J}(x_2)]&=& {2\pi} \ti  \delta_{I I'} \delta(x_1-x_2), \label{eq:commutation}\\\;\;
[\phi_I(x_1),\Pi_{J}(x_2)]&=&  \frac{1}{2} \ti  \delta_{IJ} \delta(x_1-x_2).
\eea

\subsection{Approach I:  Mass gap for gapping zero energy modes} \label{sec-app-gap-zero}

We provide the first approach to show that the anomaly-free edge states can be gapped under the properly-designed gapping terms.
Here we explicitly calculate the mass gap for the zero energy mode and its higher excitations. The generic theory is
\bea \label{eq:Ap-sine-Gordon}
S_{\partial}&=& \frac{1}{4\pi} \int \dd t \; \dd x \; ( K_{IJ} \partial_t \Phi_{I} \partial_x \Phi_{J} -V_{IJ}\partial_x \Phi_{I}   \partial_x \Phi_{J}) \nonumber\\
&+& \int \dd t \; \dd x\;  \sum_{a} g_{a}  \cos(\ell_{a,I}^{} \cdot\Phi_{I}).
\eea
We will consider the even-rank symmetric $K$ matrix, so the full edge theory has an even number of modes 
and thus potentially be gappable.
In the following we shall determine under what conditions that the edge states can obtain a mass gap.
Imagining at the large coupling $g$, the $\Phi_{I}$ field get trapped at the minimum of the cosine potential with small fluctuations.
We will perform an 
expansion of $\cos(\ell_{a,I}^{} \cdot\Phi_{I}) \simeq 1 - \frac{1}{2}(\ell_{a,I}^{} \cdot\Phi_{I})^2 +\dots$ to a quadratic order 
and see what it implies about the mass gap.
We can diagonalize the Hamiltonian, 
\be
H \simeq (\int^L_0 \dd x\; V_{IJ}\partial_x \Phi_{I}   \partial_x \Phi_{J} )+\frac{1}{2} \sum_a g_a (\ell_{a,I}^{} \cdot\Phi_{I})^2 L +\dots 
\ee
under a complete $\Phi$ mode expansion, and find the energy spectra for its eigenvalues.
To summarize the result, we find that:\\

\noindent
{\bf (E-1).} \emph{If and only if} we include all the gapping terms allowed by {\bf Boundary Full Gapping Rules}, we can open the mass gap above zero modes ($n=0$) as well as Fourier modes (non-zero modes $n\neq 0$). Namely,
the energy spectrum is in the form of
\be \label{eq:ap-mass-gap-st}
E_n= \big( \sqrt{ \Delta^2 + \# (\frac{2\pi n}{L})^2 } + \dots \big),
\ee
where $\Delta$ is the mass gap. Here we emphasize the energy of Fourier modes($n\neq 0$) behaves towards zero modes 
at long wave-length low energy limit ($L \to \infty$). Such spectra become continuous at $L \to \infty$ limit, which is the expected energy behavior. \\

\noindent
{\bf (E-2).} \emph{If} we include the \emph{incompatible} Wilson line operators, such as $\ell$ and $\ell'$ where $\ell K^{-1} \ell' \neq 0$, 
while the interaction terms contain \emph{incompatible} gapping terms $g \cos(\ell_{} \cdot \Phi) +g' \cos(\ell_{}' \cdot \Phi)$, we find the \emph{unstable} energy spectra
\be \label{eq:ap-mass-gap-unst}
E_n= \big( \sqrt{ \Delta^2 + \# (\frac{2\pi n}{L})^2+ g \,g' (\frac{L}{n})^2 \dots+\dots } + \dots \big),
\ee
The energy spectra shows an \emph{instability} of the system, because at low energy limit ($L \to \infty$), the spectra become discontinuous (from $n=0$ to $n \neq 0$) and jump to infinity as long as 
there are incompatible gapping terms(namely, $g \cdot g' \neq 0$). 
Such disastrous behavior of $(L/n)^2$ implies the quadratic expansion analysis may not account for the whole physics.
In that case, the disastrous behavior invalidates the trapping of $\Phi$ field at a local minimum, thus 
invalidates the mass gap, and the \emph{unstable} system potentially seeks to be \emph{gapless phases}.

Below we demonstrate the result explicitly for the simplest rank-2 $K$ matrix, while the case for higher rank $K$ matrix can be straightforwardly generalized.
The most general rank-2 $K$ matrix is
\be 
K\equiv {\begin{pmatrix} 
k_1 &k_3 \\
k_3 & k_2 \
\end{pmatrix}} \equiv {\begin{pmatrix} 
k_1 &k_3 \\
k_3 & (k_3^2-p^2)/k_1 \
\end{pmatrix}},
\,\;\;  
V={\begin{pmatrix} v_1 &v_2 \\
  v_2 & v_1 \\
\end{pmatrix}}, 
\ee
while the $V$ velocity matrix is chosen to be rescaled as the above. (Actually the $V$ matrix is immaterial to our conclusion.)
Our discussion below holds for both $k_3=\pm |k_3|$ cases.
We define $k_2=(k_3^2-p^2)/k_1$, so that
$
\det(K)=-p^2
$
We find that only when 
$$\sqrt{|\det(K)|} \equiv p \in \mathbb{Z},$$ 
$p$ is an integer, we can find gapping terms allowed by Boundary Fully Gapping Rules. 
(A side comment is that $\det(K)=-p^2$ implies its bulk can be constructed as a \emph{quantum double} or a \emph{twisted quantum double model} on the lattice.)
For the above rank-2 $K$ matrix, we find two independent sets, $\{\ell_{1}=(\ell_{1,1},\ell_{1,2})\}$ and $\{\ell_{1}'=(\ell_{1,1}',\ell_{1,2}')\}$, each set has only one $\ell$ vector.
Here the $\ell$ vector is written as $\ell_{a,I}$, with the index $a$ labeling the $a$-th (linear independent) $\ell$ vector in the Lagrangian subgroup, and the index $I$ labeling the $I$-component of the $\ell_{a}$ vector.
Their forms are:
\bea
\frac{\ell_{1,1}^{}}{\ell_{1,2}^{}}&=&\frac{k_1}{k_3+p}=\frac{k_3-p}{k_2},\\
\frac{{\ell_{1,1}'}}{{\ell_{1,2}'}}&=&\frac{k_1}{k_3-p}=\frac{k_3+p}{k_2}.
\eea

We denote the cosine potentials spanned by these  $\ell_1$ and  $\ell_1'$ vectors in Eq.(\ref{eq:Ap-sine-Gordon}) as:
\be
g \cos(\ell_{1} \cdot \Phi) +g' \cos(\ell_{1}' \cdot \Phi).
\ee
From our understanding of Boundary Full Gapping Rules, these two $\ell_1$, $\ell_1'$ vectors are \emph{not compatible to each other}.
In this sense, \emph{we shall not include both terms if we aim to fully gap the edge states}.

Now we focus on computing the mass gap of our interests for the bosonic $K$ matrix 
$K^{b}_{2\times 2} =\bigl( {\begin{smallmatrix}
0 &1 \\
1 & 0
\end{smallmatrix}}  \bigl)$ 
and the fermionic $K$ matrix
$K^{f}_{2\times 2} =\bigl( {\begin{smallmatrix}
1 & 0 \\
0 & -1
\end{smallmatrix}}  \bigl)$.
We use both the Hamiltonian or the Lagrangian formalism to extract the energy, for both zero modes ($n=0$) and Fourier modes(non-zero modes $n\neq 0$).
For both the Hamiltonian and Lagrangian formalisms, we obtain the consistent result for energy gaps $E_n$:

\begin{widetext}
\noindent
{\bf 1st Case: Bosonic} 
$ K^{b}_{2\times 2} =\bigl( {\begin{smallmatrix}
0 &1 \\
1 & 0
\end{smallmatrix}}  \bigl)$:
\be
E_n= \sqrt{2 \pi (g+g') v_1 +(\frac{2\pi n}{L})^2 v_1^2 + g\, g' (\frac{L}{n})^2} \pm (\frac{2\pi n}{L})v_2 .
\ee
\noindent
{\bf 2nd Case: Fermionic} 
$ K^{f}_{2\times 2}=\bigl( {\begin{smallmatrix}
1 & 0 \\
0 & -1
\end{smallmatrix}}  \bigl)$: 
\be
E_{n}=\sqrt{4 \pi g (v_1-v_2) +4 \pi g' (v_1+v_2) +(\frac{2\pi n}{L})^2 (v_1^2-v_2^2)+ (\frac{2 L}{n})^2 g\, g' }.
\ee
\end{widetext}
Logically, for a rank-2 $K$ matrix, we have shown that: \\

\noindent
$\bullet$ \emph{If} we include the gapping terms allowed by {Boundary Full Gapping Rules}, either (i) $g \neq  0, g' =  0$, or 
(ii) $g  = 0, g'  \neq 0$, then we have the \emph{stable} form of the mass gap in Eq.(\ref{eq:ap-mass-gap-st}).
Thus we show the \emph{if}-statement in {\bf (E-1)}. \\

\noindent
$\bullet$ \emph{If} we include incompatible interaction terms (here $\ell_1 K^{-1} \ell_1' \neq 0$), such that both $g \neq 0$ and $g' \neq 0$,
then the energy gap is of the \emph{unstable} form in Eq.(\ref{eq:ap-mass-gap-unst}). Thus we show the statement in {\bf (E-2)}.\\

\noindent
$\bullet$ Meanwhile, this {\bf (E-2)} implies that if we include \emph{more} interaction terms allowed by {Boundary Full Gapping Rules}, we have an unstable energy gap, thus it may
drive the system to the gapless states due to the instability.
Moreover, if we include \emph{less} interaction terms allowed by {Boundary Full Gapping Rules} (i.e. if we do not include all allowed \emph{compatible} gapping terms), then
we cannot fully gap the edge states (For $1$-left-moving mode and $1$-right-moving mode, we need at least $1$ interaction term to gap the edge.)
Thus we also show the \emph{only-if}-statement in {\bf (E-1)}.

This approach works for a generic even-rank $K$ matrix thus can be applicable to show the above statements {\bf (E-1)} and {\bf (E-2)} hold in general.  
More generally, for a rank-$N$ $K$ matrix Chern-Simons theory, 
with the boundary $N/2$-left-moving modes and $N/2$-right-moving modes, we need \emph{at least and at most} $N/2$-linear-independent interaction terms to gap the edge.
{\bf If one includes more terms than the allowed terms (such as the numerical attempt in Ref.\onlinecite{CGP1247}),
it may} \emph{drive the system to the gapless states due to the instability from the unwanted quantum fluctuation.} {\bf This can be one of the reasons why Ref.\onlinecite{CGP1247} 
fails to achieve gapless fermions by gapping mirror-fermions}.\\

\noindent
{\bf 3rd Case: General even-rank $K$ matrix:}  
Here we outline another view of the energy-gap-stability for the edge states, for a generic rank-$N$ $K$ matrix Chern-Simons theory with multiplet-chiral-boson-theory edge states. 
We include the full interacting cosine term for the lowest energy states - zero and winding modes:
\be \label{eq:cos}
\cos(\ell_{a,I}^{} \cdot\Phi_{I}) \to  
\cos(\ell_{a,I}^{} \cdot ({\phi_{0}}_{I}+K^{-1}_{IJ} P_{\phi_J} \frac{2\pi}{L}x) ),
\ee
while we drop the higher energy Fourier modes.
(Note when $L \to \infty$, the kinetic term $H_{kin}=\frac{(2\pi)^2}{4\pi L}  V_{IJ} K^{-1}_{I l1} K^{-1}_{J l2} P_{\phi_{l1}} P_{\phi_{l2}}$ has an order $O(1/L)$ so is negligible, thus the cosine potential Eq.\;(\ref{eq:cos}) dominates.
Though to evaluate the mass gap, we keep both kinetic and potential terms.) 
The stability of the mass gap can be understood from \emph{under what conditions} we can safely expand the cosine term to extract the leading quadratic terms 
by only keeping the zero modes
via $\cos(\ell_{a,I}^{} \cdot\Phi_{I}) \simeq 1 - \frac{1}{2}(\ell_{a,I}^{} \cdot\phi_{0I})^2 +\dots$.
(If one does not decouple the winding mode term, there is a complicated $x$ dependence in $P_{\phi_J} \frac{2\pi}{L}x$ along the $x$ integration.)
The challenge for this cosine expansion is rooted in the \emph{non-commuting} algebra from $[{\phi_{0}}_{I},  P_{\phi_J}]=\ti \delta_{IJ}$.
This can be resolved by requiring $\ell_{a,I}^{}  {\phi_{0}}_{I}$ and $\ell_{a,I'}^{}  K^{-1}_{I'J} P_{\phi_J}$ \emph{commute} 
in Eq.(\ref{eq:cos}),
\bea
[\ell_{a,I}^{}  {\phi_{0}}_{I}, \;\ell_{a,I'}^{}  K^{-1}_{I'J} P_{\phi_J}] &=&\ell_{a,I}^{} K^{-1}_{I'J} \ell_{a,I'}^{}  \; (\ti\delta_{IJ}) \nonumber\\
&=&(\ti)(\ell_{a,J}^{} K^{-1}_{I'J} \ell_{a,I'}^{} )=0.\;\;\;\;\;\;
\eea  
This is indeed the {Boundary Full Gapping Rules (1)}, the trivial 
statistics rule among the Wilson line operators for the gapping terms.  
Under this \emph{commuting condition} (we can interpret that there is \emph{no unwanted quantum fluctuation}), 
we can thus expand Eq.(\ref{eq:cos}) using the trigonometric identity for c-numbers as
\bea
&&\cos(\ell_{a,I}^{} {\phi_{0}}_{I}) \cos(\ell_{a,I}^{} K^{-1}_{IJ} P_{\phi_J} \frac{2\pi}{L}x) \nonumber \\
&&-\sin(\ell_{a,I}^{} {\phi_{0}}_{I}) \sin(\ell_{a,I}^{} K^{-1}_{IJ} P_{\phi_J} \frac{2\pi}{L}x) 
\eea
and then we safely integrate over $L$. 
Note that 
both $\cos(\dots x)$ and $\sin(\dots x)$ are periodic in the region $[0,L)$, so both $x$-integrations vanish unless when $\ell_{a,I}^{} \cdot K^{-1}_{IJ} P_{\phi_J} =0$ such that
$\cos(\ell_{a,I}^{} K^{-1}_{IJ} P_{\phi_J} \frac{2\pi}{L}x)=1$.
We thus obtain
\be
 g_{a} \int_0^{L} \dd x\; \text{Eq}.(\ref{eq:cos})=g_{a} L \; \cos(\ell_{a,I}^{} \cdot {\phi_{0}}_{I}) \delta_{(\ell_{a,I}^{} \cdot K^{-1}_{IJ} P_{\phi_J} ,0)}.
\ee
The Kronecker-delta-condition $\delta_{(\ell_{a,I}^{} \cdot K^{-1}_{IJ} P_{\phi_J} ,0)}=1$ 
implies that there is a nonzero value if and only if $\ell_{a,I}^{} \cdot K^{-1}_{IJ} P_{\phi_J} =0$. 
This is also consistent with the \emph{Chern-Simons quantized lattice} as the Hilbert space of the ground states.
Here $P_{\phi}$ forms a discrete quantized lattice because its conjugate  ${\phi_{0}}_{}$ is periodic. %
This result can be applied to count the ground state degeneracy of Chern-Simons theory on a closed manifold or a compact manifold with gapped boundaries.\cite{Wang:2012am,{Kapustin:2013nva}}

In short, we have shown 
that when 
$\ell^T K^{-1} \ell=0$, we have the desired cosine potential expansion via the zero mode quadratic expansion at large $g_a$ coupling,
$  g_{a} \int_0^{L}\dd x\cos(\ell_{a,I}^{} \cdot\Phi_{I}) \simeq  - g_{a} L \frac{1}{2}(\ell_{a,I}^{} \cdot\phi_{0I})^2 +\dots$.
The nonzero mass gaps of zero modes can be readily shown by solving the quadratic simple harmonic oscillators of both the kinetic and the leading-order of the potential terms:
\be
\frac{(2\pi)^2}{4\pi L}  V_{IJ} K^{-1}_{I l1} K^{-1}_{J l2} P_{\phi_{l1}} P_{\phi_{l2}} +\sum_a g_{a} L  \frac{1}{2}(\ell_{a,I}^{} \cdot\phi_{0I})^2.
\ee
The mass gap is independent of the system size, the order one finite energy gap 
\be
\Delta_{} \simeq O(\sqrt{2\pi\, g_a \ell_{a,l1} \ell_{a,l2} V_{IJ} K^{-1}_{I l1} K^{-1}_{J l2} }), 
\ee
which the mass matrix can be properly diagonalized, since there are only conjugate variables $\phi_{0I}$ and $P_{\phi,J}$ in the quadratic order.

We again find that the above statements consistent with {\bf (E-1)} and {\bf (E-2)} for a generic even-rank $K$ matrix.\\

\subsection{Mass Gap for Klein-Gordon fields and non-chiral Luttinger liquids under sine-Gordon potential} \label{subsec:RG analysis}
First, we recall the two statements {\bf (E-3)},{\bf (E-4)} that: \\

\noindent
{\bf (E-3)} A \emph{scalar boson theory} of a Klein-Gordon action with a sine-Gordon potential:
\be \label{eq:E-3}
S_{\partial}= \int \dd t \, \dd x \; \frac{\kappa }{2}( \partial_t \varphi_{} \partial_t \varphi_{} - \partial_x \varphi_{}   \partial_x \varphi_{}) 
+  g_{}  \cos(\beta\varphi_{}).
\ee
at strong coupling $g$ can induce the mass gap for the scalar mode $\varphi$.\\

\noindent
{\bf (E-4)} A \emph{non-chiral Luttinger liquids} (non-chiral in the sense of equal left-right moving modes, but can have U(1)-charge-chirality with respect to a U(1) symmetry) with $\phi$ and $\theta$ dual scalar fields with 
a sine-Gordon potential for $\phi$ field:
\bea  \label{eq:E-4}
S_{\partial} &=& \int \dd t \, \dd x \; \Big( \frac{1}{4\pi}( (\partial_t \phi_{} \partial_x \theta_{} +\partial_x \phi_{} \partial_t \theta_{}) -V_{IJ} \partial_x \Phi_{I}   \partial_x \Phi_{J}) \nonumber \\
&+&  g_{}  \cos(\beta \; \theta_{}) \Big).
\eea
at strong coupling $g$ can induce the mass gap for \emph{all} the scalar mode $\Phi \equiv (\phi, \theta)$. 

Indeed, the statement {\bf (E-3)}  and {\bf (E-4)} are related because Eq.(\ref{eq:E-3}) and Eq.(\ref{eq:E-4}) are identified by the canonical conjugate momentum relation:
\be
\partial_t \phi_{} \sim  \partial_x \theta_{}, \;\;\; \partial_t \theta \sim  \partial_x  \phi_{},
\ee
up to a normalization factor and up to some Euclidean time transformation. 

There are immense and broad amount of literatures demonstrating {\bf (E-3)},{\bf (E-4)} are true,
and we recommend to look for Ref.\onlinecite{{W},{Altland:2006si},{Giamarchi}}.

\subsection{Approach II: Map the anomaly-free theory with gapping terms to the decoupled non-chiral Luttinger liquids with gapped spectrum} \label{app:approachII}

Here we provide the second approach to show that the anomaly-free edge states can be gapped under the properly-designed gapping terms.
The key step is that we will
map the $N$-component anomaly-free theory with properly-designed gapping terms to $N/2$-decoupled-copies of non-chiral Luttinger liquids of the statement {\bf (E-4)}, each copy has the gapped spectrum.
(This key step is logically the same as the proof in Appendix A of Ref.\onlinecite{Wang:2013vna}.)
Thus, by the equivalence mapping, we can prove that the anomaly-free edge states can be fully gapped.
We include this proof\cite{Wang:2013vna} to make our claim self-contained.

 We again consider the generic theory of Eq.(\ref{eq:Ap-sine-Gordon}):
\bea  
&&S_{\partial}(\Phi,K, \{ \ell_{a}\})= \frac{1}{4\pi} \int \dd t \; \dd x \; ( K_{IJ} \partial_t \Phi_{I} \partial_x \Phi_{J}  \nonumber\\
&&-V_{IJ}\partial_x \Phi_{I}   \partial_x \Phi_{J})  + \int \dd t \; \dd x\;  \sum_{a} g_{a}  \cos(\ell_{a,I}^{} \cdot\Phi_{I}),\nonumber
\eea
where $\Phi$, $K$, $\{ \ell_{a}\}$ are the data for this 1+1D action $S_{\partial}(\Phi,K, \{ \ell_{a}\})$, while the velocity matrix is not universal and is immaterial to our discussion below.
In Appendix \ref{appendixD}, 
we had shown that the $N$-component anomaly-free theory guarantees the $N/2$-linear-independent gapping terms of boundary gapping lattice (Lagrangian subgroup)  
$\Gamma^\partial$ satisfying:
\be \label{eq:ellKell}
\ell_{a,I} K^{-1}_{IJ} \ell_{b,J}=0
\ee
for any $\ell_a, \ell_b \in \Gamma^\partial$.
In our case (both bosonic and fermionic theory), all the $K$ is invertible due to $\det(K) \neq 0$, thus one can define a dual vector as in Ref.\onlinecite{Wang:2013vna}, 
$\ell_{a,I}= K_{I I'} \eta_{a,I'}$, such that Eq.(\ref{eq:ellKell}) becomes
\be \label{eq:etKet}
\eta_{a,I'} K_{IJ} \eta_{b,J'}=0.
\ee
The data of action becomes $S_{\partial}(\Phi,K, \{ \ell_{a}\}) \to S_{\partial}(\Phi,K, \{ \eta_{a}\})$.
In our proof, we will stick to the data $S_{\partial}(\Phi,K, \{ \ell_{a}\})$. We can construct a $N \times (N/2)$-component integer-valued matrix $\mathbf{L}$: 
\be
\mathbf{L} \equiv \Big(\ell_{1}, \ell_{2}, \dots,  \ell_{N/2}  \Big)
\ee
with $N/2$ column vectors, and each column vector is $\ell_{1}, \ell_{2}, \dots,  \ell_{N/2}$.
We can write $\mathbf{L}$ base on the integer-valued Smith normal form, so 
$\mathbf{L}= V D W$,
with $V$ is a $N \times N$ integer-valued matrix and $W$ is a $(N/2) \times (N/2)$ integer-valued matrix.
Both $V$ and $W$ have a determinant $\det(V)=\det(W)=1$.
The $D$ is a $N \times (N/2)$ integer-valued matrix:
\be
D\equiv \begin{pmatrix} \bar{D}\\ 0\end{pmatrix} \equiv
\begin{pmatrix} d_1& 0& \dots & 0\\ 
0 &d_2& \dots & 0\\
\vdots & \vdots & \vdots & \vdots\\
0 &0 & \dots & d_{N/2}\\
\vdots & \vdots & \vdots & \vdots\\
0 & 0 & \vdots & 0
\end{pmatrix},
\ee
with $\bar{D}$ is a diagonal integer-valued matrix.
Since $\mathbf{L}$ has $N/2$-linear-independent column vectors, thus $\det(\bar{D}) \neq 0$, and all entries of $\bar{D}$ are nonzero.

\noindent
{\bf 1st Mapping} -
We do a change of variables:
\bea
\Phi' &=&V^{T} \Phi \nonumber\\ 
\ell' &=& V^{-1} \ell\nonumber\\ 
K' &=& V^{-1} K (V^T)^{-1} \nonumber\\ 
S_{\partial}(\Phi,K, \{ \ell_{a}\}) &\to& S_{\partial}(\Phi',K', \{ \ell_{a}'\}).\;\;\;\; \nonumber
\eea
This makes the $\mathbf{L}'$ form simpler:
\bea 
\mathbf{L}'=V^{-1}\mathbf{L}=V^{-1}(VDW)= \begin{pmatrix} \bar{D} W\\ 0\end{pmatrix}.
\eea
Here is the key step: due to Eq.(\ref{eq:ellKell}), we have the important equality,
\bea \label{eq:LKL}
\boxed{\mathbf{L}^T K^{-1} \mathbf{L}=0},
\eea
thus
\bea
&&(VDW)^T K^{-1} V D W=0\\
&&= W^T D^T K'^{-1} DW=0\\
&&=(\bar{D} W, 0) K'^{-1} \begin{pmatrix} \bar{D} W\\ 0\end{pmatrix}=0.
\eea
Hence, $K'^{-1}$ can be written as the following four blocks of $N \times N$ matrices $\tF,\tG$ ($\tF,\tG$ can have fractional values):
\be
K'^{-1}=\begin{pmatrix} 
0 & \tF\\
\tF^T & \tG
\end{pmatrix},
\ee
with $\det(\tF)\neq 0$ and $\tG$ is symmetric.
Thus the \emph{integer} $K'$ matrix has the form
\be
K'=\begin{pmatrix} 
-(\tF^T)^{-1} \tG \tF^{-1} & (\tF^T)^{-1}\\
\tF^{-1} & 0
\end{pmatrix}.
\ee
We notice that, 
\frm{{\bf Lemma 1}: Due to $K'$ matrix is an \emph{integer} matrix, the three matrices $-(\tF^T)^{-1}\tG \tF^{-1}$, $\tF^{-1}$ and $(\tF^T)^{-1}$ are \emph{integer matrices}. 
Therefore, $\tF$ and $\tG$ can be \emph{fractional matrices}.}

\noindent
{\bf 2nd Mapping} - To obtain the final mapping to $N/2$-decoupled-copies of  non-chiral Luttinger liquids, we do
another change of variables:
\bea
\Phi'' &=&U \Phi' \nonumber\\ 
\ell'' &=& (U^{-1})^T \ell' \nonumber\\ 
K'' &=& (U^T)^{-1} K' (U)^{-1} \nonumber\\ 
S_{\partial}(\Phi', K', \{ \ell_{a}'\}) &\to& S_{\partial}(\Phi'',K'', \{ \ell_{a}''\}).\;\;\;\; \nonumber
\eea
With the goal in mind to make the new $K$ matrix 
$K''=(K'')^{-1}=
\begin{pmatrix} 
0 & \mathbf{1}\\
\mathbf{1} & 0
\end{pmatrix}$ and $\mathbf{1}$ is the $N \times N$ identity matrix.
This constrains $U$, and we find
\bea
&&
(K'')^{-1}=U (K')^{-1} U^T=\begin{pmatrix} 
0 & \mathbf{1}\\
\mathbf{1} & 0
\end{pmatrix}\\
&&\Rightarrow U=\begin{pmatrix} 
-\frac{1}{2}(\tF^T)^{-1}\tG \tF^{-1} & (\tF^T)^{-1}\\
\mathbf{1} & 0
\end{pmatrix}.
\eea

Importantly, due to {{\bf Lemma 1}, we have $(\tF^T)^{-1}$ and $-(\tF^T)^{-1} \tG \tF^{-1}$  are \emph{integer matrices}, 
so U is at most a matrix taking half-integer values (almost an integer matrix). 
 
In the new $\Phi''$ basis, 
we define the $N$-component column vector
$$\Phi''=( \bar{\phi}_1, \bar{\phi}_2, \dots, \bar{\phi}_{N/2}, \bar{\theta}_1, \bar{\theta}_2, \dots,  \bar{\theta}_{N/2}).$$
Based on Appendix \ref{app:canonical-quant}, the canonical-quantization in the new basis becomes
\bea \label{eq:non-chiral-Lut}
&&[\Phi_I''(x_1), \partial_x \Phi_{J}''(x_2)]= {2\pi} \ti   ({K''}^{-1})_{IJ}  \delta(x_1-x_2), \nonumber\\ \;\;
&&[\bar{\phi}_I(x_1), \partial_x \bar{\phi}_{J}(x_2)]=[\bar{\theta}_I(x_1), \partial_x \bar{\theta}_{J}(x_2)]=0,\;\; \nonumber \\ \;
&&[\bar{\phi}_I(x_1), \partial_x \bar{\theta}_{J}(x_2)]={2\pi} \ti   \delta_{IJ} \delta(x_1-x_2).
\eea
This is exactly what we aim for the decoupled non-chiral Luttinger liquids as the form of $N/2$-copies of {\bf (E-4)}.
However, the cosine potential in the new basis is not yet fully decoupled due to
\bea
&&\ell''^T \Phi''
= \ell^T (V^{-1})^{T} (U^{-1}) \Phi''\nonumber\\
&&\Rightarrow \mathbf{L}''^T = \mathbf{L}^T(V^{-1})^{T} (U^{-1})\nonumber=(W^T D^T) (U^{-1})\\
&&\Rightarrow \mathbf{L}''^T = \begin{pmatrix} W^T \bar{D}, 0\end{pmatrix}  
\begin{pmatrix} 
0 & \mathbf{1} \\
\tF^{T} & \frac{1}{2} \tG \tF^{-1}
\end{pmatrix}
=
\begin{pmatrix} 0, W^T \bar{D}\end{pmatrix}.
\nonumber
\eea
We obtain the cosine potential term as
\be \label{eq:cosine-map}
g_{a}  \cos(\ell_{a,I}^{} \cdot\Phi_{I})
=g_a\cos( W_{Ja} d_{J} \bar{\theta}_{J}).
\ee
If $W_{Ja}$ 
is a diagonal matrix, the non-chiral Luttinger liquids are decoupled into $N/2$-copies also in the interacting potential terms.
In general, $W_{Ja} d_{J}$ may not be diagonal, but the charge quantization and the
large coupling $g_a$ of the cosine potentials cause 
$$
\sum_J W_{Ja} d_{J} \bar{\theta}_{J} =2 \pi n_I,\;\; I=1,\dots, N/2, \;\;n_I \in \Z
$$
locked to the minimum value. Equivalently, due to both $W$ and $W^{-1}$ are integer matrices with $\det(W)=1$, 
we have
\be \label{eq:locked}
d_{J} \bar{\theta}_{J} =2 \pi n_J',\;\; J=1,\dots, N/2, \;\;n_J' \in \Z.
\ee

The last step is to check the constraint on the $\bar{\phi}_{I}$ and $\bar{\theta}_{J}$. The original particle number quantization constraint changes from
$
\frac{1}{2\pi} \int^L_0  \dd x {\partial_x \Phi_I}=\zeta_I 
$ with an integer $\zeta_I \in \Z$, to
\be \label{eq:frac-quant}
\left\{ 
\begin{array}{l}
 \int^L_0 \dd x \frac{{\partial_x \bar{\phi}_I} }{2\pi} =-\frac{1}{2}((\tF^T)^{-1}\tG \tF^{-1}  V^T)_{Ij} \zeta_j +\overset{N/2}{\underset{j=1}{\sum}} (\tF^T)^{-1}_{I,j} \zeta_{N/2+j},\\
\int^L_0  \dd x \frac{{\partial_x \bar{\theta}_{I}}}{2\pi}   =\overset{N/2}{\underset{j=1}{\sum}} V^T_{I,I+j} \zeta_j.
\end{array}
\right. 
\ee

Again, from {{\bf Lemma 1}, we have $(\tF^T)^{-1}$ and $-(\tF^T)^{-1} \tG \tF^{-1}$  are \emph{integer matrices}, and $V$ is an integer matrix,
so at least the particle number quantization of $ \int^L_0 \dd x \frac{{\partial_x \bar{\phi}_I} }{2\pi}$ takes as multiples of half-integer values, due to
the half-integer valued matrix term $\frac{1}{2}((\tF^T)^{-1}\tG \tF^{-1}  V^T)$.
Meanwhile, $\int^L_0  \dd x \frac{{\partial_x \bar{\theta}_{I}}}{2\pi}$ must have integer values.

In the following, we verify that the physics at strong coupling $g$ of cosine potentials 
still render the decoupled non-chiral Luttinger liquids with integer particle number quantization
regardless a possible {\bf half-integer quantization} at Eq.(\ref{eq:frac-quant}). 
The reason is that, at large $g$, the cosine potential $g_a\cos( W_{Ja} d_{J} \bar{\theta}_{J})$
effectively acts as $g_a\cos(  d_{a} \bar{\theta}_{a})$.
In this way, $\bar{\theta}_{a}$ is locked, so $\partial_x\bar{\theta}_{a}=0$ and that constrains $\int^L_0  \dd x \frac{{\partial_x \bar{\theta}_{I}}}{2\pi}   =0$ with no instanton tunneling. 
This limits Eq.(\ref{eq:frac-quant})'s $\zeta_j=0$ for $j=1,\dots,N/2$. 
And Eq.(\ref{eq:frac-quant}) at large $g$ coupling becomes
\bea \label{eq:interger-quant}
\left\{ 
\begin{array}{l}
 \int^L_0 \dd x \frac{{\partial_x \bar{\phi}_I} }{2\pi} =\overset{N/2}{\underset{j=1}{\sum}} (\tF^T)^{-1}_{I,j} \zeta_{N/2+j} \in \Z. \\
\int^L_0 \dd x \frac{{\partial_x \bar{\theta}_{I}}}{2\pi}   =0.
\end{array}
\right. 
\eea
We now conclude that, the allowed Hilbert space at large $g$ coupling is the same as the Hilbert space of $N/2$-decoupled-copies of non-chiral Luttinger liquids.

Though we choose a different basis for the gapping rules than Ref.\onlinecite{Wang:2013vna}, 
we still reach the same conclusion 
as long as the key criteria Eq.(\ref{eq:LKL}) 
holds. Namely, with
$\mathbf{L}^T K^{-1} \mathbf{L}=0$,
we can derive these three equations 
Eq.(\ref{eq:non-chiral-Lut}),(\ref{eq:locked}),(\ref{eq:interger-quant}), thus 
we have mapped the theory with gapping terms (constrained by $\mathbf{L}^T K^{-1} \mathbf{L}=0$)  
to the $N/2$-decoupled-copies of  non-chiral Luttinger liquids with $N/2$ number of effective decoupled gapping terms $\cos(d_{J} \bar{\theta}_{J})$ with $J=1,\dots, N/2$.
This maps to $N/2$-copies of non-chiral Luttinger liquids {\bf (E-4)}, and we have shown that each {\bf (E-4)} has the gapped spectrum. We prove the mapping:

\frm{\center{the $K$ matrix multiplet-chirla boson theories with gapping terms $\mathbf{L}^T K^{-1} \mathbf{L}=0$ \\
$\to$ \;\;\;\;\\ 
 $N/2$-decoupled-copies of  non-chiral Luttinger liquids of {\bf (E-4)} with energy gapped spectra. (Q.E.D.)}}

Since we had shown in Appendix \ref{appendixD} that for the U(1) {theory} of totally even-$N$ left/right chiral Weyl fermions, 
only the {\bf anomaly-free} theory 
can provide the $N/2$-gapping terms 
with $\mathbf{L}^T K^{-1} \mathbf{L}=0$, this means that we have established the map:
\frm{\center{the U(1)$^{N/2}$ anomaly-free theory ($\mathbf{q} \cdot {K}^{-1} \cdot \mathbf{q} = \mathbf{t} \cdot  {K}^{} \cdot  \mathbf{t}=0$) with gapping terms $\mathbf{L}^T K^{-1} \mathbf{L}=0$ \\
$\to$ \;\;\;\;\\ 
 $N/2$-decoupled-copies of non-chiral Luttinger liquids of {\bf (E-4)} with gapped energy spectra.}}
This concludes the second approach proving the 1+1D U(1)-anomaly-free theory can be gapped by adding properly designed interacting gapping terms with $\mathbf{L}^T K^{-1} \mathbf{L}=0$. (Q.E.D.)

\subsection{Approach III: Non-Perturbative statements of Topological Boundary 
Conditions,
Lagrangian subspace, and the exact sequence } \label{sec:TQFT-proof}

In this subsection, from a TQFT viewpoint, we provide another non-perturbative proof of Topological Boundary Gapping Rules (which logically follows
Ref.\onlinecite{Kapustin:2010hk}) 
\be
\boxed{\mathbf{L}^T K^{-1} \mathbf{L}=0},
\ee
with
\be
\mathbf{L} \equiv \Big(\ell_{1}, \ell_{2}, \dots,  \ell_{N/2}  \Big)
\ee
with $N/2$ column vectors, and each column vector is $\ell_{1}, \ell_{2}, \dots,  \ell_{N/2}$;
the 
even-$N$-component left/right chiral Weyl fermion theory with Topological Boundary Gapping Rules
must have
$N/2$-linear independent gapping terms of {\bf Boundary Gapping Lattice(Lagrangian subgroup)}  $\Gamma^\partial$ satisfying:
$
\ell_{a,I} K^{-1}_{IJ} \ell_{b,J}=0
$
for any $\ell_a, \ell_b \in \Gamma^\partial$.

Here is the general idea: For any field theory, 
a boundary condition is defined by a Lagrangian submanifold in the
space of Cauchy boundary condition data on the boundary. If we want a boundary condition
which is topological (namely with a mass gap without gapless modes), then importantly it must treat all directions on
the boundary in the equivalent way. 
So, for a gauge theory, we end up choosing 
a Lagrangian subspace 
in the Lie algebra of the gauge group. 
A subspace is {\bf Lagrangian} \emph{if and only if} it is both {\bf isotropic and coisotropic}. 
For a finite-dimensional vector space $\bfV$, 
a Lagrangian subspace is an isotropic one whose dimension is half that of the vector space. 

More precisely,
for $\mathbf{W}$ be a linear subspace of a finite-dimensional vector space $\mathbf{V}$. Define the symplectic complement of $\bfW$ to be the subspace $\bfW^{\perp}$ as
$
\bfW^{\perp} = \{v\in \bfV \mid \omega(v,w) = 0, \;\;\; \forall w\in \bfW\}.
$
Here $\omega$ is the symplectic form, in the commonly-seen matrix form is
$
\omega=(\begin{smallmatrix} 0 & \mathbf{1} \\ -\mathbf{1} & 0\end{smallmatrix})
$
with $0$ and $\mathbf{1}$ are the block matrix of the zero and the identity.
In our case, $\omega$ is related to the fermionic $K=K^{f}$ and bosonic $K=K^{b0}$ matrices.
The symplectic complement $\bfW^{\perp}$ satisfies:
$(\bfW^{\perp})^{\perp} = \bfW$ and
$\dim \bfW + \dim \bfW^\perp = \dim \bfV.$
{\bf Isotropic, coisotropic, Lagrangian} means the following:\\

\noindent
$\bullet$ $\bfW$ is isotropic if $\bfW \subseteq \bfW_{\perp}$. This is true if and only if $\omega$ restricts to $0$ on $\bfW$. \\

\noindent
$\bullet$  $\bfW$ is coisotropic if $\bfW_{\perp} \subseteq  \bfW$. 
$\bfW$ is coisotropic if and only if $\omega$ has a non-degenerate form on the quotient space $\bfW/\bfW_{\perp}$. 
Equivalently $\bfW$ is coisotropic if and only if $\bfW_{\perp}$ is isotropic.\\

\noindent
$\bullet$  $\bfW$ is Lagrangian if and only if it is both isotropic and coisotropic, namely, 
 if and only if $\bfW = \bfW_{\perp}$. 
In a finite-dimensional $\bfV$, a Lagrangian subspace $\bfW$ is an isotropic one whose dimension is half that of $\bfV$. \\

With this understanding, following Ref.\onlinecite{Kapustin:2010hk},
we consider a U(1)$^N$ Chern-Simons theory, whose bulk action is 
$
S_{bluk}=\frac{K_{IJ}}{4\pi} \int_{\mathcal{M}}  a_I \wedge d a_J.
$
and the boundary action for a manifold $\cM$ with a boundary ${\partial \cM}$ (with the restricted $a_{\parallel,I}$ on ${\partial \cM}$ ) is
$
S_{\partial}=\delta S_{bluk}=\frac{K_{IJ}}{4\pi} \int_{ \cM}  (\delta a_{\parallel,I}) \wedge d a_{\parallel,J}.
$
The symplectic form $\omega$ is given by the $K$-matrix via the differential of this 1-form $\delta S_{bluk}$
$
\omega=\frac{K_{IJ}}{4\pi} \int_{ \cM}  (\delta a_{\parallel,I}) \wedge d (\delta a_{\parallel,J}).
$

The gauge group U(1)$^N$ can be viewed as the torus $\mathbb{T}_\Lambda$, as the quotient space of $N$-dimensional vector space $\bfV$ by a subgroup $\Lambda \cong \Z^N$. Namely,
$
\text{U(1)}^N  \cong \mathbb{T}_\Lambda \cong (\Lambda \otimes \mathbb{R})/(2 \pi \Lambda) \equiv  \mathbf{t}_\Lambda/(2 \pi \Lambda).
$
Locally the gauge field $a$ is a 1-form,
which has values in the Lie algebra of $\mathbb{T}_\Lambda$, we will denote this Lie algebra $\mathbf{t}_\Lambda$ as the vector space $\mathbf{t}_\Lambda =\Lambda \otimes \mathbb{R}$.

A self-consistent {\bf boundary condition} must %
define a Lagrangian submanifold with
respect to this symplectic form $\omega$ and must be local. 
(For example, the famous chiral boson theory has $a_{\bar{z}}=0$ along the complex coordinate $\bar{z}$. This 
defines a 
consistent boundary condition, but it is gapless not topologically gapped.)

In addition, a {\bf topological boundary gapping condition} must be invariant
in respect of
the orientation-preserving diffeomorphism of $\cM$.
A local diffeomorphism-invariant constraint on the Lie algebra $\mathbf{t}_\Lambda$-valued 1-form $a_{\parallel,I}$ demands it to 
live in the subspace of $\mathbf{t}_\Lambda$. 
This corresponds to the if and only if conditions that: \\

\noindent
$\bullet(i)$  The subspace is isotropic with respect to the symmetric bilinear form $K$. \\
\noindent 
$\bullet(ii)$  The subspace dimension is a half of the dimension of $\mathbf{t}_\Lambda$. \\
\noindent 
$\bullet(iii)$ The signature of $K$ is zero. This means that $K$ has the same number of positive and negative eigenvalues.\\

We notice that $\bullet(ii)$ is true for our boundary gapping lattice, $\mathbf{L} \equiv \Big(\ell_{1}, \ell_{2}, \dots,  \ell_{N/2}  \Big)$, where there are $N/2$-linear independent gapping terms.
And $\bullet(iii)$ is true for our bosonic $K_{b0}$ and fermionic $K_{f}$ matrices. Importantly, for {\bf topological gapped boundary conditions},
$a_{\parallel,I}$ lies in a Lagrangian subspace of $\mathbf{t}_\Lambda$ implies that the {\bf boundary gauge group}
is a {\bf Lagrangian subgroup}.
(Here we consider the boundary gauge group is connected and continuous; one can read Section 6 of Ref.\onlinecite{Kapustin:2010hk} 
for the case of more general disconnected or discrete boundary gauge group.)

The {\bf bulk gauge group} is $\mathbb{T}_\Lambda$, 
and we denote the {\bf boundary gauge group} as $\mathbb{T}_{\Lambda_0}$, which $\mathbb{T}_{\Lambda_0}$ is a {\bf Lagrangian subgroup} of $\mathbb{T}_\Lambda$
for {\bf topological gapped boundary conditions}.

Here the torus $\mathbb{T}_\Lambda$ can be decomposed into a product of 
$\mathbb{T}_{\Lambda_0}$ and other torus.
$\Lambda \cong \Z^N$ contains the subgroup $\Lambda_0$, and $\Lambda$ contains a direct sum of $\Lambda_0$. 
These form an exact sequence:
$
0 \to \Lambda_0 \overset{\mathbf{h}}{\to} \Lambda \to \Lambda/\Lambda_0 \to 0.
$
Here $0$ means the trivial Abelian group with only the identity,
or the zero-dimensional vector space.
The exact sequence means that a sequence of maps $\text{f}_i$ from domain $A_{i}$ to $A_{i+1}$:
$
\text{f}_i: A_{i} \to A_{i+1}
$
satisfies a relation between the image and the kernel:
$
\text{Im}(f_{i}) = \text{Ker}(f_{i+1}).
$

Here we have $\mathbf{h}$ as an injective map from $\Lambda_0$ to $\Lambda$: 
$\Lambda_0 \overset{\mathbf{h}}{\to} \Lambda.$
Since $\Lambda$ is a rank-$N$ integer matrix generating a $N$-dimensional vector space, 
and $\Lambda_0$ is a rank-$N/2$ integer matrix generating a $N/2$-dimensional vector space; 
we have $\mathbf{h}$ as an integral matrix of $N \times (N/2)$-components.

The transpose matrix $\mathbf{h}^T$ is an integral matrix of $(N/2) \times N$-components. $\mathbf{h}^T$ is a surjective map: 
$\Lambda^* \overset{\mathbf{h}^T}{\to} \Lambda_0^*.$
Some mathematical relations are 
$\Lambda_0=H_1(\mathbb{T}_{\Lambda_0},\Z)$,
$\Hom(\mathbb{T}_{\Lambda_0}, \U(1))=\Lambda_0^*$ and
$\Hom(\mathbb{T}_{\Lambda}, \U(1))=\Lambda^*$.
Here $H_1(\mathbb{T}_{\Lambda_0},\Z)$ is the first homology group of $\mathbb{T}_{\Lambda_0}$ with a $\Z$ coefficient.
$\Hom(X,Y)$ is the set of all module homomorphisms from the module $X$ to the module $Y$.

Furthermore, for $\mathbf{t}_{\Lambda}^*$ being the dual of the Lie algebra $\mathbf{t}_{\Lambda}$, 
one can properly define the Topological Boundary 
Conditions by restricting the values of boundary gauge fields (taking values in Lie algebra $\mathbf{t}_{\Lambda}^*$ or $\mathbf{t}_{\Lambda}$), and 
one can obtain the corresponding exact sequence by choosing the following splitting of the vector space $\mathbf{t}_{\Lambda}^*$:\cite{Kapustin:2010hk}
$
0 \to \mathbf{t}_{(\Lambda/\Lambda_0)}^* \to  \mathbf{t}_{\Lambda}^* \to \mathbf{t}_{\Lambda_0}^* \to  0.
$



In Sec.\ref{sec:topo-nonp-proof}, these above conditions $\bullet(i)$,$\bullet(ii)$,$\bullet(iii)$ are shown to be equivalent to the {\bf boundary full gapping rules}.
To summarize, in this subsection, we provide a third approach from a non-Perturbative TQFT viewpoint 
to prove that, for  $\U(1)^N$-Chern-Simons theory, {\bf Topological Boundary 
Conditions} hold \emph{if and only if} the
boundary interaction terms satisfy {\bf Topological Boundary Fully Gapping Rules}.(Q.E.D.)

\color{black}

\section{More about Our Lattice Hamiltonian Chiral Matter Models} \label{AppendixF} 

\subsection{More details on our Lattice Model producing nearly-flat Chern-bands} \label{sec:App-Chern-band}

We fill more details on our lattice model presented in Sec.\ref{numeric} for the free-kinetic part.
The lattice model shown in Fig.\ref{chiral-pi-flux} has two sublattices: $a$ (black dots) and $b$ (white dots). In momentum space, we have a generic pseudospin form of Hamiltonian $H(\mathbf{k})$,
\be
H(\mathbf{k})=B_0(\mathbf{k}) + \vec{B}(\mathbf{k}) \cdot \vec{\sigma}. 
\ee
$\vec{\sigma}$ are Pauli matrices $(\sigma_x,\sigma_y,\sigma_z)$. In this model $B_0(\mathbf{k})=0$ and
$\vec{B}$ have three components:
\bea
B_x(\mathbf{k})&=&2 t_1 \cos (\pi /4) (\cos(k_x a_x)+\cos ( k_y a_y)) \nonumber \\
B_y(\mathbf{k})&=& 2 t_1 \sin (\pi /4) (\cos(k_x a_x)-\cos ( k_y a_y)) \\
B_z(\mathbf{k})&=&-4 t_2 \sin ( k_x a_x ) \sin (k_y  a_y). \nonumber
\eea

In Fig.\ref{kx_chiral}(a), the energy spectrum $\E(k_x)$ is solved from putting the system on a 10-sites width ($9a_y$-width) cylinder.
Indeed the energy spectrum $\E(k_x)$ in Fig.\ref{kx_chiral}(b) is as good when putting on a smaller size system such as the ladder (Fig.\ref{chiral-pi-flux}(c)).
The cylinder is periodic along $\hat{x}$ direction so $k_x$ momentum is a quantum number, while $\E(k_x)$ has real-space $y$-dependence along the finite-width $\hat{y}$ direction.
Each band of $\E(k_x)$ in Fig.\ref{kx_chiral} is solved by exactly diagonalizing $H(k_x,y)$ with $y$-dependence.
By filling the lower energy bands and setting the chemical potential at zero, we have Dirac fermion dispersion at $k_x=\pm\pi$ for the edge state spectrum, shown as the blue curves in Fig.\ref{kx_chiral}(a)(b).

In Fig.\ref{kx_chiral}(c), we plot the density $\langle f^\dagger f \rangle$ of the edge eigenstate on the ladder (which eigenstate is the solid blue curve in Fig.\ref{kx_chiral}(b)),
for each of two edges A and B, and for each of two sublattice $a$ and $b$.
One can fine tune $t_2/t_1$ such that the edge A and the edge B have the least mixing.
The least mixing implies that the left edge and right edge states nearly decouple.
The least mixing is very important for the interacting $G_1,G_2 \neq 0$ case, so we can impose interaction terms on the right edge B only as in Eq.(\ref{H3540}), decoupling from the edge A.
We can explicitly make the left edge A density $\langle f_{\A}^\dagger f_{\A} \rangle$ dominantly locates in $k_x<0$,
the right edge B density $\langle f_{\B}^\dagger f_{\B} \rangle$ dominantly locates in $k_x>0$.
The least mixing means the eigenstate is close to the form $| \psi (k_x) \rangle = | \psi_{k_x<0} \rangle_A  \otimes | \psi_ {k_x>0} \rangle_B$.
The fine-tuning is done with $t_2/t_1=1/2$ in our case.
Interpret this result together with Fig.\ref{kx_chiral}(b),
we see the solid blue curve at $k_x<0$ has negative velocity along $\hat{x}$ direction,
and at $k_x>0$ has positive velocity along $\hat{x}$ direction.
Overall it implies the chirality of the edge state on the left edge A moving along $-\hat{x}$ direction, and on the right edge B moving along $+\hat{x}$ direction 
--- the clockwise chirality as in Fig.\ref{chiral-pi-flux}(b), consistent with the earlier result
$C_{1,-}=-1$ of Chern number.}

An additional bonus for this ladder model is that the density $\langle f^\dagger f \rangle$ distributes equally on two sublattice $a$ and $b$ on either edges, shown in Fig.\ref{kx_chiral}(c).
Thus, it will be beneficial for the interacting model in Eq.(\ref{H3540}) when turning on interaction terms $G_1,G_2\neq 0$,
we can universally add the same interaction terms for both sublattice $a$ and $b$.

For the free kinetic theory, all of the above can be achieved by a simple ladder lattice, which is effectively as good as 1+1D because of finite size width. 
To have mirror sector becomes gapped and decoupled without interfering with the gapless sector, we propose to design the lattice with length scales of Eq.(\ref{eq:energyscale}). 

\subsection{Explicit lattice chiral matter models} \label{app:more_models} 
For model constructions, we will follow the four steps introduced earlier in Sec.\ref{model}.

\subsubsection{1$_L$-(-1$_R$) chiral fermion model \label{1-(-1)model}} 

The most simplest model of fermionic model 
suitable for our purpose is, {\bf Step 1},
$K^f_{2 \times 2}=({\begin{smallmatrix}
1 &0 \\
0 & -1
\end{smallmatrix}} )$ in Eq.(\ref{CSbulk}),(\ref{CSboundary}). We can choose, {\bf Step 2},
 $\mathbf{t}=(1,-1)$, so this model satisfies Eq.(\ref{Anomalyf}) as anomaly-free. 
It also satisfies 
 the total U(1) charge chirality $\sum q_{L} -\sum q_{R} =2 \neq 0$ as {\bf Step 3}.
As {\bf Step 4}, we can fully gap out one-side of edge states by a gapping term Eq.(\ref{eq:Sgap}) with $\ell_a=(1,1)$, which preserves U(1) symmetry by Eq.(\ref{eq:gapsym}).
Written in terms of $\mathbf{t}$ and $\mathbf{L}$ matrices: 
\be
\mathbf{t}=\left(
\begin{array}{cc}
 1 \\
 -1 
\end{array}
\right)
\Longleftrightarrow
\mathbf{L}=\left(
\begin{array}{cc}
1 \\
 1 
\end{array}
\right).
\ee
Through its U(1) charge assignment $\mathbf{t}=(1,-1)$,
we name this model as 1$_L$-(-1$_R$) chiral fermion model.
It is worthwhile to go through this 1$_L$-(-1$_R$) chiral fermion model in more details, where its bosonized low energy action is 
\bea 
S_{\Phi} 
&=&\frac{1}{4\pi}  \int \dd t \dd x  \big(K^{\A}_{IJ}  \partial_t \Phi^{\A}_I   \partial_x \Phi^{\A}_{J} -V_{IJ}  \partial_x \Phi^{\A}_I   \partial_x \Phi^{\A}_{J}\big)\;\;\;\nonumber \\
&+&\frac{1}{4\pi}  \int \dd t \dd x \big(K^{\B}_{IJ}  \partial_t \Phi^{\B}_I   \partial_x \Phi^{\B}_{J} -V_{IJ}  \partial_x \Phi^{\B}_I   \partial_x \Phi^{\B}_{J} \big) \label{b1-1}        \;\;\;\nonumber\\
&+&\int \dd t \dd x  \; g_{1}  \cos( \Phi^{\B}_{1}+\Phi^{\B}_{-1}). \;\;\;\;\;\;\;
 \eea
Its fermionized action (following the notation as Eq.(\ref{Lf3-5-4-0}), {with a \emph{relevant} interaction term of $g_1$ coupling) is} 
\bea
&S_\Psi&=\int  \dd t \; \dd x \; \bigg( \ti\bar{\Psi}_{\A} \Gamma^\mu  \partial_\mu \Psi_{\A}+\ti\bar{\Psi}_{\B} \Gamma^\mu  \partial_\mu \Psi_{\B}  \nonumber  \label{} \\
 &\;&+\tilde{g}_{1} \big( \tilde{\psi}_{R,1}  \tilde{\psi}_{L,-1}  +\text{h.c.} \big) \bigg).   
\eea
We propose that a lattice Hamiltonian below (analogue to 
Fig.\ref{3540}'s) realizes this 1$_L$-(-1$_R$) chiral fermions theory non-perturbatively, 
\bea 
\label{H1-1}
H&=&\sum_{q=1,-1} 
\bigg(  \sum_{\langle i, j \rangle}
\big(t_{ij,q}\; \hat{f}^\dagger_{q}(i)
\hat{f}_{q}(j)+\text{h.c.}\big) \\
&+& \sum_{\langle\langle i, j
\rangle\rangle} \big(  t'_{ij,q}
\;\hat{f}^\dagger_{q}(i) \hat{f}_{q}(j)+\text{h.c.}\big) \bigg) \nonumber
\\ 
&+& G_{1} \sum_{j \in \B} \bigg(
\big(\hat{f}_1(j)_{pt.s.}\big)
\big(\hat{f}_{-1}(j)_{pt.s.}\big)
+\text{h.c.}
\bigg).\;\; 
\nonumber 
\eea
This Hamiltonian is in a perfect quadratic form. 
We can solve it exactly by writing down Bogoliubov-de Gennes
(BdG) Hamiltonian in the Nambu space form, on a cylinder (in Fig.\ref{3540}),
\be 
H=\frac{1}{2}\sum_{k_{x},p_{x}} (f^{\dagger},f)
 {\begin{pmatrix} 
H_{\text{kinetic}} &  \mathcal{G}^\dagger(k_x,p_x)\\
\mathcal{G}(k_x,p_x) &-H_{\text{kinetic}}  
\end{pmatrix}}
{\begin{pmatrix} 
f^{}\\
f^{\dagger}
\end{pmatrix}}.
\ee
Here $f^\dagger= (f^{\dagger}_{1,k_x},f^{\dagger}_{-1,p_x})$, $f=( f^{}_{1,k_x},f^{}_{-1,p_x})$, while $H_{\text{kinetic}}$ is the hopping term and $\mathcal{G}$ is from the $G_1$ interaction term. Here momentum $k_x, p_x$ (for charge 1 and -1 fermions) along the compact direction x are good quantum numbers. 
Along the non-compact y direction, we use the real space basis instead. We diagonalize this BdG Hamiltonian exactly and find out the edge modes on the right edge B become fully gapped at large 
$G_1$. For example, at $ |G_1| \simeq  10^4$, the edge state density on the edge B is $\langle f^\dagger_B f_B \rangle \leq 5 \times 10^{-8}$.\cite{JWunpublished} We also check that the low energy spectrum realizes the 1-(-1) chiral fermions on the left edge A,\cite{JWunpublished}
\bea
\label{cf2}
S_{\Psi_{\A},free}&=&\int  \dd t\; \dd x \; \Big(
\ti \psi^\dagger_{L,1} (\partial_t-\partial_x) \psi_{L,1} \nonumber \\
&+& \ti \psi^\dagger_{R,-1} (\partial_t+\partial_x) \psi_{R,-1} 
\Big).
\eea
Thus Eq.(\ref{H1-1}) defines/realizes 1$_L$-(-1$_R$) chiral fermions non-perturbatively on the lattice. 

The 1$_L$-(-1$_R$) chiral fermion model provides a wonderful example that we can confirm, both numerically and analytically, the mirror fermion idea and our model construction will work. 

However, unfortunately the {\bf 1$_L$-(-1$_R$) chiral fermion model} is {\it not} strictly a chiral theory. In a sense that one can do a field redefinition,
$$
\psi_{1} \to \psi_{1},\;\; \text{and} \;\;\psi_{-1} \to \psi_{1'}^\dagger, 
$$ 
sending the charge vector $\mathbf{t}=(1,-1) \to (1,1)$. So the model becomes a {\bf 1$_L$-1$_R$ fermion model} 
with one left moving mode and one right moving mode both carry the same U(1) charge 1. Here we use  $\psi_{1'}$ to indicate another fermion field carries the same U(1) charge as  $\psi_{1}$.
The 1$_L$-1$_R$ fermion model is obviously a non-chiral Dirac fermion theory, where the mirror edge states can be gapped out by forward scattering \emph{mass terms} $\tilde{g}_{1} \big( \tilde{\psi}_{R,1}  \tilde{\psi}_{L,1'}^\dagger  +\text{h.c.} \big)$,
or the $g_{1}  \cos( \Phi^{\B}_{1}-\Phi^{\B}_{1'})$ term in the bosonized language. 
Since 1$_L$-(-1$_R$) chiral fermion model is a field-redefinition of 1$_L$-1$_R$ fermion model, 
it becomes apparent that we can gap out the mirror edge of 1$_L$-(-1$_R$) chiral fermion model. 

It turns out that the next simplest U(1)-symmetry chiral fermion model, 
which violates parity 
and time reversal symmetry 
 (but strictly being chiral under any field redefinition),
is the 3$_L$-5$_R$-4$_L$-0$_R$ chiral fermion model, appeared already in Sec.\ref{sec3-5-4-0}. 


\subsubsection{3$_L$-5$_R$-4$_L$-0$_R$ chiral fermion model and others} \label{sec-app-3540}

We consider a rank-4 $K^f_{4 \times 4}=({\begin{smallmatrix}
1 &0 \\
0 & -1
\end{smallmatrix}} ) \oplus ({\begin{smallmatrix}
1 &0 \\
0 & -1
\end{smallmatrix}} )$ in Eq.(\ref{CSbulk}),(\ref{CSboundary}) for {\bf Step 1}. We can choose $\mathbf{t}_a=(3,5,4,0)$ to construct a 3$_L$-5$_R$-4$_L$-0$_R$ chiral fermion model
in Sec.\ref{sec3-5-4-0} for {\bf Step 2}.
One can choose the gapping terms in Eq.(\ref{eq:Sgap}) with $\ell_a=(3,-5,4,0), \ell_b=(0,4,-5,3)$.
Another U(1)$_{\text{2nd}}$ symmetry is allowed, which is $\mathbf{t}_b=(0,4,5,3)$.
By writing down the chiral boson theory 
of Eq.(\ref{CSboundary}), (\ref{eq:Sgap}) on a cylinder with two edges A and B as in Fig.\ref{3540}, it becomes a multiplet chiral boson theory with an action 
\begin{widetext}
\bea \label{b3540app}
S_{\Phi}=S_{\Phi^{\A}_{free}}+S_{\Phi^{\B}_{free}}+S_{\Phi^{\B}_{interact}}=&&
\frac{1}{4\pi}  \int \dd t \dd x  \big(K^{\A}_{IJ}  \partial_t \Phi^{\A}_I   \partial_x \Phi^{\A}_{J} -V_{IJ}  \partial_x \Phi^{\A}_I   \partial_x \Phi^{\A}_{J}\big)+\big(K^{\B}_{IJ}  \partial_t \Phi^{\B}_I   \partial_x \Phi^{\B}_{J} -V_{IJ}  \partial_x \Phi^{\B}_I   \partial_x \Phi^{\B}_{J} \big) \label{b3540App}        \;\;\;\nonumber\\
&& +\int \dd t \dd x  \bigg( g_{1}  \cos( 3\Phi^{\B}_{3}-5\Phi^{\B}_{5}+4\Phi^{\B}_{4})+  g_{2}  \cos( 4\Phi^{\B}_{5}-5\Phi^{\B}_{4}+3\Phi^{\B}_{0}) \bigg). \;\;\;\;\;\;\;
 \eea
After fermionizing 
Eq.(\ref{b3540}) by
$\Psi \sim e^{i \Phi}$,
we match it to Eq.(\ref{Lf3-5-4-0}).\cite{fermionization2}
\bea \label{Lf3-5-4-0app}
&S_\Psi&=S_{\Psi_{\A},free}+S_{\Psi_{\B},free}+S_{\Psi_{\B},interact}=\int  \dd t \; \dd x \; \bigg( \ti \bar{\Psi}_{\A} \Gamma^\mu  \partial_\mu \Psi_{\A}+ \ti \bar{\Psi}_{\B} \Gamma^\mu  \partial_\mu \Psi_{\B}    \label{Lf3-5-4-0App} \\
 &\;&+\tilde{g}_{1} \big( (\tilde{\psi}_{R,3} \nabla_x \tilde{\psi}_{R,3} \nabla^2_x\tilde{\psi}_{R,3})  (\tilde{\psi}_{L,5}^\dagger  \nabla_x \tilde{\psi}_{L,5}^\dagger \nabla^2_x \tilde{\psi}_{L,5}^\dagger \nabla^3_x \tilde{\psi}_{L,5}^\dagger \nabla^4_x \tilde{\psi}_{L,5}^\dagger)
 ( \tilde{\psi}_{R,4} \nabla_x \tilde{\psi}_{R,4}  \nabla^2_x \tilde{\psi}_{R,4} \nabla^3_x \tilde{\psi}_{R,4} ) +\text{h.c.} \big)  \nonumber \\
&\;& + \tilde{g}_{2}   \big( (\tilde{\psi}_{L,5} \nabla_x \tilde{\psi}_{L,5}  \nabla^2_x \tilde{\psi}_{L,5} \nabla^3_x \tilde{\psi}_{L,5})  (\tilde{\psi}_{R,4}^\dagger  \nabla_x \tilde{\psi}_{R,4}^\dagger \nabla^2_x \tilde{\psi}_{R,4}^\dagger  \nabla^3_x \tilde{\psi}_{R,4}^\dagger \nabla^4_x \tilde{\psi}_{R,4}^\dagger )( \tilde{\psi}_{L,0} \nabla_x \tilde{\psi}_{L,0} \nabla^2_x \tilde{\psi}_{L,0} )+\text{h.c.} \big)\bigg), \nonumber
\eea

Our 3-5-4-0 fermion model satisfies Eq.(\ref{eq:gapsym}), Eq.(\ref{Anomalyf}) and boundary fully gapping rules, 
and also violates parity and time-reversal symmetry, 
so the lattice version of the Hamiltonian 
\bea 
\label{H3540App}
H&=&\sum_{q=3,5,4,0} 
\bigg(  \sum_{\langle i, j \rangle}
\big(t_{ij,q}\; \hat{f}^\dagger_{q}(i)
\hat{f}_{q}(j)+\text{h.c.}\big) + \sum_{\langle\langle i, j
\rangle\rangle} \big(  t'_{ij,q}
\;\hat{f}^\dagger_{q}(i) \hat{f}_{q}(j)+\text{h.c.}\big) \bigg)
\\
&+& G_{1} \sum_{j \in \B} \bigg(
\big(\hat{f}_3(j)_{pt.s.}\big)^3
\big(\hat{f}^\dagger_5(j)_{pt.s.}\big)^5
\big(\hat{f}_4(j)_{pt.s.}\big)^4
+\text{h.c.} \bigg) + G_{2} \sum_{j \in \B} \bigg(
\big(\hat{f}_5(j)_{pt.s.}\big)^4
\big(\hat{f}^\dagger_4(j)_{pt.s.}\big)^5
\big(\hat{f}_0(j)_{pt.s.}\big)^3
+\text{h.c.}
\bigg),\;\;
\nonumber
\eea
\end{widetext}
provides a non-perturbative anomaly-free chiral fermion model on the gapless edge A when putting on the lattice.
We notice that the choices of gapping terms 
with $\ell_a=(3,-5,4,0), \ell_b=(0,4,-5,3)$ of the model
in Eq.(\ref{b3540app}),(\ref{Lf3-5-4-0app}),(\ref{H3540App}) here are distinct from the version of gapping terms $\ell_a=(1,1,-2,2)$, $\ell_b=(2,-2,1,1)$ of the model 
Eq.(\ref{Lf3-5-4-0}),
(\ref{b3540}),
(\ref{H3540}) in the main text. 
This is rooted in the \emph{different} choice of basis for the \emph{same} vector space of column vectors of $\mathbf{L},\mathbf{t}$ matrices, and the dual structure shown in Eq.(\ref{eq:Lt3540dual}).
Both ways (or other linear-independent linear combinations) will produce a 
3$_L$-5$_R$-4$_L$-0$_R$ model.

In Sec.\ref{app:approachII}, we outline that our anomaly-free chiral model can be mapped to decoupled 
Luttinger liquids of Eq.(\ref{eq:E-4}). Here let us explicitly find out the outcomes of mapping.
Based on the Smith normal form $\mathbf{L}= V D W$ shown in Sec.\ref{app:approachII}, we can rewrite the gapping term matrices $\mathbf{L}$. 
From Eq.(\ref{eq:cosine-map}),
the original cosine term $g_{a}  \cos(\ell_{a,I}^{} \cdot\Phi_{I})$ in the old basis will be mapped to
$g_a\cos( W_{Ja} d_{J} \bar{\theta}_{J})$. Namely, given the model of Eq.(\ref{b3540app}), 
\bea
&&\left(
\begin{array}{cc}
 3 & 0 \\
 -5 & 4 \\
 4 & -5 \\
 0 & 3
\end{array}
\right)=
\left(
\begin{array}{cccc}
 3 & -1 & 0 & 1 \\
 -5 & 3 & 0 & -2 \\
 4 & -3 & 1 & 2 \\
 0 & 1 & 0 & 0
\end{array}
\right).\left(
\begin{array}{cc}
 1 & 0 \\
 0 & 3 \\
 0 & 0 \\
 0 & 0
\end{array}
\right).\left(
\begin{array}{cc}
 1 & 1 \\
 0 & 1
\end{array}
\right)\nonumber\\
&& 
\Rightarrow  g_a\cos( \bar{\theta}_{1})+g_b\cos(\bar{\theta}_{1}+ 3\bar{\theta}_{2}). 
\eea
On the other hand, given the model of Eq.(\ref{H3540}), we have
\bea
&& \left(
\begin{array}{cc}
 1 & 2 \\
 1 & -2 \\
 -2 & 1 \\
 2 & 1
\end{array}
\right)=
\left(
\begin{array}{cccc}
 1 & 2 & 0 & -1 \\
 1 & -2 & 0 & 0 \\
 -2 & 1 & 1 & 1 \\
 2 & 1 & 0 & -1
\end{array}
\right).\left(
\begin{array}{cc}
 1 & 0 \\
 0 & 1 \\
 0 & 0 \\
 0 & 0
\end{array}
\right).\left(
\begin{array}{cc}
 1 & 0 \\
 0 & 1
\end{array}
\right) \nonumber\\ 
&&\Rightarrow g_a\cos(  \bar{\theta}_{1})+g_b\cos( \bar{\theta}_{2}).
\eea

One reason that we choose Eq.(\ref{H3540}) for our model in the main text, instead of Eq.(\ref{H3540App})
is that the interaction terms for the model of Eq.(\ref{H3540}) has the order of 6-body interaction among each gapping term, which is easier to simulate
than the 12-body interaction among each gapping term for the model of Eq.(\ref{H3540App}).


  We list down another three similar chiral fermion models of $K^f_{4 \times 4}$ matrix, with different choices of $\mathbf{t}$, such as:\\
\noindent
{\bf (i)} 1$_L$-5$_R$-7$_L$-5$_R$ chiral fermions: 
$\mathbf{t}_a=(1,5,7,5)$, $\mathbf{t}_a=(0,3,5,4)$, with gapping terms
$\ell_a=(1, -5, 7, -5)$, $\ell_b=(0, 3, -5, 4)$.  Written in terms of $\mathbf{t}$ and $\mathbf{L}$ matrices: 
\be
\mathbf{t}=\left(
\begin{array}{cc}
 1 & 0 \\
 5 & 3 \\
 7 & 5 \\
 5 & 4
\end{array}
\right)
\Longleftrightarrow
\mathbf{L}=\left(
\begin{array}{cc}
 1 & 0 \\
 -5 & 3 \\
 7 & -5 \\
 -5 & 4
\end{array}
\right).
\ee
\noindent
{\bf (ii)} 1$_L$-4$_R$-8$_L$-7$_R$ chiral fermions: $\mathbf{t}_a=(1,4,8,7)$, $\mathbf{t}_b=(3, -3, -1, 1)$, with gapping terms
$\ell_a=(1, -4, 8, -7)$, $\ell_b=(3, 3, -1, -1)$.
\be
\mathbf{t}=\left(
\begin{array}{cc}
 1 & 3 \\
 4 & -3 \\
 8 & -1\\
 7 & 1
\end{array}
\right)
\Longleftrightarrow
\mathbf{L}=\left(
\begin{array}{cc}
 1 & 3 \\
 -4 & 3 \\
 8 & -1\\
 -7 & -1
\end{array}
\right).
\ee
\noindent
{\bf (iii)} 2$_L$-6$_R$-9$_L$-7$_R$ chiral fermions: $\mathbf{t}_a=(2,6,9,7)$, $\mathbf{t}_b=(2, -2, -1, 1)$ with gapping terms
$\ell_a=(2, -6, 9, -7)$, $\ell_b=(2, 2, -1, -1)$.
\be
\mathbf{t}=\left(
\begin{array}{cc}
 2 & 2 \\
 6 & -2 \\
 9 & -1\\
 7 & 1
\end{array}
\right)
\Longleftrightarrow
\mathbf{L}=\left(
\begin{array}{cc}
 2 & 2 \\
 -6 & 2 \\
 9 & -1\\
 -7 & -1
\end{array}
\right).
\ee
Indeed, there are infinite many possible models just for $K^f_{4 \times 4}$ matrix-Chern Simons theory construction. One can also construct a higher rank $K^f$ theory with infinite more models
of U(1)$^{N/2}$-anomaly-free chiral fermions.

\noindent
\subsubsection{Chiral boson model} 

Similar to fermionic systems, we will follow the four steps introduced earlier for bosonic systems.
The most simple model of bosonic SPT suitable for our purpose is, {\bf Step 1}, $K^b_{2 \times 2}=({\begin{smallmatrix}
0 &1 \\
1 & 0
\end{smallmatrix}} )$ in Eq.(\ref{CSbulk}),(\ref{CSboundary}). We can choose, {\bf Step 2}, $\mathbf{t}=(1,0)$,
so this model satisfies Eq.(\ref{Anomalyb}) as anomaly-free, and violates parity and time-reversal symmetry as  {\bf Step 3}.  
As {\bf Step 4}, we can fully gap out one-side of edge states by gapping term Eq.(\ref{eq:Sgap}) with $\ell_a=(0,1)$, which preserves U(1) symmetry by Eq.(\ref{eq:gapsym}).
Written in terms of $\mathbf{t}$ and $\mathbf{L}$ matrices: 
\be
\mathbf{t}=\left(
\begin{array}{cc}
 1 \\
 0 
\end{array}
\right)
\Longleftrightarrow
\mathbf{L}=\left(
\begin{array}{cc}
0 \\
 1 
\end{array}
\right).
\ee

For $K^{b0}_{4 \times 4}=({\begin{smallmatrix}
0 &1 \\
1 & 0
\end{smallmatrix}} )\oplus ({\begin{smallmatrix}
0 &1 \\
1 & 0
\end{smallmatrix}} )$, we list down two models:\\
{\bf (i)} 2$_L$-2$_R$-4$_L$-$(-1)_R$ chiral bosons: $\mathbf{t}_a=(2,2,4,-1)$, $\mathbf{t}_b=(0, 2, 0, -1)$ with gapping terms
$\ell_a=(2,2,-1,4)$, $\ell_b=(2, 0, -1, 0)$.
\be
\mathbf{t}=\left(
\begin{array}{cc}
 2 & 0 \\
 2 & 2 \\
 4 & 0\\
 -1 & -1
\end{array}
\right)
\Longleftrightarrow
\mathbf{L}=\left(
\begin{array}{cc}
 2 & 2 \\
 2 & 0 \\
 -1 & -1\\
 4 & 0
\end{array}
\right).
\ee
{\bf (ii)} 6$_L$-6$_R$-9$_L$-$(-4)_R$ chiral bosons:  $\mathbf{t}_a=(6,6,9,-4)$, $\mathbf{t}_b=(0, 3, 0, -2)$ with gapping terms
$\ell_a=(6,6,-4,9)$, $\ell_b=(3, 0, -2, 0)$.
\be
\mathbf{t}=\left(
\begin{array}{cc}
 6 & 0 \\
 6 & 3 \\
 9 & 0\\
 -4 & -2
\end{array}
\right)
\Longleftrightarrow
\mathbf{L}=\left(
\begin{array}{cc}
 6 & 3 \\
 6 & 0 \\
-4 & -2\\
 9 & 0
\end{array}
\right).
\ee
Infinite many chiral boson models can be constructed in the similar manner.

\color{black}

\onecolumngrid


\begin{thebibliography}{99}
\bibitem{WZW}
In the case of $K_{N\times N}$ matrix Chern-Simons theory with U(1)$^N$ symmetry, the Wess-Zumino-Witten model reduces to a $K$ matrix chiral bosons theory.
We like to clarify that the $K$ matrix chiral bosons are the bosonic phases field $\Phi$ in the bosonization method.
The $K$ matrix chiral bosons does not conflict with the chiral fermions(or other chiral matters) model we have in 1+1D.
The matter field content $\Psi$ of edge theory is $\Psi \sim :e^{{\rm i} \Phi_I K_{IJ} L_J}:$, with some integer vector $L_J$, where $\Psi$ is
what we mean by the chiral matter field content in our non-perturbative anomaly-free chiral matter model.
\bibitem{fermionization1}
By doing fermionization or bosonization, one can recover the chiral matter content in the field theory language, roughly 
$\psi_I \sim :e^{{\rm i} K_{IJ}\phi_J}:$.
One can construct the lattice model by adding several layers of the zeroth Landau levels (precisely, several layers of the first Chern bands), as described in Sec.\ref{sec3-5-4-0},\ref{numeric}.
\bibitem{particle lattice}
Here 
the non-fractionalized $\Gamma_e$ means that the composition of quasi-particles forms non-fractionalized physical particles. 
\end{thebibliography}
\end{document}